\documentclass[english]{article}
\usepackage[utf8]{inputenc}
\usepackage[T1]{fontenc}
\usepackage{lmodern}
\usepackage[a4paper]{geometry}
\usepackage[english]{babel}
\usepackage{csquotes}
\usepackage{tabularx,booktabs}

\usepackage{amsmath}
\usepackage{amssymb}
\usepackage{stmaryrd}
\usepackage{empheq}
\usepackage{bm}
\usepackage{slashed}
\usepackage{microtype}

\usepackage{mathrsfs}

\usepackage[dvipsnames]{xcolor}
\usepackage{soulutf8}
\newcommand{\com}[1]{\textcolor{red}{[\hl{#1}]}}
\renewcommand{\com}[1]{}

\usepackage{tikz-cd}

\newtheorem{lemma}{Lemma}

\newcommand{\setR}{\mathbb{R}}
\newcommand{\setZ}{\mathbb{Z}}
\newcommand{\setC}{\mathbb{C}}
\newcommand{\setK}{\mathbb{K}}
\newcommand{\setH}{\mathbb{H}}
\newcommand{\h}{{\mathfrak h}}
\newcommand{\bh}{{\bar{\h}}}

\newcommand{\balpha}{{\boldsymbol{\alpha}}}

\newcommand{\bi}{\mathbf{i}}
\newcommand{\ba}{\mathbf{a}}

\newcommand{\eAB}{{\epsilon^A_B}}

\newcommand{\Dirac}{\slashed{\nabla}}
\newcommand{\DirGD}{{\overleftrightarrow{\Dirac}}}
\newcommand{\nabGD}{{\overleftrightarrow{\nabla}}}
\renewcommand{\div}{\operatorname{div}}
\newcommand{\hwedge}{{\hat\wedge}}

\newcommand{\Sp}{\Sigma}
\newcommand{\bSp}{\Sp^*}
\newcommand{\bs}{{\bar s}}

\newcommand{\bp}{{\bar\psi}}
\newcommand{\bPh}{{\bar\Phi}}

\newcommand{\met}{{\eta}}

\newcommand{\Lie}{\mathcal L}

\newcommand{\thewc}{\Theta_{EC}}
\newcommand{\bthewc}{\overline{\thewc}}
\newcommand{\thdir}{\Theta_{Dirac}}
\newcommand{\bthdir}{\overline{\thdir}}
\newcommand{\thcons}{\Theta^{cons}}
\newcommand{\thconsdir}{\thcons_{Dirac}}
\newcommand{\thconsewc}{\thcons_{EC}}
\newcommand{\Lag}{\mathscr{L}}
\DeclareMathOperator{\EL}{EL}
\newcommand{\ELcons}{\EL^{cons}}
\newcommand{\Econs}{E^{cons}}

\newcommand{\ExT}{\mathsf{\Lambda}}

\renewcommand{\d}{{\operatorname d}} 
\newcommand{\dl}{\d^\lambda\!}

\newcommand{\dll}{\Lambda}
\newcommand{\dom}{\d^\omega}

\newcommand{\dd}{\partial}
\newcommand{\der}[2]{\frac{\partial#2}{\partial#1}}
\newcommand{\ddl}[2]{\dd_{\lambda^{#1}_#2}}
\newcommand{\ddp}[2]{\dd_{p^{#1}_#2}}
\newcommand{\lhuit}[1]{\lambda^{(8)}_{#1}}
\newcommand{\lneuf}[1]{\lambda^{(9)}_{#1}}
\newcommand{\lsept}[1]{\lambda^{(7)}_{#1}}
\newcommand{\ldix}{\lambda^{(10)}}
\newcommand{\lmtrois}[1]{\lambda^{\m(3)}}
\newcommand{\lmquatre}{\lambda^{\m(4)}}
\newcommand{\omsix}{\omega^{(6)}}
\newcommand{\aquatre}{\alpha^{(4)}}
\newcommand{\vsept}[1]{{\varpi^{(7)}_{#1}}}
\newcommand{\vhuit}[1]{{\varpi^{(8)}_{#1}}}
\newcommand{\vneuf}[1]{{\varpi^{(9)}_{#1}}}
\newcommand{\vdix}{{\varpi^{(10)}}}

\newcommand{\nonbc}[1]{{\widehat{#1}}}

\newcommand{\Glie}{\mathfrak g}

\newcommand{\lr}{\lrcorner\,}

\newcommand{\Jun}{\mathcal{J}^1}

\newcommand{\vol}{\mathrm{vol}}
\newcommand{\ve}{\varepsilon}

\newcommand{\lp}{\left(}
\newcommand{\rp}{\right)}

\newcommand{\wb}[2]{\left[#1\wedge#2 \right]}
\newcommand{\fwb}[2]{\frac{1}{2}\wb{#1}{#2}}

\newcommand{\U}{\mathcal U}

\DeclareMathOperator{\T}{T}
\DeclareMathOperator{\Riem}{Riem}
\DeclareMathOperator{\Ric}{Ric}
\DeclareMathOperator{\Scal}{Scal}

\DeclareMathOperator{\tr}{tr}
\DeclareMathOperator{\Ad}{Ad}
\DeclareMathOperator{\ad}{ad}
\DeclareMathOperator{\End}{End}
\DeclareMathOperator{\Iso}{Iso}
\DeclareMathOperator{\Mat}{M}
\DeclareMathOperator{\Spin}{Spin}
\DeclareMathOperator{\Pin}{Pin}

\DeclareMathOperator{\SL}{SL}

\DeclareMathOperator{\Hom}{Hom}
\DeclareMathOperator{\Stab}{Stab}

\DeclareMathOperator{\Id}{Id}

\renewcommand{\varphi}{\phi}
\newcommand{\tot}{\mathcal P}

\DeclareMathOperator{\SO}{SO}
\renewcommand{\O}{\operatorname{O}}
\newcommand{\SOp}{\SO^{+}}
\DeclareMathOperator{\SOpL}{{\SOp_{1,3}}}
\newcommand{\Spinp}{\Spin^+}
\renewcommand{\so}{\operatorname{\mathfrak{so}}}
\newcommand{\spin}{\operatorname{\mathfrak{spin}}}
\DeclareMathOperator{\soL}{{\so_{1,3}}}
\DeclareMathOperator{\Mink}{\m}
\DeclareMathOperator{\Cl}{Cl}
\newcommand{\Clpq}{\Cl_{p,q}}

\newcommand{\FrameB}{{\SOp(\ET)}}
\newcommand{\SpB}{{\Spinp(\ET)}}
\newcommand{\aux}{V}
\newcommand{\FrameV}{{\SOp(\aux)}}

\newcommand{\Lor}{\mathfrak{L}}
\renewcommand{\lor}{\mathfrak l}
\let\lbar\l
\renewcommand{\l}{\lor} 
\newcommand{\p}{\mathfrak p}
\newcommand{\m}{\mathfrak m}
\newcommand{\ET}{{\mathcal E}}

\DeclareMathOperator{\eucl}{{\mathfrak{iso}_4}}
\DeclareMathOperator{\soq}{{\mathfrak{so}_4}}

\DeclareMathOperator{\sopq}{\so_{p,q}}
\DeclareMathOperator{\Rpq}{\setR^{p,q}}
\newcommand{\Sppq}{\Sp_{p,q}}

\newcommand{\hTh}{\widehat{\Theta}}

\newcommand{\Isototp}{\Iso(T\tot,\p)}
\newcommand{\momQ}{\lp \m\wedge\l \oplus \l\wedge\l \rp \otimes \p^*}

\newcommand{\sprod}[2]{\langle #1 | #2 \rangle}

\newcommand{\SpX}{\underline{X}}
\newcommand{\uvpi}{\underline{\varpi}}
\renewcommand{\emptyset}{\varnothing}

\newcommand{\lel}[3]{#1\leqslant #2 \leqslant #3}

\newcommand{\Ax}{\mathcal{A}}

\usepackage{breakurl}
\usepackage[style=alphabetic, isbn=false, doi=false]{biblatex}
\title{Dirac Spinors on Generalised Frame Bundles : a frame bundle formulation for Einstein-Cartan-Dirac theory}
\author{Jérémie Pierard de Maujouy
	\thanks{ \textit{Institut Mathématiques de Jussieu-Paris Rive Gauche (IMJ-PRG)}
	 	\textit{Université de Paris}, France, \newline
	 	{jeremie.pierard-de-maujouy@imj-prg.fr}
	 }
}

\date{\today}

\begin{document}

\maketitle

\begin{abstract}
We clarify the structure obtained in Hélein and Vey's proposition for a variational principle for the Einstein-Cartan gravitation formulated on a frame bundle starting from a structure-less differentiable $10$-manifold \cite{LFB}. The obtained structure is locally equivalent to a frame bundle. We show that under global hypotheses the formalism allows to produce variational principles which induce a frame bundle structure with a fibration over a base manifold and a variational principle on the base manifold. Our perspective on the model allows us to extend the Einstein-Cartan model to the Einstein-Cartan-Dirac theory which includes a Dirac spinor field coupled to the torsion field.
\end{abstract}

\tableofcontents
\section{Introduction}
\subsection{Introduction}

The theory of General Relativity models spacetime as a $4$-dimensional differentiable manifold. The gravitational field is encoded in a linear connection and gives \emph{geometrical structure} to the manifold. In the original approach, a Lorentzian metric models the gravitational potential and the gravitational field corresponds to the associated Levi-Civita connection. The variational formulation of the theory is due to Einstein and Hilbert; the dynamical field is the metric and the Lagrangian density is simply the scalar curvature (multiplied by the pseudo-Riemannian density).

As research in General Relativity progressed and the framework of differential geometry kept developing alternative formulations of General Relativity were discovered. The \emph{Palatini formulation} of General Relativity relaxes the relation between the linear connection and the metric. This allows for a first-order formulation of gravity. A variant of this formulation uses the \emph{tetrad formalism} : the metric field is replaced by a frame field which defines the metric for which it is orthonormal. Since the metric field is a quadratic function of the frame field, the tetrad is sometimes called \enquote{square-root of the metric}. Because the metric is defined through a tetrad field, the theory gains $\O_{1,3}$'s worth of gauge freedom.

The \emph{Einstein-Cartan theory} is the case where there is a Lorentzian metric and the connection is required to be metric but torsion is allowed. The corresponding geometry on spacetime is suitably understood in the framework of \emph{Cartan geometry}~\cite{CartanWise}.
Differences with Einstein's original theory manifest themselves when the gravitation is coupled to spinor fields. The spinor fields act as a source for the torsion field. One simple example is the \emph{Einstein-Cartan-Dirac} theory which couples the Einstein-Cartan gravitation with a Dirac spinor. The usual variational treatment of spinor field theories in a dynamical spacetime uses a tetrad in order to allow variations of the metric structure while the spinor fields remain \enquote{constant}. When a tetrad is used instead of a metric, the theory is called \emph{Sciama-Kibble} theory \cite{STandFields}.

In~\cite{LFB} Hélein and Vey, following~\cite{CFTRefFrames}, proposed an action defined on a (structure-less) $10$-dimensional manifold $\tot$ such that under some hypotheses a solution of the associated Euler-Lagrange equations defines :
\begin{itemize}
\item A $4$-dimensional manifold $\ET$ over which $\tot$ is fibred.
\item A Riemannian metric and an orientation on $\ET$ 
\item An identification from the fibration bundle $\tot\to \ET$ to the orthonormal frame bundle
\item A metric connection on $\ET$ satisfying the (Riemannian) Einstein-Cartan field equations
\end{itemize}
Their model can be used with a Lorentzian signature (and space-and-time orientations) but in this case both the construction of the fibration and mechanism giving the Einstein-Cartan field equations fail in general. The reason lies in the non-compactness of the (connected) Lorentz group. We argue that there is another obstruction to the construction of the fibration $\tot\to\ET$, even in Riemannian signature. Indeed in this case the $4$-dimensional base space $\ET$, obtained as an orbit space, can have singularities and is not \emph{a priori} a smooth manifold. Our second point is that their derivation of the field equations involves heavy computations on which we pretend a more geometrical perspective can shed some light.

The aim of the present paper is twofold : first, we give an analysis of the mechanism according to which the variational equations on the bundle total space factor to the base manifold of the fibration ; second, we show how it is possible to use the same principle to build an Einstein-Cartan-Dirac Lagrangian on a structure-less $10$-manifold.

The principal bundle structure is constructed from a nondegenerate $1$-form $\omega\oplus\alpha$ with value in the Poincaré Lie algebra. The Poincaré Lie algebra can be decomposed as a semi-direct product $\lor\ltimes \Mink$ with $\lor$ the Lie algebra of the Lorentz group and $\Mink$ the Minkoswki space seen as an abelian Lie algebra. The $1$-form obeys the following equations : 
\begin{subequations}\label{eqno:introCartForm}
\begin{align}
	\d \omega^i + \fwb{\omega}{\omega}^i &= \frac12 \Omega^i_{bc} \alpha^b\wedge\alpha^c\\
	\d \alpha^a + \wb{\omega}{\alpha}^a &= \frac12 \Omega^a_{bc} \alpha^b\wedge\alpha^c
\end{align}
\end{subequations}
From these equations is it possible to build a Lie algebra action on the manifold. In fact, this mechanism is well known from the group manifold approach to supergravity~\cite{GeoSUGRA,SUGRAI}. However more hypotheses are needed for the action to integrate to a Lie group action, and for the fibration over the orbit space to form a principal bundle fibration. The first hypothesis is a completeness requirement, which is stated in~\cite{LFB}, but it is not sufficient. These missing details are presented in Appendix~\ref{annMC} and will be the subject of an upcoming paper~\cite{CartInt}.

In~\cite{LFB} field equations on the base manifold are obtained in the following way : one first uses the Euler-Lagrange equations associated to the variation of an extra field that plays the role of Lagrange multipliers. They allow to construct the frame bundle structure on the $10$-manifold. Then in order to deal with the remaining Euler-Lagrange equations, associated to variations of the fields $\omega$ and $\alpha$, a change of fibre coordinates is applied that amounts to choosing a local frame on the four dimensional base manifold. It allows converting equivariant quantities on the frame bundle to \emph{invariant quantities}. In these coordinates part of the Euler-Lagrange equations manifestly becomes an exact divergence and an other part is invariant under the group action. Integration along fibres allows one to conclude that the invariant part of the Euler-Lagrange equations has to vanish, which turns out to be equivalent to the Einstein field equations on spacetime.

Once part of Euler-Lagrange equations is used to obtain the (local) frame bundle structure, we approach the problem of the field equations from a different perspective. Instead of considering the Euler-Lagrange equations corresponding to field variations associated with a specific set of coordinates, it is enough to consider variations of the fields which are \emph{equivariant} with respect to the action of the structure group. The calculations are in a sense equivalent but our perspective makes it clear why and how the Euler-Lagrange equations split in two terms that respectively give an exact term and the usual field equations on the base manifold. With this insight it is simple to extend the theory to include spinor fields. One peculiarity of the spinor theory thus constructed is that instead of starting from a spacetime and a tetrad which connects the spacetime to a reference bundle in which spinor fields live, we start from the tentative frame bundle on which the spinor fields live and from there construct back the spacetime.

The paper is organised as follows. In Section~\ref{annLegTrans} we give a brief introduction to the aspects of multisymplectic field theory we will be using throughout the paper. Section~\ref{secno:EC} is a short presentation of the Lagrangian of the Einstein-Cartan theory of gravitation. We show in Section \ref{secno:FrameB} how the Einstein-Cartan Lagrangian can be lifted to the frame bundle.

In Section \ref{secno:FBdyn} we get to the crux of Hélein and Vey's model : the frame bundle structure can be \emph{locally} characterised by a $1$-form satisfying Equations~\eqref{eqno:introCartForm}. The frame bundle is dropped and we only work with the $1$-form which defines what we call a \enquote{generalised frame bundle structure}. The Equations are then translated into a Lagrange multiplier term which is added to the Lagrangian. We perform 	the Legendre transform and obtain the same Poincaré-Cartan form as in~\cite{LFB}, with Lagrange multipliers identified with specific components of the momentum. Compared to the non-holonomic Legendre transformation used in~\cite{LFB}, we think our construction is more systematic. Next in Section~\ref{VEqGR} we derive the associated Euler-Lagrange equations. However their treatment is deferred until Section~\ref{secno:EuclFE}. Section~\ref{spinor} is the equivalent of Sections~\ref{secno:EC}-\ref{secno:FrameB} for the Dirac Lagrangian : we present the Lagrangian and lift it to the \emph{spinor frame bundle}, then express it in terms of the generalised frame bundle structure. The corresponding Euler-Lagrange equations are derived in Section~\ref{VEqSp}.

In Section~\ref{LagMult} we present a general framework for Lagrange multipliers in multisymplectic field theory. We then show how for the Lagrangians from Sections~\ref{secno:FBdyn} and~\ref{spinor} it is possible to obtain Euler-Lagrange equations with the dependence in the Lagrange multipliers reduced to an exact term. This is applied in Section~\ref{secno:EuclFE} to the Einstein-Cartan-Dirac theory formulated over a $10$-dimensional manifold as described in Section~\ref{spinor}. Assuming the generalised frame structure integrates into a standard frame structure, we derive from the Euler-Lagrange equations the usual Einstein-Cartan-Dirac field equations on the $4$-dimensional base manifold. There is however part of the Euler-Lagrange equations on the $10$-dimensional manifold involving the Lagrange multipliers which we fail to express on the base manifold. Finally, we give a very brief analysis of the Einstein-Cartan-Dirac equations by identifying the value of the torsion, how it relates to energy-momentum, and an explicit comparison to the Einstein-Dirac theory in which the Dirac operator is associated to the torsion-free Levi-Civita connection.

\subsection{Notations and conventions}\label{secno:notconv}

We will make free use of the Einstein convention :
\[ A_iB^i := \sum_i A_i B^i \]
as well as of the so-called \enquote{musical} isomorphisms raising and lowering indices through a metric that will generally be implicit (as long as there is no ambiguity). The convention for the order of the indices will be to keep the relative order of upper indices and lower indices, and place upper indices \emph{before lower indices}. For example given a tensor $T^\pi{}_{\mu\nu}$ the corresponding totally covariant and totally contravariant tensors are
\begin{align*}
	T_{\tau\mu\nu} &:= g_{\tau\pi}T^\pi{}_{\mu\nu}\\
	T^{\pi\tau\rho} &:= g^{\tau\mu}g^{\rho\nu}T^\pi{}_{\mu\nu}
\end{align*}
given a metric $g$ on the space in which the indices $\tau,\mu,\nu$ live. The metric and the inverse metric will be both written with the same symbol, but we will generally use the Kronecker delta $\delta^\mu_\nu = g^{\mu\tau}g_{\tau\nu}$ for the identity which is the corresponding endomorphism of both covariant and contravariant vectors.

We will write $\met_{ab}$ for the Minkowski metric on the Minkowski space $\m$, used at the same time as a pseudo-Euclidean affine space, as an abelian Lie group and as an abelian Lie algebra. Our convention for the Lorentzian signature is $(+---)$ and for the Clifford algebras $u\cdot v + v\cdot u = -2\sprod u v$ (more detail in Appendix~\ref{annspin}). We will be working with the connected \emph{proper orthochronous Lorentz group} $\Lor=\SOpL$ which we will just call \emph{Lorentz group}. Its Lie algebra is $\lor=\soL$.
The \emph{Poincaré group} is isomorphic to the semi-direct product of $\m$ with $\Lor$ : 
\[
\mathfrak{P} \simeq \Lor\ltimes \m	
\]
and there is an isomorphism between the associated Lie algebras :
\[
\p \simeq \lor\ltimes \m
\]

If not explicitly mentioned (for example in Appendix~\ref{annMC}), we will consider \emph{finite dimensional second countable Hausdorff} differentiable manifolds.

\section{Multisymplectic Field Theory}\label{annLegTrans}
Before dealing with the actual Lagrangians we are interested in, we give a very brief introduction to the multisymplectic formalism for (classical) fields theories. The reader is free to skip to the next section and come back to check the details of this section whenever is needed. A general reference for multisymplectic field theory is \cite{GIMMSY1} ; more on the geometrical aspects can be found is~\cite{Multisympl}.

\subsection{The multisymplectic formalism}

Lagrangian Field Theory deals with critical points of actions that are defined by the integral of a local Lagrangian which is at each point of the integration space a local function of the physical fields involved. We will discuss \emph{first order theories}. Let us consider a fibre bundle $Q\xrightarrow{\pi}M$ representing a configuration space fibred over a $n$-dimensional manifold $M$ (that will be assumed oriented to avoid density considerations). The \emph{first order jet bundle} (or $1$-jet bundle for short) is written
\[\Jun(\pi) \xrightarrow{\pi^1} Q \xrightarrow{\pi} M\]
or without mentioning the fibration $\Jun(Q)$. Over every open subset $\U\subset M$, to each section of $Q$ can be associated its \emph{first order prolongation} ($1$-jet for short) : 
\[
	\phi\in\Gamma(\U,Q|_{\U}) \mapsto j\phi \in\Gamma(\U,\Jun(Q)|_{\U})
\]
A (first order) Lagrangian is a section $\Lag\in (\pi\circ\pi^1)^*(\ExT^n T^*M)$. The associated action is
\[
	S : \phi \mapsto S[\phi] = \int \Lag(\phi,\d\phi) = \int j\phi^*\Lag
\]
The integration domain is intentionally not written : the integral may be ill-defined over non-compact domains, but it does not pose problem in (classical) field theory. What defines the equation of motion is the \emph{variation} of the action with respect to \emph{compactly-supported} variations which is well defined over each compact subset of $M$.

The usual derivation of the Euler-Lagrange equations (in coordinates) proceeds by taking arbitrary variations of the field $\phi$ and performing an integration by parts in order to obtain an integrand linear in the 0-order variation of $\phi$ which is free. In the multisymplectic formalism this computation is carried in a more geometrical language. It involves the multisymplectic \emph{Legendre transformation}.

\subsection{The Legendre transformation}\label{ssecno:LegTrans}

The Legendre transformation in covariant field theory is a mapping from the jet bundle to the dual jet bundle. We will here again restrict our considerations to first order Lagrangians. In the case of mechanics (1-dimension) the transform takes the form $(t,q,\dot q)\mapsto -H(t,q,\dot q)\d t + p_i(t,q,\dot q)\d q^i$. The usual Hamiltonian then takes the form of an \emph{hamiltonian section}
\[h : 
	\begin{cases}
	\setR\times T^*Q\simeq T^*(\setR\times Q)/\setR\d t &\to T^*(\setR\times Q)\\
	\qquad\qquad (t,q^j,p_i\d q^i) &\mapsto (t,q^j,-H(t,q,p)\d t+ p_i\d q^i)
	\end{cases}
\]

We get back to the fibre bundle $Q\xrightarrow{\pi}M$. In the so-called \emph{de Donder-Weyl} theory, the momentum space is the \emph{affine dual jet bundle} with values in $\ExT^n T^*M$ \cite{Multisympl} :
\begin{equation}\label{eqno:DualAff}
	\ExT^n_1 T^*Q = \{a\in \ExT^n T^*Q \,|\ (VQ\wedge VQ)\lr a = 0 \}
\end{equation}
where $VQ$ is the vertical tangent bundle of the fibration $Q\to M$. Let us record one useful property here : for any two fibre bundles $Q_1,Q_2\to M$, there is a natural isomorphism
\[
	\ExT^n_1 T^*(Q_1\times_M Q_2) \simeq \ExT^n_1 T^*Q_1 \times_M \ExT^n_1 T^*Q_2
\]

The Legendre transform is realized by the Poincaré-Cartan form $\Theta$, a section of $(\pi^1)^*\ExT^n_1 T^*Q$, which be describe now. It can be seen as a \enquote{corrected} extension of the Lagrangian $\Lag$ from sections of $\pi$ to arbitrary sections of $\Jun (Q)$. It is defined by the following properties :
\begin{align}
	j\phi^*\Theta &= j\phi^*\Lag\qquad \forall\phi\in\Gamma(M,Q) \label{PCcontact}\\
	j\phi^*\lp i_X\d \Theta \rp&= 0\qquad \forall\phi\in\Gamma(M,Q),\forall X\in V(\Jun(Q)\to Q)\label{PC0jet}
\end{align}
so that it defines the same Lagrangian form as $\Lag$ while its first variation is (locally) trivial in the first-order directions : it depends only on the variation of the \emph{0-jet}. From this perspective, the Legendre transformation is the geometrical formulation of the integration by parts producing the Euler-Lagrange term (the variational derivative) coupled to the 0-jet variation in the usual coordinate-explicit derivation of the Euler-Lagrange equations.

In order to obtain an explicit expression, one considers the \textit{contact forms} which generate the differential forms of $\Jun(Q)$ vanishing under the pullback by any first order prolongation \cite{ExSysEL,GIMMSY1}. They are described below.
As we want explicit formulas, we will use (local) coordinates. Let $(x^i)$ be coordinates on $M$, let $(y^A)$ be the vertical coordinates on $Q$ with respect to a local trivialisation and let $(v^A_i)$ be the corresponding $1$st-order coordinates, so that $v^A_i(j\phi) = \dd_i \phi^A$.

A basis for the contact $1$-forms is given by 
\[\theta^A = \d y^A -v^A_i\d x^i\]
Hence the contact elements of $\ExT^n_1 T^*Q$ are of the form $\theta^A \wedge p_A$ for $p_A\in\pi^*\ExT^{n-1}T^*M$ (we use the Einstein summation convention). To the $x^i$ coordinates is associated a local volume form $\vol=\d x^1\wedge \cdots \wedge \d x^n$ so that $\Lag$ decomposes as $L(x,y,v)\vol$. We write 
\[\Theta = \Lag+\theta^A \wedge p_A\]
The components $p_A$ are then uniquely determined (the condition~\eqref{eqno:DualAff} fixing any further contact term) : write $\vol_i$ for the $(n-1)$-form dual to $\d x^i$ and decompose $p_A = p_A^i\vol_i$ :
\begin{equation*}\begin{aligned}
	\dd_{v^A_i}\lr\d(\Lag + \theta^Bp_B) &=
		(\dd_{v^A_i}L)\vol
		+ (\dd_{v^A_i}\lr\d\theta)^B \wedge p_B - (\dd_{v^A_i}\lr\theta^B) \wedge \d p_B\\
			&\qquad	+ \d\theta^B \wedge \dd_{v^A_i}\lr p_B - \theta^B \wedge \dd_{v^A_i}\lr \d p_B\\
		&= \dd_{v^A_i}L\vol + (\dd_{v^A_i}\lr\d\theta)^B \wedge p_B 
			- \theta^B \wedge \dd_{v^A_i}\lr \d p_B\\ 
		&= \dd_{v^A_i}L\vol + (\dd_{v^A_i}\lr(-\d v^B_i\d x^i))^B \wedge p_B
			- \theta^B \wedge \dd_{v^A_i}\lr \d p_B \\  
		&= \dd_{v^A_i}L\vol - \d x^i \wedge p_A
			- \theta^B \wedge \dd_{v^A_i}\lr \d p_B\\
			&= (\dd_{v^A_i}L - p^i_A)\vol
			- \theta^B \wedge \dd_{v^A_i}\lr \d p_B
\end{aligned}\end{equation*}
The $p^i_A$ are then defined by 
\[ (\dd_{v^A_i}L - p^i_A)\vol - \theta^B \wedge \dd_{v^A_i}\lr \d p_B
		\equiv 0 \mod (\theta^B)\]
so that
\begin{equation}
	p^i_A = \dd_{v^A_i} L
\end{equation}

We obtain the usual expression for the Poincaré-Cartan form :
\begin{equation}\label{LTform}
	\Theta = \Lag + \lp \dd_{v^A_i} L \rp \theta^A\wedge \vol_i
\end{equation}

The action can equivalently be defined using $\Theta$ : 
\[
	 S[\phi] = \int j\phi^*\Theta
\]
The corresponding \emph{Euler-Lagrange equation} is
\begin{equation}
	\forall X \in \phi^*TQ,\qquad
		\boxed{
			\phi^*(i_{jX}\d\theta) = 0}
\end{equation}
with $jX$ the $1$st order prolongation of $X$. The $(n+1)$-form $\d\Theta$ is called a \emph{premultisymplectic form}. We will call the $n$-forms $i_{jX}\d\Theta$ \emph{Euler-Lagrange forms} and write them $EL_X$ : 
\begin{equation}
	\EL_X := i_{jX}\d \Theta
\end{equation}
\com{Inspiré de la notation $E(L)$ utilisée par exemple par Anderson}

If we consider more general Lagrangian forms that are sections of ${\pi^1}^*(\ExT^n T^*Q)$, the Poincaré-Cartan form can still be defined. According to \cite{ExSysEL} (our $\Lag$ is their $\Lambda$ and their $\Pi$ is our $\d\Theta$; they call \emph{Betounes form} what we call Poincaré-Cartan form), $\theta^A\wedge p_A$ is uniquely defined such that
\begin{equation}\label{eqno:PCgen}
	\d(\Lag + \theta^B\wedge p_B) = \theta^A\wedge E_A
\end{equation}
The $E_A$ forms then correspond to the Euler-Lagrange terms. The coefficient forms $p_A$ themselves are then uniquely defined if we require a condition such as $p_A = p^I_A \vol_I$ (for a suitable subset of Lagrangians). In any case the Poincaré-Cartan form itself is well defined.

In particular, for forms $\Lag$ that are sections of $\pi^*(\ExT^n T^*M)$ (that is, which have coefficients independent of the $1$st order coordinates), the Poincaré-Cartan form is $\Lag$ itself. Indeed $\d\Lag$ is a section of $\ExT^{n+1}_1 T^*Q$, which has $\sum_A \theta^A\wedge {\pi^1}^*(\ExT^n T^*Q)$ as pullback bundle to $\Jun(Q)$  (because there are only $n$ horizontal directions and $\theta^A\wedge \vol = \d y^A\wedge \vol$).

\section{The Einstein-Cartan theory of gravitation}\label{secno:EC}
We start with a very succinct presentation of the Einstein-Cartan theory of gravitation. In Einstein's original theory of gravitation, the gravitation is modelled by \emph{the curvature} of a Lorentzian metric on the differentiable manifold modelling the spacetime. More accurately, the Newtonian gravitational field is replaced by the torsion-free Levi-Civita connection associated to the metric, which defines the geodesic equation governing inertial trajectories. Newton's formula for the interaction between massive bodies and the Poisson equation for the gravitational potential are replaced by Einstein's field equation, which is a second order differential equation. It was then realised that a first order formulation is possible if one was to consider as possible field a couple gathering the metric itself and an a priori independent affine connection. This is called the \emph{Palatini formalism} and exist in several variations, requiring the connection to be metric or not (see \cite{MetricAffine1,MetricAffine2,Palatini}). In the case the connection is neither assumed to be metric nor torsion-free an extra gauge freedom appears, the so-called projective symmetry \cite{MetricAffine1,Palatini}.

Within the Palatini formalism, it is possible to make explicit a Lorentzian gauge symmetry: this is the so-called \emph{tetradic Palatini formalism}, in which the metric field is replaced by a field of linear frames, or equivalently (in our case) \emph{coframes}. The linear frame is called a \emph{tetrad}, or \emph{vierbein} (\emph{vielbein} in general dimension). The \emph{Einstein-Cartan theory} (more precisely Einstein-Cartan-Sciama-Kibble) is the case in which the connection is assumed to be metric but not torsion-free. We will see in Section~\ref{secno:EuclFE} how torsion is involved in the dynamics of spinors in a curved spacetime.

We start with a differentiable 4-manifold $\ET$ that is meant to embody spacetime. The fields will be :
\begin{itemize}
\item A \emph{generalized tetrad} $e$ which identifies the tangent bundle with a reference oriented lorentzian vector bundle $(\aux,\eta)$. We do not assume $\aux$ to be trivial as it may implicitly be the case in standard tetrad formalism. 
\item A dynamic \emph{metric connection} $\tilde\nabla$ on $\aux$, or equivalently a connection $\nabla$ on $T\ET$ compatible with the inverse pullback metric.
\end{itemize}
The tetrad is a (nondegenerate) $\aux$-valued 1-form written locally as 
$(e^a)_{0\leqslant a\leqslant 3}$.
in which the $a$ superscript corresponds to any reference basis of $\aux$. With $(x^\mu)$ a local coordinate system of $\ET$ the vielbein decomposes with components $e^a=e^a_\mu \d x^\mu$

The (adimensional) \emph{Einstein-Cartan action} (hereafter EC action) with no cosmological constant is then given by~\cite{LFB,PalatiniLag}
\begin{equation}\label{SEWC}
	S_{EC}[e,\tilde\nabla]
		= \int_\ET \Lag_{EC}[e,\tilde\nabla]
		= \int_\ET \widetilde{\Riem^a_c}\eta^{cb} \wedge e^{(2)}_{ab}
\end{equation}
with $\widetilde{\Riem}$ the Riemannian curvature $2$-form of $\widetilde\nabla$, $\eta^{ac}$ representing the \emph{inverse} metric and $e^{(2)}_{ab}$ the dual exterior 2-forms as described in Appendix~\ref{anndual}. In the case $\ET$ is noncompact, so that the integral can be ill-defined, the action has to be understood as a motivation for the corresponding \emph{variational} (Euler-Lagrange) equation, which is a local equation on the fields: even when the action is not globally defined it is locally, as well as its first variation.

\section{General Relativity formulated on the frame bundle}\label{secno:FrameB}
We now wish to formulate the Einstein-Cartan theory on the \emph{frame bundle of $\aux$}, or to be precise on the \emph{direct orthonormed orthochronous frame bundle of $\aux$}, which we write $\FrameV$ (hereafter referred to as the frame bundle). The tetrad field on spacetime is represented by a \emph{solder form} on $\FrameV$ (defined below). In this way the tetrad data is integrated in the geometrical setting and we consider all coframes in an equivariant manner. Anticipating on Section~\ref{spinor}, working on a frame bundle will allow us to consider spinor fields with value in a trivialised bundle, in the same way the tetrad does. For this purpose, we will lift the Lagrangian form defining the action~(\ref{SEWC}) to the frame bundle.

\subsubsection*{The space of fields}

The frame bundle of $V$ depends on the metric, orientation and time-orientation structures on $\ET$; the connection is then an extra structure on the frame bundle. 
Write $\pi:\FrameV\to\ET$ the principal fibration. Vectors transverse to the fibration are called \emph{vertical} and differential forms which have a trivial contraction with all vertical vectors are called \emph{horizontal}. 

The frame bundle has the structure of a $\Lor$-principal bundle on $\ET$ provided with a \emph{solder form} $\alpha$ : a nondegenerate $\Lor$-equivariant horizontal 1-form with values in $\Mink$  (more detail in Appendix~\ref{annFrameB}). The solder form establishes an isomorphism between the associated vector bundle of fibre $\Mink$ and $T\ET$, or between the corresponding frame bundles. Hence choosing an isomorphism class for $\aux$ amounts to selecting a frame bundle and $\alpha$ plays the role of the tetrad $e$. The Lorentz gauge symmetry on $e$ is geometrically realised as the principal action of the Lorentz group on the frame bundle.

A metric connection on $\aux$ corresponds to a nondegenerate $\lor$-equivariant $\lor$-valued $1$-form $\omega$ on $\FrameV$ which is normalized with respect to the action of the Lorentz algebra $\lor$. Normalisation means that representing $\h\in\lor\hookrightarrow\bh \in \Gamma(T\FrameV)$, the following holds :
\[  \forall \h\in\lor, \quad \omega(\bh)=\h \]
Its kernel is a horizontal distribution which is the corresponding \emph{Ehresmann connection}. The data of a metric connection together with the solder form gives rise to a \emph{Cartan connection} $\omega\oplus\alpha\in\Gamma(\FrameV,\lor\oplus\Mink)^\Lor$ (the superscript $\Lor$ means restricting to the $\Lor$-equivariant forms). The space of fields can then be described as the set of $\lor\oplus\m$-valued Cartan connections on the principal bundle $\FrameV$.

\subsubsection*{Lifting the Lagrangian}

We want to express the action~(\ref{SEWC}) lifted to $\FrameV$ as a function of the Cartan connection. As the signature is non-Riemannian, the structure group $\Lor = \SOpL$ is noncompact hence so is the bundle space. We shall consider the action as a formal integral and a motivation to derive variational equations over local variations, and forget about any domain of integration. As a remark note that a structure group reduction to maximal compact subgroups $\SO_3$ always \emph{exists} but would involve extra physical data (although topologically trivial): a nowhere vanishing timelike vector field (it is called a \emph{field of observers} in~\cite{ObsSp}).

The curvature 2-form associated with the connection is expressed as
\[ \Omega = \d\omega + \frac12\wb\omega\omega \]
It takes value in the Lie algebra $\lor$ but we will allow ourself to consider the associated $\End(\Mink)$-valued 2-form without changing the notation. We write $\pi^*$ for the pullback to $\FrameV$ of any tensorial-valued differential form, which is then identified with a (horizontal) differential form of the same degree with values in a \emph{constant} trivial bundle and which is equivariant.

In order to relate the curvature $2$-form to the curvature tensor, we introduce $\rho^b_{i,d}$ the components of the action $\lor\to \End(\setR^{1,3})$ so that \[(\h_i\cdot x)^b = \rho^b_{i,d}x^d\]
for $x\in\setR^{1,3}$. They satisfy the antisymmetry relation 
\[\rho^b_{i,d}\eta^{dc} + \rho^c_{i,d}\eta^{db} = 0\]
Then the curvature tensor $\Riem$ is defined so that
\[(\pi^*\widetilde\Riem)^a_c = \Omega^i \rho^a_{i,c} \]
as explained in Appendix~\ref{annFrameB}, using implicitly the identification \[\pi^*\End(TM)\simeq \Rpq\otimes\Rpq^*\times\FrameV\]
We equip the Minkowski vector space $\Mink$ with a spacetime orientation in addition to its metric structure, so that a duality operator $\ExT^k\Mink\simeq \ExT^{4-k} \Mink^*$ is defined (as in Appendix~\ref{anndual}). We can then lift the EC Lagrangian form~(\ref{SEWC}):
\begin{equation}
	\pi^* \lp \widetilde{\Riem^a_c}\eta^{cb}\wedge e^{(2)}_{ab} \rp
	= \Omega^i \rho^a_{i,c} \eta^{cb}\wedge \alpha^{(2)}_{ab}
\end{equation}
which gives a horizontal (scalar-valued) $4$-form on the frame bundle. It is not a top form on the bundle space yet, to turn it into a $\Lor$
-invariant top form it has to be wedge-multiplied with a (right-)invariant volume form on $\Lor$. 
Such a volume form is not unique, but we can specify it in a consistent way for all fibres using the Lie algebra action, as we explain now.

As a $\lor$-valued $1$-form, $\omega$ establishes a map $\lor^*\to \Omega^1_{vert}(\FrameV)^\Lor$ which extends to an graded algebra morphism 
\begin{equation}
	\omega^* : \ExT^\bullet(\lor^*) \to \Omega_{vert}^\bullet(\FrameV)^\Lor
\end{equation}
so that specifying a volume element in $\ExT^{6}\lor^*$ gives a vertical $6$-form on $\FrameV$, which we will write $\omega^{(6)}$. Although $\omega^{(6)}$ effectively depends on the connection, detailed computations show that $h \wedge \omega^{(6)}$ does not for $h$ any specified \emph{horizontal} $4$-form. We thus define the following Lagrangian form on $\FrameV$:
\begin{equation}
	\Lag =  \Omega^i \rho^a_{i,c} \eta^{cb}\wedge\alpha^{(2)}_{ab}\wedge\omega^{(6)}
\end{equation}
Even if we do not explicitly specify the volume element of $\lor^*$, it is still possible to discuss about coupling constants and relative signs when considering a Lagrangian with matter components as long as the same volume element is used for all terms. These considerations remain however out of the scope of this paper. Using the whole Cartan connection form we can in a similar way consider the morphism
\begin{equation}
	(\omega \oplus \alpha)^* : \ExT^\bullet((\lor \oplus \Mink)^*) \to \Omega^\bullet(\FrameV)^\Lor
\end{equation}
and providing $\lor$ with an arbitrary volume element we use the vector-forms duality (as described in Appendix~\ref{anndual}) on \emph{$\lor\ltimes\Mink$} so that we can write
\begin{equation}\label{FrameLag}
	\Lag =  \Omega^i \rho^a_{i,c} \eta^{cb}\wedge (\omega\oplus\alpha)^{(8)}_{ab}
\end{equation}

We can now formulate the variational problem on $\FrameV$: the field is a Cartan connection $1$-form $\omega\oplus\alpha$, in other words a $1$-form with value in $\lor\oplus\Mink$, more precisely as a Lie algebra $\lor\ltimes\Mink$, which is nondegenerate (i.e. of constant rank $10$), normalised on the principal action of $\lor$ and equivariant. It has to be an extremal point of the locally defined action
\begin{equation}\label{SEWCFrame}\boxed{
	S_{EC}[\alpha,\omega] =
	\int \Omega^i \rho^a_{i,c}
	\eta^{cb}\wedge (\omega\oplus\alpha)^{(8)}_{ab}
}\end{equation}
for compactly-supported variations. The integral is taken with respect to the orientation given by $\alpha^{(4)}\wedge\omega^{(6)}$. The constraint of being a Cartan connection $1$-form on $\omega\oplus\alpha$ can be written in the following way, writing $\bh$ for the (left-invariant) vector field representing $\h\in\lor$ and $R_g$ for the action of $g\in\Lor$ :
\begin{empheq}[left=\empheqlbrace]{gather*}
	\alpha^{(4)}\wedge\omega^{(6)}\text{is nowhere vanishing} \\
	i_\bh(\omega\oplus\alpha) = \h\oplus 0	\\
	R_g^*(\omega\oplus\alpha) + g \cdot(\omega\oplus\alpha) = 0
\end{empheq}
Now as the group $\Lor$ is connected, $\lor$-equivariance is equivalent to $\Lor$-equivarlence so that the constraint can be written as local equations as follows : 
\begin{subequations}\label{eqno:omalcons}
\begin{empheq}[left=\empheqlbrace]{gather}
	\alpha^{(4)}\wedge\omega^{(6)}\text{is nowhere vanishing} \label{eqno:nondegen}\\
	i_\bh(\omega\oplus\alpha) = \h\oplus 0	\label{normalisation}\\
	\Lie_\bh (\omega\oplus\alpha) + \h\cdot(\omega\oplus\alpha) = 0 \label{equivconsnorm}
\end{empheq}
\end{subequations}

We want to derive the variational Euler-Lagrange equation for~(\ref{SEWCFrame}) under the constraints~(\ref{eqno:nondegen}-\ref{equivconsnorm}) but as the equivariance constraint~(\ref{equivconsnorm}) is non-holonomic, and actually non-local, the usual derivation does not directly apply. Moreover, the action~(\ref{SEWCFrame}) cannot be used as such as the domain is non compact (and requiring equivariance along the noncompact fibres definitely prevents any nontrivial field variation from having a compact support, or even from decaying at infinity). The central question of this paper is the derivation and the treatment of the variational equations under such constraints. This will involve translating the constraints into Lagrange multipliers terms (as presented in Section~\ref{LagMult}) in the Lagrangian and is explained in the next section.
%

\section{Generalised frame bundle structure with connection as a dynamical field}\label{secno:FBdyn}

In Section~\ref{secno:FrameB} we described a formulation of Einstein-Cartan gravitation with one field $\varpi$ which is defined over the frame bundle of spacetime.
The idea of the model proposed by Hélein and Vey in \cite{LFB} is to forget any a priori structure of frame bundle and simply study the field $\varpi$ defined over a structure-less $10$-dimensional manifold. Indeed, turning around the constraints~(\ref{normalisation},\ref{equivconsnorm}), they can be seen as defining a \enquote{generalised frame bundle structure} from $\varpi$, as we define below. A similar mechanism assuming only part of the fibration structure is studied in~\cite{ObsSp}.

\subsubsection*{Generalised frame bundles}

Given a $\p$-valued coframe $\varpi=\omega\oplus\alpha$, \eqref{normalisation} can be understood as defining the infinitesimal action
of $\lor$ :
\[ \bh := \varpi^{-1}(\h\oplus 0) \]
The $\omega$ component of \eqref{equivconsnorm} then implies that the vectors $\bh$ form a representation of the Lie algebra $\lor$ (as explained in detail in Appendix~\ref{annCartanConnForm}). Equation~\eqref{equivconsnorm} then means that $\varpi$ is equivariant under this Lie algebra action. In the case this Lie algebra action corresponds to the infinitesimal action of $\lor$ on a $\Lor$-principal bundle fibration (over some $4$-manifold), $\alpha$ is a solder form (defined in Appendix~\ref{annFrameB}) so that the $\Lor$-principal bundle can be identified with a frame bundle. 

The structure of principal bundle is in a sense a global structure so that it cannot be entirely characterised by local equations like~\eqref{eqno:omalcons}. But up to global topological aspects (to which Appendix~\ref{annMC} is dedicated), these equations encapsulate the \enquote{local structure} of the frame bundle with a metric connection. It is this insight which motivates the generalisation of the Einstein-Cartan theory formulated over a \enquote{blank} $10$-manifold.

We call a $\p$-valued coframe $\varpi$ which satisfies 
\begin{equation}\tag{\ref{equivconsnorm}}
	\Lie_\bh\varpi + \ad_\h \varpi = 0
\end{equation}
a \emph{generalised Cartan connection} (simply called Cartan connection in~\cite{DiffGeoCartan}), or a \emph{Cartan $1$-form for short}. We say that the manifold $\tot$ equipped with a generalised Cartan connection has the structure of a \emph{generalised frame bundle}, which is to be understood as a abbreviation for \enquote{generalised frame bundle with connection}. As $\varpi$ defines an action of $\lor$, we can define equivariant sections on a generalised frame bundle. In particular, we will use the following notions, which generalise the usual ones on a standard frame bundle and more generally make sense on foliated manifolds.

A \emph{basic vector field} is a $\lor$-equivariant application with value in $\Mink$. More generally, a \emph{(basic) tensor field} is a $\lor$-equivariant application with value in some tensor product $\Mink^{\otimes m} \otimes \Mink^{*\otimes n}$.\\
A \emph{basic differential form} is a $\lor$-equivariant differential form with purely horizontal components, that is to say along products of $\alpha^a$. One defines similarly basic differential forms with value in a vector space, or in an equivariant vector bundle. See Appendix~\ref{annFrameB} for more detail. One example which will be omnipresent throughout the article is the curvature form associated to $\varpi$ : 
\begin{equation*}
	\Omega = \d\varpi + \fwb\varpi\varpi = \frac12\Omega_{ab}\alpha^a\wedge\alpha^b
\end{equation*}
which is an equivariant $\p$-valued basic $2$-form (equivariant here means that $\p$ is used as a equivariant trivial vector bundle).

The structure of generalised frame bundle is the one we wish to obtain from the Euler-Lagrange equations.

\subsubsection*{Lagrange multiplier terms}

We motivated dropping the constraint \eqref{normalisation} so as to take it as a definition of the fields $\bh$ instead. Constraint~\eqref{eqno:nondegen} is an open (and algebraic) condition so we will simply keep it as restraining the configuration space to an open subset. Our manipulations will actually be meaningful on all of $T^*\tot\otimes \p$ including degenerated points, with due adjustments since $T\tot$ and $\p$ are no longer identified.

However Equation \eqref{equivconsnorm} is different. It is both a differential constraint and a closed constraint. We want to incorporate it into the Lagrangian by means of Lagrange multipliers. For this, the convenient formulation of \eqref{equivconsnorm} is 
\begin{equation*}
	\d \varpi^A + \fwb\varpi\varpi^A = \frac12\Omega^A_{bc}\alpha^b\wedge\alpha^c 
\end{equation*}
with $\Omega_{bc}^A$ arbitrary (non-constant) coefficients which are antisymmetric in $b,c$ (the derivation is given in Appendix~\ref{annCartanConnForm}). As the coefficients $\Omega^A_{bc}$ are arbitrary, the equation only means that the components along $\omega\wedge\omega$ and along $\alpha\wedge\omega$ vanish. According to Appendix~\ref{anndual} if we use again the notation $\varpi^{(8)}_{BC}$ for the $8$-form dual to $\varpi^B\wedge\varpi^C$, we can rewrite the equation as
\begin{subequations}\label{seqno:equivvarpi8}
\begin{align}
	\lp \d\varpi + \fwb\varpi\varpi\rp \wedge \varpi^{(8)}_{jk} = 0 \label{equivconsij}\\
	\lp \d\varpi + \fwb\varpi\varpi\rp \wedge \varpi^{(8)}_{bk} = 0 \label{equivconsai}
\end{align}
\end{subequations}
with the index $b$ corresponding to a basis of $\m$ and indices $j,k$ corresponding to a basis of $\lor$.

Under this form, it is straightforward to impose the conditions~\eqref{seqno:equivvarpi8} using Lagrange multipliers : we add to the theory free fields $P_A^{jk}$ and $P_A^{bk}$ and consider the following term to add to the Lagrangian : 
\begin{equation}\label{multvirg}
	\lp \d\varpi + \fwb\varpi\varpi\rp^A \wedge \frac12 P^{jk}_A \varpi^{(8)}_{jk} 
	+ \lp \d\varpi + \fwb\varpi\varpi\rp^A \wedge P^{bk}_A \varpi^{(8)}_{bk}
\end{equation}
which would impose the constraint~\eqref{seqno:equivvarpi8} through the equations of motion corresponding to variations of $P^{jk}_A$ and $P^{bk}_A$.
Note the fundamental difference with \emph{holonomic} Lagrange multipliers which would be coupled to a term such as $f(\varpi,v)\varpi^{(10)}$. The Lagrange multipliers we use serve to impose an (exterior) differential constraint. Such Lagrange multipliers are presented in more detail in Section~\ref{LagMult}, which explains how to use the variational terms derived from them.

The $\varpi^{(8)}_{BC}$ forms being antisymmetric in $BC$, only the antisymmetric part of $P_A^{BC}$ is involved in the term \eqref{multvirg}. We thus constrain the multipliers $P_A^{BC}$ (with $BC=bk$ or $BC=jk$) to be antisymmetric in $BC$, effectively using the $8$-forms
\[
	\frac12 P_A^{BC}\lp \d\varpi + \fwb\varpi\varpi\rp^A \wedge \varpi^{(8)}_{BC}
\]
as Lagrange multiplier fields. 
If one wanted to impose a torsion-freeness constraint on the connection, one could in a similar fashion use a free $p^{bc}_a$ term, as is for example done in three dimensions in~\cite{D3ECD}.

\subsubsection*{The Lagrangian}

The Lagrangian~\eqref{FrameLag} is
\[
	\Lag[\varpi] = \Omega^i \rho_{i,d}^b \eta^{dc}\wedge (\omega\oplus\alpha)^{(8)}_{bc}
		= \Omega^i \rho_{i,d}^b \eta^{dc}\wedge \varpi^{(8)}_{bc}
\]
so that the Lagrangian~\eqref{FrameLag} takes the form
\[
	\lp \d\varpi + \fwb\varpi\varpi \rp^i
		\rho_{i,d}^b\eta^{dc} \wedge \varpi^{(8)}_{bc}
\]
Note how as a linear function of the curvature $2$-form it is very similar to the terms~\eqref{multvirg}.

Let $\tot$ a $10$-manifold and $\p\simeq \lor\ltimes\m$ the Poincaré Lie algebra. Let us denote $\Isototp \subset T^*\tot\otimes \p$ the subbundle of \emph{$\p$-valued coframes}. We consider the following configuration bundle over $\tot$ : 
\begin{equation*}
	Q = \underbrace{\Isototp}_{\varpi^A}
		\times_\tot \underbrace{\lp \m\wedge\lor \oplus \ExT^2\lor \rp\otimes\p^*}_{P^{bk}_A,\, P^{jk}_A}
\end{equation*}
with $\m\wedge\lor \subset \ExT^2\p$ the image of $\m\otimes\lor \subset \p^{\otimes 2}$. 
We call $\lambda$ the canonical $\p$-valued $1$-form on $T^*\tot\otimes\p$. On $\Iso(T\tot,\p)$ it can be identified with a solder form (defined in Appendix~\ref{annFrameB}). We will use the notation $\lambda^{(10-k)}_{A_1\cdots A_k}$ for the dual $(10-k)$-forms defined according to Appendix~\ref{anndual}. We also use the notations $p_A^{bk}$ and $p_A^{jk}$ for the (trivial) fibre coordinates of the component in $ \lp \m\otimes\lor \oplus \ExT^2\lor \rp\otimes\p^*$, in order to establish a clear distinction with the corresponding components $P_A^{bk}$ and $P_A^{jk}$ of \emph{sections} of $Q$

The Lagrangian form gathering both the lifted Einstein-Cartan Lagrangian and the Lagrange multiplier fields is
\begin{multline}\label{eqno:ECLagomp}
	\Lag[\omega,P] = \rho_{i,d}^b\eta^{dc}\lp \d\varpi + \fwb\varpi\varpi \rp^i
	 \wedge \varpi^{(8)}_{bc} \\
	+ \frac12 P^{jk}_A \lp \d\varpi + \fwb\varpi\varpi\rp^A  \wedge \varpi^{(8)}_{jk} 
	+ P_A^{bk} \lp \d\varpi + \fwb\varpi\varpi\rp^A  \wedge \varpi^{(8)}_{bk}
\end{multline} 
To obtain a local expression in terms of coordinates, let $(z^I)$ be a local system of coordinates on $\tot$. It induces local coordinates $\lambda^A_I$ on $T^*\tot\otimes\p$ and fibre coordinates on the $1$-jet bundle $\Jun(T^*\tot\otimes \p)$ : we write them $v^A_{I,J}$, defined so that
\[
	v^A_{I,J}(\varpi) = \dd_J \lambda^A_I(\varpi)
\]
Then the Lagrangian \eqref{eqno:ECLagomp} can be expressed as a $10$-form on $\Jun(Q)$ in terms of the local coordinates (there is no $1$st order contribution from $P$) :
\begin{multline}\label{eqno:LagvIJ}
	\Lag = \rho_{i,d}^b\eta^{dc}\lp v_{I,J}\d z^J\wedge\d z^I + \fwb\lambda\lambda \rp^i
	 \wedge \lhuit{bc} \\
	+ \frac12 p^{jk}_A \lp v_{I,J}\d z^J\wedge\d z^I + \fwb\lambda\lambda\rp^A  \wedge \lambda^{(8)}_{jk} 
	+ p_A^{bk} \lp v_{I,J}\d z^J\wedge\d z^I + \fwb\lambda\lambda\rp^A \wedge  \lambda^{(8)}_{bk}
\end{multline}
In this form, there is no constraint extraneous to the Lagrangian other than the open constraint of nondegeneracy of $\varpi$. Hence we can use the usual Legendre transform formula~(\ref{LTform}) (described in Section~\ref{annLegTrans}) to compute the Poincaré-Cartan form. If we define 
\begin{equation}\label{EWCmultcons} p^{bc}_A = 2\delta^i_A \rho_{i,d}^b\eta^{dc} \end{equation}
then $\Lag$ takes the concise form 
\begin{equation}
	\Lag = \frac12 p_A^{BC}
		\lp v_{I,J}\d z^J\wedge\d z^I + \fwb\lambda\lambda\rp^A \wedge  \lambda^{(8)}_{BC}
\end{equation}
We can thus consider that we have a field $p_A = p_A^{BC}\lhuit{BC}$ which is subject to the holonomic constraint \eqref{EWCmultcons}.

\subsubsection*{The Poincaré-Cartan form}

Introduce the following notation : for a $p$-form $u$ on $T^*\tot\otimes\p$ with values in a $\p$-module we will write
\begin{equation}
	\dl u:=\d u + \lambda \wedge u
\end{equation}
with $\lambda\wedge u$ including the action of $\p$ (in this paper it will mainly be about products of adjoint and coadjoint representations of $\lor\simeq \p/\m$). Define also
\begin{equation}
	\dll := \d\lambda + \fwb\lambda\lambda = \dl\lambda -\fwb\lambda\lambda
\end{equation}

The operator $\dl$ is meant to model a covariant differential while $\dll$ models a universal curvature 2-form. They satisfy the expected equations (proved in Appendix~\ref{anncalc})
\begin{align}
	\dl\dl u &= \dll\wedge u \label{dldls}\\
	\dl\dll &= 0 \label{annBianchi}\\
	\dl(u^A \wedge v_A) &= (\dl u^A) \wedge v_A + (-1)^{|u|}u^A \wedge \dl v_A \label{eq:dlcontr}
\end{align}
for $u^A$ and $v_A$ homogeneous differential forms with values in dual $\p$-modules.

To compute the Poincaré-Cartan form we will use the following formula from Section~\ref{annLegTrans}: 
\[
	\dd_{v^A_{I,J}}\lr\d \lp \Lag + \pi^J_B \wedge \chi^B_J \rp = 0 \mod [\theta^A_J]
\]
with $\pi^J_B$ $9$-form fields on $\tot$ and $\chi^A_J = \d\lambda^A_J - v^A_{I,J}\d z^I$. Using this formula rather than Formula~\eqref{LTform} will save us some back and forth between the coframes $\d z^I$ and $\lambda^A$. We determine the value of $\pi^J_B$ as a function of $p_A^{BC}$ : 
\begin{equation*}
\begin{aligned}
	\dd_{v^A_{I,J}}\lr\d \lp \Lag + \pi^J_B \wedge \chi^A_J \rp
		&= \dd_{v^A_{I,J}}\lr\d \lp 
			\frac12 p_D^{BC}
					\lp v_{K,L}\d z^L\wedge\d z^K + \fwb\lambda\lambda\rp^D \wedge  \lambda^{(8)}_{BC}
		+ \pi^K_D \wedge \chi^D_K \rp\\
		&= \frac12 p_A^{BC}
					\lp \d z^J\wedge\d z^I - z^I\wedge\d z^J\rp \wedge  \lambda^{(8)}_{BC}\\
		&\qquad+ \lp \dd_{v^A_{I,J}}\lr\d\pi^J_D \rp \wedge \chi^D_J
		- \pi^K_D \wedge \dd_{v^A_{I,J}}\lr \lp - \d v^D_{K,L}\wedge \d z^L \rp \\
		&= p_A^{BC} \d z^J\wedge\d z^I \wedge  \lambda^{(8)}_{BC}
		+ \pi^I_A \wedge \d z^J
		+ \lp \dd_{v^A_{I,J}}\lr\d\pi^J_D \rp \wedge \chi^D_J\\
		&= \d z^J\wedge (- p_A^{BC} \d z^I \wedge \lhuit{BC} + \pi^I_A)
		+ \lp \dd_{v^A_{I,J}}\lr\d\pi^J_D \rp \wedge \chi^D_J
\end{aligned}
\end{equation*}
As the term $\lp \dd_{v^A_{I,J}}\lr\d\pi^J_D \rp \wedge \chi^D_J$ is a contact term the momentum forms $\pi^I_A$ are defined by
\[
	\d z^J\wedge (- p_A^{BC} \d z^I \wedge \lhuit{BC} + \pi^I_A) = 0 
\]
Since $\d z^J$ form a basis of $1$-forms on $\tot$, we conclude that the momenta forms are directly parameterised by the Lagrange multipliers $p_A^{BC}$
\begin{equation*}
	 \pi^I_A = p_A^{BC} \d z^I \wedge \lhuit{BC}
\end{equation*}
and the corresponding contact term in the Poincaré-Cartan form is
\begin{equation}
	p_A^{BC} \d z^I \wedge \lhuit{BC} \wedge \lp \d\lambda^A_I - v^A_{J,I}\d z^J \rp 
		= p_A^{BC} \wedge \lhuit{BC} \wedge \lp \d\lambda^A - v^A_{J,I}\d z^I \wedge \d z^J \rp
\end{equation}
The Poincaré-Cartan form is 
\begin{equation*}\begin{aligned}
	\Lag + p_A^{BC} \wedge \lhuit{BC} \wedge \lp \d\lambda^A - v^A_{J,I}\d z^I \wedge \d z^J \rp
		&=  \frac12 p_A^{BC}
				\lp v_{I,J}\d z^J\wedge\d z^I + \fwb\lambda\lambda\rp^A \wedge  \lambda^{(8)}_{BC}\\
		&\quad 	+ p_A^{BC} \wedge \lhuit{BC} \wedge \lp \d\lambda^A - v^A_{J,I}\d z^I \wedge \d z^J \rp\\
		&= \frac12 p^{BC}_A (\d\lambda + \fwb\lambda\lambda)^A \wedge \lhuit{BC}
\end{aligned}\end{equation*}

We write the Poincaré-Cartan as follows : 
\begin{equation}\label{PCEWC}\boxed{
	\bthewc = \frac12 p^{BC}_A\dll^A \wedge \lhuit{BC}
}\end{equation}
It goes with the holonomic constraint~\eqref{EWCmultcons} on the $p_A^{bc}$. We identify two components : 
\begin{align}
 	\thewc	&= \frac12 2\rho_{i,d}^b\eta^{dc}\dll^i  \wedge \lhuit{bc}\\
 	\thconsewc &= \frac12 p^{jk}_A\dll^A \wedge \lhuit{jk} + p_A^{bk} \dll^A \wedge \lhuit{bk}
\end{align} 
Note that as 
\[
	\fwb\lambda\lambda^i \wedge \lhuit{ab} = \frac12 c^i_{DE} \lambda^D\wedge\lambda^E \lhuit{ab} = c^i_{ab} \ldix = 0
\]
the form $\thewc$ can also be expressed with $\d \lambda$ replacing $\dll$. However the expression $\ExT$ has an interpretation as the curvature, furthermore if we generalise from $\p \simeq \lor\ltimes \m$ to other Lie algebras then $\fwb\lambda\lambda^i\wedge\lhuit{ab}$ may not vanish.

We introduce the notation $\nonbc{BC}$ for pairs of indices of $\p$ \emph{except pairs which correspond to $\m\otimes\m$} (i.e. only pairs which correspond to $(\lor\otimes\lor) \oplus (\m\otimes\lor)\oplus(\lor\otimes\m)$). The Poincaré-Cartan form can then be expressed as
\begin{equation}
	\bthewc = 
	\underbrace{
		\delta^i_A \rho^a_{i,c}\eta^{bc}\dll^A \wedge \lhuit{ab}
		}_{\thewc}
	+ \underbrace{
		\frac12 p^{\nonbc{BC}}_A \dll^A \wedge \lhuit{\nonbc{BC}}
		}_{\thconsewc}
\end{equation}

The Poincaré-Cartan form takes value in the \emph{affine dual} of $\Jun(Q)$ (described in Section~\ref{ssecno:LegTrans}). It is the fibre bundle of $10$-forms on $Q$ that have a vanishing contraction with all $2$-vectors of $Q$ that are purely vertical with respect to the fibration above the source space $\tot$. This affine dual is usually written :
\[
	\ExT^{10}_1 T^*Q
\]

Note that the coefficients of Poincaré-Cartan form have no dependency on the first order component of the $1$-jet, so that the only dependency comes from the factor \[\dll^A=\d\lambda^A+\fwb\lambda\lambda^A\]
We can therefore restrict the momentum space from the whole affine dual to $Q$ itself, on which $\bthewc$ can still be expressed as a $10$-form, along with the supplementary constraint~(\ref{EWCmultcons}) (for details on the equivalence between the Lagrangian and the Hamiltonian formulations see the last section of \cite{ConsFT} on affine Lagrangians). One can also say that $\frac12 p^{\nonbc{BC}}_A\lhuit{\nonbc{BC}}$ correspond to specific elements of the linear dual space of $\Jun(Q)$ of the form $\dll^A \wedge \frac12 p_A^{\nonbc{BC}} \lhuit {\nonbc{BC}}$, according to~(\ref{multvirg}). In fact, the Lagrange multiplier terms are derived in \cite{LFB} as free momenta obtained by a Legendre transformation under constraint.

In fact it could be argued that \eqref{PCEWC} is the natural formulation on $\ExT^8 T^*\tot \otimes \p^* \times_\tot Q$ of the Lagrangian given in \eqref{eqno:ECLagomp}, rather than \eqref{eqno:LagvIJ}. We chose to go for the \enquote{naive} formulation of the Lagrangian on the $1$-jet bundle to illustrate the systematic approach to the Legendre transformation.

We derive the variational equations in the next section.

\section{Variational equations for gravitation}\label{VEqGR}
In this section we compute the variational equations corresponding to the Poincaré-Cartan form~(\ref{PCEWC}). The \emph{Euler-Lagrange form} is defined for vertical vector fields $X$ on $Q$ as
\begin{equation}
	\EL_X := i_X\d \bthewc
\end{equation}
As the Poincaré-Cartan form is defined on $Q$ there is no need to use the $1$st prolongation of $X$ as described in Section~\ref{annLegTrans}.

The Euler-Lagrange equations consist in the vanishing of all its components (with respect to $X$) under the pullback by a field $(\varpi,P) : \tot \to \Isototp \times_\tot \momQ$ :
\[
	\forall X,\quad (\varpi,P)^*\EL_X = 0
\]
A local basis of the vertical vector fields is given by the integrable fields $\ddl{A}{I}$ and $\ddp{\nonbc{BC}}{A}$. For convenience, we define $\ddl A B = \lambda^B_I\ddl A I$ so that
\begin{equation}
	\ddl A B \lr \dll^C
		= \ddl A B \lr \d\lambda^C + \ddl A B \lr \fwb\lambda\lambda^C
		= (\lambda^B_I \ddl A I) \lr \d\lambda^C + 0
		= \delta^C_A\lambda^B_I dz^I 
		= \delta^C_A\lambda^B
\end{equation}
We impose on the field space that $(\lambda^A)$ form a nondegenerate family (coframe) at each point of $\tot$, consequently we can substitute $\ddl A B$ for $\ddl A I$ in the basis of vertical vector fields. One can interpret $\ddl A B$ as representing the infinitesimal action of $\End(\p)$ on $\Isototp$.

We will make use of the following identities, proved in Appendix~\ref{anncalc} :
\begin{align*}
	\d \ldix 		&= \dll^D  \wedge \lneuf D	 			\\
	\d \lneuf A		&= \dll^D \wedge \lhuit{AD}	 			\\
	\d\lhuit {AB} &= 	\dll^D \wedge \d\lsept{ABD}-c^D_{AB}\lneuf D	 \\
	\dl \dll^A &= 	0				
\end{align*}
\subsection{\texorpdfstring{Variation of the multiplier $P^{\nonbc{BC}}_A$}
							{Variation of the multiplier P}}

We check that the variation of the Lagrange multipliers yields the expected constraint equations. Recall that we do not allow $BC$ to take the form $bc$ in $\ddp{\nonbc{BC}}A$. We use the notation $(\EL_{EC})^A_{BC}$ for $(\EL_{EC})_{\dd^{BC}_A}$ :
\begin{equation}\begin{aligned}\label{eqno:grcons}
	(\EL_{EC})^A_{\nonbc{BC}} &= \ddp{\nonbc{BC}}{A} \lr \d \lp \frac12 p^{EF}_D\dll^D \wedge \lhuit{EF} \rp	
	= \frac12 \lp \ddp{\nonbc{BC}}{A} \lr \d p^{EF}_D\rp \dll^D \wedge \lhuit{EF}
	= \frac12\dll^A \wedge \lhuit{\nonbc{BC}}
\end{aligned}\end{equation}
We write $\Omega:=\varpi^*\dll$ which has as components
\[
	\Omega^A = \frac12\Omega^A_{BC} \varpi^B\wedge\varpi^C
\]
A critical field $(P,\varpi)$ satisfies
\begin{equation} \Omega^A_{\nonbc{BC}} = 0  \label{ELMC}\end{equation}
which is equivalent to Equations \eqref{seqno:equivvarpi8}. Thus a solution $\varpi$ of the Euler-Lagrange equations defines a generalised Cartan connection and a generalised frame bundle structure on $\tot$.

\subsection{\texorpdfstring{Variation of the coframe $\varpi$}
							{Variation of the coframe}}

Instead of using $\ddl A B$, it will be convenient to use a vertical vector field $X$ on $\Isototp$ which has variable coefficients $\eAB$ as follows :
\[
	X = \eAB\ddl A B
\]
We also gather these coefficients into a $\p$-valued $1$-form :
\[
	\epsilon^A = \epsilon^A_B\lambda^B
\]
so that 
\begin{equation}
	X\lr\d\lambda^A = \Lie_X\lambda^A = \epsilon^A
\end{equation}
The correspondence $X\leftrightarrow \epsilon$ is the usual identification between the vertical tangent bundle to a vector bundle and the Whitney sum of the vector bundle with itself.

We mention one more identity from Appendix~\ref{anncalc} :
\[
\d \lp \bs_{[A]} s^{[A]} \rp = (\dl \bs)_\alpha \wedge s^{[A]} + \bs_\alpha \dl s^{[A]}
\]
for $[A]$ any index, or index list, in a $\lor$-module. 
We now compute $\EL_X$ : 
\begin{equation*}\begin{aligned}
		X\lr \d \lp \frac12 p^{BC}_A\dll^A \wedge \lhuit{BC} \rp
		&= X\lr \lp \dl \dll^D  \wedge \frac12 p^{BC}_A\lhuit{BC}
		+ \dll^A  \wedge \dl \lp \frac12 p^{BC}_A\lhuit{BC} \rp \rp\\
		&= X \lr \lp 0 + \dll^A  \wedge \dl \lp \frac12 p^{BC}_A\lhuit{BC} \rp \rp\\
		&= \lp X \lr \dll^A \rp \wedge  \dl \lp \frac12 p^{BC}_A\lhuit{BC} \rp
		+ \dll^A \wedge X \lr \dl \lp \frac12 p^{BC}_A\lhuit{BC} \rp		
\end{aligned}\end{equation*}
On one hand
\begin{equation*}
	X\lr \dll = X\lr \lp \d\lambda + \fwb\lambda\lambda \rp = \epsilon + 0
\end{equation*}
and on the other hand
\begin{equation*}
\begin{aligned}
	X\lr \lp \dl \lp \frac12 p^{BC}_D\lhuit{BC} \rp \rp 
		&= X\lr \lp \lp \dl \frac12 p^{BC}_D \rp \wedge \lhuit{BC} \rp 
		+ X\lr \frac12 p^{BC}_D \lp \dl \lhuit{BC} \rp\\
		&= 0 + X \lr  \frac12 p^{BC}_D  \lp \dll^A \wedge \lsept{BCA} \rp \\
		&= \epsilon^A\wedge \frac12 p^{BC}_D \lsept{BCA}
\end{aligned}
\end{equation*}
Gathering the two terms, we obtain
\begin{equation}\label{eqno:grdyn}
\boxed{
	\EL_X =
		\epsilon^A\wedge \lp 
			\dll^D \wedge \frac12 p^{BC}_D \lsept{BCA}
		+
			\dl \lp \frac12 p^{BC}_A\lhuit{BC} \rp \rp
}\end{equation}

The corresponding Euler-Lagrange equations on a field $(\varpi,P)\in\Gamma(\tot,Q)$ are then :
\begin{equation*}
	\forall \epsilon\in\Omega^1(\tot,\p), \qquad
		\epsilon^A\wedge \lp
			\Omega^D \wedge \frac12 P^{BC}_D \varpi^{(7)}_{ABC}
		+
			\d^\varpi \lp \frac12 P^{BC}_A\varpi^{(8)}_{BC} \rp \rp
		=0
\end{equation*}
where $\epsilon^A_B$ are defined over $\tot$ since we consider $X$ which are variations of the field $\varpi$. The equations are equivalent to
\begin{equation}\label{eqno:EqGRvarpi}
			\Omega^D \wedge \frac12 P^{BC}_D \varpi^{(7)}_{ABC}
		+
			\d^\varpi \lp \frac12 P^{BC}_A\varpi^{(8)}_{BC} \rp
		=0
\end{equation}

\subsubsection*{The Einstein term}

We now explain how the usual Einstein tensor can be identified in \eqref{eqno:EqGRvarpi}. First, we assume that Equation~\eqref{ELMC} is satisfied, so that $\varpi$ defines a generalised frame bundle structure. We will identify tensors built out of $\varpi$ which correspond to the various curvature and the torsion tensors in the standard frame bundle case. For more detail on curvature on the frame bundle, see Appendix~\ref{annFrameB}.

Let us isolate the part depending the fixed momenta $p_D^{bc}=2 \delta^l_D \rho_{l,e}^b\eta^{ec}$ : we obtain
\begin{equation*}\begin{aligned}
			\Omega^l \wedge \frac12 p^{bc}_l \varpi^{(7)}_{Abc}
		+
			\delta^i_A \d^\varpi \lp \frac12 p^{bc}_i\varpi^{(8)}_{bc} \rp
		&=
			\Omega^l \wedge \frac12 p^{bc}_l \varpi^{(7)}_{Abc}
		+
			\delta^i_A \frac12 p^{bc}_i \d^\varpi \lp \varpi^{(8)}_{bc} \rp\\
		&=
			\Omega^i \wedge \frac12 p^{bc}_i \varpi^{(7)}_{Abc}
		+
			\delta^i_A \frac12 p^{bc}_i \Omega^D \wedge \varpi^{(7)}_{bcD}\\
		&= \lp
			\delta^D_A\Omega^i  
		+
			\delta^i_A \Omega^D 
			\rp \wedge \frac12 p^{bc}_i \varpi^{(7)}_{bcD}
\end{aligned}\end{equation*}
Now assuming that Equation~\eqref{ELMC} is satisfied so that $\Omega^A = \frac12\Omega^A_{bc}\alpha^a\wedge\alpha^b$, we compute the wedge product:
\begin{equation*}\begin{aligned}
		&\lp
			\delta^D_A\Omega^i  
		+
			\delta^i_A \Omega^D 
			\rp \wedge \frac12 p^{bc}_i \varpi^{(7)}_{bcD}\\
		={}& 
		\frac12 p^{bc}_i \lp
			\delta^D_A \lp 
				 \Omega^i_{cD} \vneuf b - \Omega^i_{bD} \vneuf c + \Omega^i_{bc} \vneuf D 
			\rp
		+
			\delta^i_A \lp
				\Omega^D_{cD} \vneuf b - \Omega^D_{bD} \vneuf c + \Omega^D_{bc} \vneuf D
			\rp\rp\\
		={}& 
		\frac12 p^{bc}_i \lp
			2 \Omega^i_{Ab} \vneuf c + \Omega^i_{bc} \vneuf A
		+
			\delta^i_A \lp
			 2 \Omega^d_{db} \vneuf c + \Omega^D_{bc} \vneuf D
			\rp\rp
\end{aligned}\end{equation*}
We want to separate the terms according to their dependency on $A$ (of type $a$ or $i$) and $\vneuf{}$ (with a subscript $c$ or $j$) : 
\begin{multline*}
		\lp
			\delta^D_A\Omega^i  
		+
			\delta^i_A \Omega^D 
			\rp \wedge \frac12 p^{bc}_i \vsept{bcD}
		=
			 \lp \delta^a_A p_i^{bc}\Omega^i_{ab}
			 	+ \delta^c_a \frac12 p_i^{de}\Omega^i_{de} \rp \vneuf c\\
		+
			\delta^i_A \lp
				\lp p_i^{bc} \Omega^d_{db} 
					+ \frac12 p_i^{de} \Omega^c_{de} \rp \vneuf c
				+ \lp \frac12 p_i^{de} \Omega^j_{de}
				+ \frac12 p^{bc}_k \Omega^k_{bc} \delta^j_i \rp \vneuf j
			\rp
\end{multline*}
Now we just have to rewrite the factors in front of $\vneuf c$ so as to get rid of $p$. Recall the definition
\[p_i^{bc} = 2\rho^{b}_{i,d} \eta^{dc}\]
The first term is
\begin{align*}
	p_i^{bc}\Omega^i_{ab} + \delta^c_a \frac12 p_i^{de}\Omega^i_{de}
		= 2\rho^{b}_{i,f} \eta^{fc} \Omega^i_{ab} + \delta^c_a \rho^{d}_{i,f} \eta^{fe} \Omega^i_{de}
\end{align*}
We recognize the contractions of $\Omega^i$ : these are the components of the tensor
\begin{equation}
	-2\Ric_{a,f} \eta^{fc} + \delta^c_a \Scal 
\end{equation}
which is (minus twice) the Einstein tensor.

For the second term, we will use the following property : for any tensor field $A$ and any list of indices $[D]$, we have
\[
	p_i^{bc}A_{bcD} = 0 \Leftrightarrow A_{bc[D]} - A_{cb[D]} = 0
\]
This is a consequence of the definition $p_i^{bc} = 2\rho^{b}_{i,e} \eta^{ec}$ interpreted as an isomorphism $\lor \xrightarrow{\sim}  \ExT^2 \Mink$. We have
\begin{align*}
	 p_i^{bc} \Omega^d_{db} + \frac12 p_i^{de} \Omega^c_{de}
	 	= p_i^{de} \lp \delta_e^c \Omega^f_{fd} + \frac12 \Omega^c_{de} \rp
	 	= \frac12 p_i^{de} \lp \Omega^c_{de} + \delta_e^c \Omega^f_{fd} - \delta^c_d \Omega^f_{fe} \rp
\end{align*}
in which the antisymmetric term
\begin{equation*}
	\Omega^c_{de} + \delta^c_e \Omega^f_{fd} - \delta^c_d \Omega^f_{fe} 
\end{equation*}
corresponds to the components of the tensor field
\begin{equation}\label{eqno:Ttrace}
	T + \tr(T)\wedge \Id
\end{equation}
This is a contraction of the torsion which is quite similar to the Ricci curvature. Its divergence is non zero and actually equates (twice) the antisymmetric part of the Ricci tensor.

These tensor are to be equated in \eqref{eqno:EqGRvarpi} with quantities dependent on the multipliers $P_A^{\nonbc{BC}}$. We will see how to get rid of the multipliers in order to extract meaningful field equations building on Section~\ref{LagMult}.

\subsubsection*{Comparison with gravitation on \enquote{soft Poincaré manifolds}}

In~\cite{SUGRAPrimer} is presented a very similar term for the so-called gravitation on \enquote{soft Poincaré manifolds}.
In our language, they take the following form :
\begin{align*}
	\frac12 p^{ab}_i \lp \delta^c_A \Omega^i \wedge \alpha^{(1)}_{cab} + \delta^i_A \Omega^c \wedge \alpha^{(1)}_{cab} \rp
\end{align*}
and its vanishing is equivalent to
\[\begin{cases}
		\frac12 p^{ab}_i \Omega^i \wedge \alpha^{(1)}_{cab} &=0\\
		\frac12 p^{ab}_i \Omega^c \wedge \alpha^{(1)}_{cab} &=0
\end{cases}\]
These equations imply the Einstein field equation, the vanishing of the term \eqref{eqno:Ttrace} \emph{as well as Equation~\eqref{ELMC}}. However, as our theory is formulated with $10$-forms, there are in our equations factors of $\omega^i$ due to the term $\lsept{bcD}$ in \eqref{eqno:EqGRvarpi}. They weaken the constraint imposed by the equation on the non-horizontal components $\Omega^A_{bk}, \Omega^A_{jk}$. For this reason, we needed to add the Lagrange multipliers $p_A^{\nonbc{BC}}$ in Section~\ref{secno:FBdyn} in order to enforce Equation~\eqref{ELMC}.

\section{Dirac spinors on the spinor frame bundle}\label{spinor}
\subsection{Building the spinor Lagrangian}

\subsubsection*{The Dirac Lagrangian}

We want to use the same method as in Section \ref{secno:FBdyn} to build a Lagrangian term on generalised frame bundles for \emph{Dirac spinors} in a dynamical spacetime. The suitable frame bundle is not the frame bundle of spacetime $\FrameB$, but the \emph{spin frame bundle} $\SpB$ instead. It is a twofold covering of the (direct orthochronous orthonormal) frame bundle which depends on a choice of spin structure, which is an extra topological structure on the spacetime. As it is a covering of the standard frame bundle, we can lift the construction from the previous section to the spin frame bundle. As stated in Appendix~\ref{annMC}, the principal bundle structure we can obtain from Equations~(\ref{ELMC}) has as structure group either $\Lor$ or its twofold cover $\Spinp_{1,3}$. The presence of a non-zero spinor field requires the group to be $\Spinp_{1,3}$.

Let $\Sp$ an irreducible complex $\Cl_{1,3}$-module, provided with a \emph{spinor metric} : a (possibly indefinite) Hermitian product for which $\setR^{1,3}$ has an antisymmetric action (see Appendix~\ref{annspin}). Recall the (adimensional) Dirac Lagrangian~\cite{WeinbergI,EMTens} :
\begin{equation}
	\frac{1}2\bp \DirGD \psi
	-m\bp \psi
\end{equation}
in which the spinor contraction is implicit : \[\bp_1\psi_2 :=\bp_{1\alpha}\psi^\alpha_2\] 
The symmetrized Dirac operator $\DirGD$ is defined by
\[\bp_1\DirGD\psi_2
	= \bp_1\gamma^\mu\nabla_\mu\psi_2 - (\nabla_\mu \bp_1)\gamma^\mu\psi_2
	= 2\Re\lp \bp_1 \Dirac \psi_2 \rp\]

Note that the $\alpha$ subscript (resp. index) corresponds to a \emph{complex} basis of $\Sp$. The absence of an $i$ factor is due to our conventions for Clifford algebra and Lorentzian signature~\cite{PinGroups}. To keep in line with the traditional treatment of spinors fields, we will work with complex indices and make use of the \enquote{holomorphic} and \enquote{anti-holomorphic} directions.

\subsubsection*{Lift to the spin frame bundle}

On the spin frame bundle, spinors are represented by $\Spinp_{1,3}$-equivariant $\Sp$-valued fields so that the configuration space of the spinor field is $\Sp\times\SpB$. We will use $s=(s^\alpha)$ as coordinates in the fibre $\Sp$ and $(\bs_{\alpha})$ for dual coordinates ; the latter can be read as the spinor metric $\bs : \Sp\to \bSp$.

We write $\gamma_a\in \End(\Sp)$ for the action of vectors $e_a\in \setR^{1,3}$, use $\sigma_i\in\End(\Sp)$ for the action of $\h_i\in\lor\simeq\spin_{1,3}$ (resp. $\bar\sigma_i\in\End(\bSp)$) and $\psi:\SpB\to\Sp$. More detail on the notation and the action of the $\Spin$ group on Clifford modules will be found in Appendix~\ref{annPinSpin}. 
Given a system $(z^I)$ of local coordinates on $\tot$, we call $\zeta^\alpha_I$ the associated coordinates on the $1$-jet bundle $\Jun(\SpB,\Sp)$ :
\[
	\zeta^\alpha_I \circ \psi = \dd_I\psi^\alpha
\]
The covariant derivative of the spinor field formulated on the frame bundle takes the form 
\[\d\psi + \varpi^i\sigma_i\cdot\psi\]
The pulled-back Dirac Lagrangian form can be expressed as :
\begin{equation}
	\Lag_{Dirac} = \lp \frac{1}2
		\lp \bs \gamma^a (\zeta_J\d z^J + \lambda^i\sigma_i s)
		- (\bar\zeta_J\d z^J+ \lambda^i\bar\sigma_i\bs) \gamma^a s \rp
		 \wedge \lambda^{\m (3)}_a
		-m\bs s \lambda^\m 
		\rp \wedge \lambda^\l
\end{equation}
where $\lambda^\m$ (resp. $\lambda^\l$) denotes the pullbacks of volume elements of $\m$ (resp. $\lor$) by the canonical form $\lambda$ and $\lambda^{\m(3)}_a$ is the $3$-form dual to $\lambda^a$ in $\m$. Note that in this expression the contribution of the term  $\lambda^i\sigma_i$ (resp. $\lambda^i\bar\sigma_i$) actually vanishes due to the $\lambda^{\m(3)}_a \wedge \lambda^\lor=\lneuf a$ factor which already selects the horizontal directions from $\zeta_{\alpha,J}\d z^J$ (resp. $\zeta^\alpha_J\d z^J$).

\subsubsection*{Dropping the frame bundle structure and Lagrange multipliers}

As in the previous section, we consider as source space for the fields a differentiable $10$-manifold $\tot$. We will consider a total Lagrangian composed of $\Lag_{Dirac}$, $\Lag_{EC}$ and Legendre multipliers terms in order to make the $\Spinp_{1,3}$ structure emerge dynamically on the space $\tot$. As mentioned earlier, the structure obtained from Equations \eqref{seqno:equivvarpi8} on $\varpi$ induce a action of the Lie algebra $\soL$, which is naturally isomorphic to $\spin_{1,3}$.

In particular, it is enough to define the equivariance of the spinor fields under $\spin_{1,3}$. Indeed, the difference between the usual linear frame bundle and spinor frame bundles only appears at the \enquote{global} level, when there are complete orbits under the group action. Thus we do not have to adapt our notion of generalised frame bundle in order to accommodate for spin structures.

A spinor field $\psi : \tot\to\Sp$ will have to satisfy the equivariance condition
\begin{equation}\label{eqno:psiequiv}
	\Lie_\bh \psi + \h\cdot\psi = 0
\end{equation}
with $\lor$ acting via $\sigma : \lor\to\End(\Sp)$, we write $\h_i\cdot\psi = \sigma_i\psi$. We will formulate the equivariance in a similar way to previously : recall the notation
\begin{subequations}\begin{align}
	\dl s  &:=\d s   + \lambda^i\sigma_i s\\
	\dl\bs &:= \d \bs + \lambda^i \bar\sigma_i \bs
\end{align}\end{subequations}
with the operators $\sigma_i$ being anti-selfadjoint. This notation allows us to write the condition \eqref{eqno:psiequiv} (writing separately $\setC$-linear and $\setC$-antilinear directions although they correspond to the same degree of freedom)

\begin{subequations}
\begin{empheq}[left=\empheqlbrace]{align}
	\psi^* \lp \dl s  \wedge \lneuf i \rp &= 0\\
	\psi^* \lp \dl \bs\wedge \lneuf i \rp &= 0
\end{empheq}
\end{subequations}
We consider the following Lagrange multiplier term (using a similar notation $\dl \bs$)
\begin{equation*}
	\frac{i}2 \lp\bar\kappa_\alpha^i \dl s^\alpha - \kappa^{\alpha i}\dl \bs \rp \wedge \lneuf i 
\end{equation*}
with $\bar\kappa^i$ conjugate to $\kappa^i$ (so that the constraint term is real), which added to $\Lag_{Dirac}$ makes the following Lagrangian
\begin{multline}\label{LagDircons}
	\overline{\Lag_{Dirac}} = \lp \frac{1}2
		\lp \bs \gamma^a (\zeta_J\d z^J + \lambda^i\sigma_i s)
		- (\bar\zeta_J\d z^J+ \lambda^i\bar\sigma_i\bs) \gamma^a s \rp
		 \wedge \lambda^{\m (3)}_a
		-m\bs s \lambda^\m 
		\rp \wedge \lambda^\l\\
		+ \frac i 2 \lp
		\bar\kappa^i (\lambda^J\dd_J + \lambda^j\sigma_j) s
		- \kappa^i (\lambda^J\dd_J + \lambda^j\bar\sigma_j) \bar s \rp \wedge  \lambda^{(9)}_i
\end{multline}
defined over 
\[\lp \underbrace{\Sp}_{s^\alpha}
	\oplus \underbrace{\Sp\otimes\lor}_{\kappa^{\alpha i}} \rp
	\times \underbrace{\Isototp}_{\lambda^A_I} \]

\subsection{The Poincaré-Cartan form}

We now compute the Poincaré-Cartan form. The Lagrangian~(\ref{LagDircons}) being affine in the 1st-order jet, the Legendre transformation is straightforward. The image of the Legendre transform, namely the \emph{momentum space}, is a subspace of :
\begin{multline*}
	\ExT^{10}_1 T^* \left[
		\lp \Sp	\oplus \Sp\otimes\lor \rp
			\times \Isototp \right]
		\\
	\simeq 
		(\Sp\oplus \Sp\otimes\lor)
		\times_\tot \left[
						(\bSp\oplus \bSp\otimes\lor^*)\otimes\ExT^{9} T^*\tot
						\oplus_\tot \ExT^{10} T^*\tot
					\right] 
		\times_\tot \ExT^{10}_1 T^* \Isototp 
\end{multline*}
We use the Legendre transformation formula~(\ref{LTform}) (see Section~\ref{annLegTrans}) :
\begin{align}
	&\begin{multlined}
		\bthdir =\overline{\Lag_{Dirac}} 
		+ \der{v^A_{I,J}}{\overline{\Lag_{Dirac}}}
		(\d\lambda^A_I - v^A_{I,L} dz^L) \wedge dz^{(9)}_J
		\\
		\qquad \qquad
		+ \der{\zeta_J^\alpha}{\overline{\Lag_{Dirac}}}
		(\d s^\alpha - \zeta_L^\alpha dz^L) \wedge dz^{(9)}_J
		+ \der{\zeta_{\alpha,J}}{\overline{\Lag_{Dirac}}}
		(\d\bs_\alpha - \zeta_{\alpha,L} dz^L) \wedge dz^{(9)}_J
	\end{multlined}\nonumber\\
	&\qquad\quad 
	=\frac{1}2 \lp
	\bs \gamma^a (\dl s)
	- (\dl\bs) \gamma^a s \rp
	 \wedge \lambda^{(9)}_a
	+ i \frac 1 2\lp \bar\kappa^i (\dl s)
	- (\dl\bs) \kappa^i \rp
	 \wedge \lambda^{(9)}_i
	-m\bs s \lambda^{(10)}
\end{align}
The momentum dual to $\kappa$ vanishes as $\kappa$ only appears in the Lagrangian at order $0$ so that the momenta have trivial component in $(\lor^*\otimes\Sp)\otimes\ExT^{9} T^*\tot$.

We are interested in a model coupling the Dirac spinor with the Einstein-Cartan gravitational fields. We thus consider a Lagrangian which is the sum of the two Lagrangians and is defined over 
\[ Q =
	\Isototp
 	\times \Sp
 	\times_\tot \big[ \momQ
 	 \oplus (\Sp\otimes\lor)\big]
\]
the whole Poincaré-Cartan form decomposes as follows
\begin{equation}\label{PoinCarSp}
	\bar\Theta 		= \thewc + \thdir  + \thconsewc + \thconsdir
				= \bthewc + \bthdir
\end{equation}
with
\begin{align}
	\bthewc &= \frac12 p^{BC}_A\dll^A\lhuit{BC}
	 \\
	\bthdir &= \frac{1}2 \lp
		\bs \gamma^a (\dl s)
		- (\dl\bs) \gamma^a s \rp
	\lambda^{(9)}_a
		+ i\frac 1 2\lp \bar\kappa^i (\dl s)
		- (\dl\bs) \kappa^{i} \rp
	\lambda^{(9)}_i
		-m\bs s \lambda^{(10)} 
\end{align}
with the line over $\overline{\theta}$ denoting the inclusion of the Lagrange multiplier terms.

Here as well, the Poincaré-Cartan form is defined on the configuration space to which we added the Lagrange multipliers (the coefficients only depend on the 0-order jet).
In accordance with the results from Section~\ref{VEqGR} the momenta dual to $\lambda^A_I$ are restricted to the subspace $\ExT^8 T^*P\otimes\p^* \times_\tot \Isototp$ of $\ExT^{10}_1 T^*\tot\otimes\p$.
For convenience working with complex spinor indices, we enlarge the component $\Sp\otimes\ExT^{9} T^*\tot$ to a factor $(\Sp\oplus \bSp)\otimes\ExT^{9} T^*\tot$. 
We define fibre coordinates on the momentum space using the components of the canonical $10$-form :
\begin{equation*}
		  j^{BC}_{A}\dll^A \wedge \lhuit{BC}
		+ \bPh^A_{\alpha}\dl s^\alpha \wedge  \lneuf{A}
		+ \Phi^{\alpha A} \dl\bs_{\alpha} \wedge  \lneuf{A}
		- h\lambda^{(10)}
\end{equation*}
Each coordinate corresponds to a factor of the momentum space, as follows :
\begin{equation*}
	\big( 
		\underbrace{\Isototp}_{\lambda^A_I}
		\oplus \underbrace{\Sp}_{s^\alpha}
		\oplus \underbrace{\Sp\otimes\lor}_{\kappa^{\alpha j}}
	\big)
	\times_\tot \underbrace{\ExT^8 T^*P\otimes\p^*}_{j^{BC}_A}
	\times_\tot 
	\big[  	\underbrace{ (\Sp \oplus \bSp )\otimes \ExT^{9} T^*\tot
						}_{\Phi^{\alpha A},\bPh_\alpha^A}
			\oplus_\tot
			\underbrace{\ExT^{10} T^*\tot}_h
		\big] 
\end{equation*}

The image of the Legendre transform is a subspace defined by (holonomic) constraints, which take the following form

\renewcommand\arraystretch{1}
\begin{center}
\begin{subequations}
\begin{tabularx}{\textwidth}{@{}|X|X|@{}}
	\hline
	\begin{equation}\label{eqno:jabc}
		j^{bc}_a = 0
	\end{equation}
	&
	\begin{equation}\label{eqno:jabi}
		j^{bc}_i = 2\rho_{id}^b \met^{dc}
	\end{equation}
	\\[-15pt]
	\midrule
	\begin{equation}
		j^{bj}_A = p^{bj}_A
	\end{equation}
	&
	\begin{equation}
		j^{ij}_A = p^{ij}_A
	\end{equation}
	\\[-15pt]
	\midrule
	\begin{equation}
		\Phi^{\alpha a} = -\frac 1 2 (\gamma^a s)^\alpha
	\end{equation}
	&
	\begin{equation}
		\Phi^{\alpha j} = -\frac i 2 \kappa^{\alpha j}
	\end{equation}
	\\[-13pt]
	\midrule
	\begin{equation}
		\bPh^a_\alpha   = -\frac 1 2 (\bar\gamma^a \bs)_\alpha
	\end{equation}
	&
	\begin{equation}
		\bPh^j_\alpha   =\frac i 2 \bar\kappa^j_\alpha
	\end{equation}
	\\[-13pt]
	\midrule
	\begin{equation}
	h = m\bs_\alpha s^\alpha
	\end{equation}
	&
	\\[-2pt] \bottomrule
\end{tabularx}
\end{subequations}
\end{center}
The momentum space can hence be identified as
\begin{equation}
	\big[ \underbrace{\Isototp}_{\lambda^A_I}
	\times \underbrace{\Sp}_{s^\alpha} \big]
	\times_\tot \big[ \underbrace{\momQ}_{p^{\nonbc{BC}}_A}
	 \oplus \underbrace{(\Sp\otimes\lor)}_{\kappa^{\alpha i}} \big]
	 = Q
\end{equation}
defined by the two constraints~(\ref{eqno:jabc},\ref{eqno:jabi}).

We compute the variational equations in the next section.

\section{Variational equations for a spinor on a generalised frame bundle}\label{VEqSp}
We already have the Euler-Lagrange forms corresponding to $\bthewc$, we have to compute the term corresponding to $\bthdir$. Recall its expression
\begin{equation}
	\bthdir =
	\frac{1}2 \lp
		\bs \gamma^a (\dl s)
		- (\dl\bs) \gamma^a s \rp
	 \wedge \lambda^{(9)}_a
	+ \frac i 2\lp \bar\kappa^i (\dl s)
		- (\dl\bs) \kappa^i \rp
	 \wedge \lambda^{(9)}_i
	-m\bs s \lambda^{(10)} 
\end{equation} 

We will make use of the following identities, proven in Appendix \ref{anncalc} :
\begin{equation}\tag{\ref{dldls}}\begin{aligned}
	\dl(\dl s)^\alpha &=	(\dll \cdot s)^\alpha	\\
	\dl(\dl\bs)_\alpha &= (\dll \cdot \bs)_\alpha
\end{aligned}\end{equation}

Let $\phi$ be a section of the phase space. We call its different components as follows :
\begin{equation}
	\big[ \underbrace{T^*\tot \otimes\p}_{\varpi^A_I}
	\times \underbrace{\Sp}_{\psi^\alpha} \big]
	\times_\tot \big[ \underbrace{(\ExT^8 T^*\tot\otimes \lor)}_{P^{BC}_A}
	 \oplus \underbrace{(\ExT^9 T^*\tot\otimes\Sp\otimes\lor)}_{K^{\alpha i}} \big]
\end{equation}

\subsection{\texorpdfstring{Variation of the multipliers $K^{\alpha i}$}
			{Variation of the multipliers K}}

We check here that the variation of the Lagrange multipliers $K_\alpha^i$ yield the expected constraint equations. There are independent variations in holomorphic directions corresponding to the index $\alpha$, and in the anti-holomorphic direction corresponding to $\bar K^{\alpha i}$.

\begin{align}\label{eqno:Spcons}\begin{split}
	\lp \EL_{Dirac} \rp_{\alpha i} \\
	=
	\dd_{\kappa^{\alpha i}}\lr
	&\d \lp \frac{1}2 \lp
	\bs \gamma^a (\dl s)
	- (\dl\bs) \gamma^a s \rp
	 \wedge \lambda^{(9)}_a
	+ \frac i 2\lp \bar\kappa^j_\beta (\dl s)^\beta
	- (\dl\bs)_\beta \kappa^{\beta j} \rp
	 \wedge \lambda^{(9)}_j
	-m\bs s \lambda^{(10)} \rp\\
	= -\frac{i}2\dl&\bs_\alpha  \wedge \lneuf i
	\end{split}
\end{align}

As $\bthdir$ is real the two Euler-Lagrange terms are conjugate under the antilinear correspondence $\Sp\to\bSp$ :
\begin{equation} \label{eqno:bSpcons}
	\lp \EL_{Dirac} \rp_i^\alpha = \frac{i}2\dl s^\alpha \wedge  \lneuf i	
\end{equation}
From now on we will only use the Euler-Lagrange terms coupled to anti-holomorphic $\dd_{\kappa^{\alpha i}}$ variations.
The corresponding Euler-Lagrange equations on a field $\phi$ are thus :
\begin{equation}\label{eqno:Spconsphi}
	\phi^* \lp \dl s^\alpha \wedge  \lneuf i \rp = \dom \psi^\alpha \wedge\vneuf i = 0
\end{equation}

Therefore, if we assume there is a principal bundle structure derived from the variational equations derived in Section \ref{VEqGR}, the vanishing of the pullback of the Euler-Lagrange form $\phi^*{\EL}^\alpha_i$ is equivalent to requiring the equivariance of the corresponding section $\psi$ of $\Sp$ with respect to the action of $\lor$ defined by $\varpi$. Namely, in the case of the spin frame bundle, it means that $\psi$ is associated to a section of the associated spinor bundle.

\subsection{Variation of the spinor field and of the coframe}

We start with the variational equations with respect to the variation on the spinor fields. We have to compute $\d\bthewc$. For the sake of clarity we will compute 
\[	\d\lp 
		\bs \gamma^a (\dl s)
		 \wedge \lambda^{(9)}_a
	+ i \bar\kappa^i (\dl s)
		 \wedge \lambda^{(9)}_i
	-m\bs s \lambda^{(10)} 
	\rp 
\]
and conclude is has $\d\bthewc$ as real part. 
Let us start with the term dependent of $\kappa^i$ : 
\begin{equation*}\begin{aligned}
	\d \lp \bar\kappa^i (\dl s)
		 \wedge \lambda^{(9)}_i \rp
	&= \lp \dl \bar\kappa^i  \wedge (\dl s)
		 + \bar\kappa^i \dl(\dl s) \rp
		 \wedge \lambda^{(9)}_i
		- \bar\kappa^i (\dl s)
		 \wedge \dl \lambda^{(9)}_i\\
	&= \lp \dl \bar\kappa^i \wedge  (\dl s)
		+ \bar\kappa^i \dll \cdot s \rp
		 \wedge \lambda^{(9)}_i
		- \bar\kappa^i (\dl s)
		 \wedge \dll^B \wedge  \lhuit {iB} \\
	&= \dl \bar\kappa^i  \wedge \dl s
		 \wedge \lambda^{(9)}_i
		+ \bar\kappa^i (\sigma_j s)
		 \wedge  \lneuf i
		- \bar\kappa^i (\dl s)
		 \wedge \dll^B  \wedge \lhuit{iB}
\end{aligned}\end{equation*}
An identical calculation replacing $\kappa^{\alpha i}\lneuf i$ with $\gamma^a s^\alpha\lneuf a$ gives
\begin{equation*}\begin{aligned}
	\d  \lp
		\bs \gamma^a (\dl s)
		 \wedge \lambda^{(9)}_a \rp
	= \lp \dl (\bs \gamma^a) \wedge \dl s
		+ \bs \gamma^a (\sigma_j s) \dll^j
		 \rp \wedge \lambda^{(9)}_a
	- \bs \gamma^a (\dl s)
		 \wedge \dll^B  \wedge \lhuit{aB}
\end{aligned}\end{equation*}
Recall that $\gamma^a$ is parallel in the following sense :
\[ (\dl \gamma^a)  \wedge \lneuf b = \lp \d\gamma^a + \lambda^j[\sigma_j,\gamma^a] \rp \wedge  \lneuf b = 0 + 0 \]
We can then compute
\begin{equation*}\begin{aligned}
	\d \lp
		\bs \gamma^a (\dl s)
		 \wedge \lambda^{(9)}_a \rp
		&= \begin{multlined}[t]
				\lp (\dl \bs) \gamma^a+\bs\dl\gamma^a \rp \wedge  \dl s
					\wedge  \lneuf a\\
						+ \bs \gamma^a (\sigma_j s) \dll^j
						 \wedge \lambda^{(9)}_a
						- \bs \gamma^a (\dl s)
						 \wedge \dll^B  \wedge \lhuit{aB}
			\end{multlined}\\
	&= \lp \dl \bs \gamma^a \wedge \dl s
 			+\bs \gamma^a \sigma_j s
 			\dll^j\rp
 		 \wedge \lambda^{(9)}_a
 		- \bs \gamma^a (\dl s)
  		 \wedge \dll^B  \wedge \lhuit{aB}
\end{aligned}\end{equation*}

Last, corresponding to the mass term
\begin{equation}\begin{aligned}
	\d\lp m\bs_\alpha s^\alpha \ldix \rp
	&= m\d(\bs_\alpha s^\alpha)  \wedge \ldix
		+ m\bs s \d\ldix\\
	&= m\lp
		(\dl\bs)_\alpha s^\alpha + \bs_\alpha \dl s^\alpha
		 \wedge \rp\ldix
		+ m\bs s \dll^B  \wedge \lneuf B
\end{aligned}\end{equation}

Recall that both $\gamma^a$ and $\sigma_i$ are anti-selfadjoint :
\begin{gather*}
	\overline{\gamma^a s} s = - \bs \gamma^a s\\
	\bs \bar\sigma_i s = - \bs \sigma_i s
\end{gather*}
We obtain the total exterior differential by taking the real part of the sum of the three terms. We use curly braces $\{\cdot,\cdot\}$ for anticommutators \com{L'anticommutateur n'est pas normalisé alors que les $[\mu\nu]$ plus loin le sont} :
\begin{multline}
	\d\bthdir
		=\frac{1}{2} \lp
				\lp 2\dl \bs \wedge  \gamma^a\dl s
				+\bs \{\sigma_j,\gamma^a\} s
				\dll^j\rp
			 \wedge \lneuf a
			- \lp \bs \gamma^a (\dl s)
				- (\dl\bs) \gamma^a s \rp
			 \wedge \dll^B \wedge  \lhuit{aB} \rp\\
		+ \frac{i}{2} \left[
				\lp \dl \bar\kappa^i \wedge \dl s
				+ (\dl\bs)  \wedge \dl\kappa^i\rp
			 \wedge \lneuf i
			+ \lp \bar\kappa^i (\sigma_j s)
				-  (\bs\bar\sigma_j) \kappa^i \rp	\dll^j  \wedge \lneuf i
			\right.\\
			\left.
				- \lp \bar\kappa^i (\dl s)
					- (\dl\bs) \kappa^i \rp
				 \wedge \dll^B  \wedge \lhuit{iB} \right] 
		-
		m\lp
			(\dl\bs) s + \bs\dl s
		\rp \wedge \ldix
		- m\bs s \dll^B \wedge  \lneuf B
\end{multline}

Here too we adopt the notation
\begin{align}
	\EL^\alpha&:=\EL_{\dd_{\bs_\alpha}}
\end{align}

Start with the Euler-Lagrange terms corresponding to variations of the spinor field :
\begin{align*}
	\begin{aligned}
		\EL_{Dirac}^\alpha
		&= \dd_{\bs_\alpha}\lr\d \bthdir \\
		&= \frac{1}{2} \lp
				2(\gamma^a(\dl s))^\alpha
				 \wedge \lambda^{(9)}_a
				+ (\gamma^as)^\alpha
				\dll^B  \wedge \lhuit{aB} \rp\\
			&\quad	+ \frac{i}{2}\lp
				\dl \kappa^{\alpha i}
				 \wedge \lambda^{(9)}_i
				+ \kappa^{\alpha i}
				\dll^B \wedge  \lhuit{iB} \rp
			- m s^\alpha \ldix
	\end{aligned}
\end{align*}
We introduce the notation $\equiv$ for equality which holds up to a \enquote{constraint} term~(\ref{eqno:grcons}, \ref{eqno:Spcons}). This is justified by the fact that the analysis in Section~\ref{secno:EuclFE} will proceed by first assuming these equations satisfied.
\begin{equation}
		\EL_{Dirac}^\alpha
		\equiv \frac{1}{2} \lp
			2 (\gamma^a\dl s)^\alpha
			 \wedge \lambda^{(9)}_a
			+ (\gamma^a s)^\alpha
			\dll^c  \wedge \lhuit{ac}
			+ i \dl \lp \kappa^i
				\lambda^{(9)}_i \rp^\alpha \rp
		- m s^\alpha \ldix
\end{equation}

We now compute the Euler-Lagrange terms corresponding to variations of the coframe $\varpi$, which will govern the interaction of the spinors with the spacetime geometry. Recall the notation \[X = \epsilon^A_B \ddl A B\] and \[\epsilon^A = \epsilon^A_B \lambda^B\]
We obtain
\begin{equation}\label{eqno:Spdyn}\begin{split}
	\lp\EL_{Dirac}\rp_X
	&= X \lr \d \bthdir \\
	&\mkern-18mu 
	 = \frac{1}{2}\lp
		\bs \{\sigma_j,\gamma^a\} s
		\epsilon^j
		 \wedge \lneuf a
		+ \lp \bs \gamma^a (\dl s)
			- (\dl\bs) \gamma^a s \rp
		 \wedge \epsilon^B \wedge \lhuit{aB} \rp\\
		&\mkern-10mu
		 + \frac{i}{2} \lp 
			\lp \bar\kappa^i (\sigma_j s)
				- (\bs\bar\sigma_j) \kappa^i \rp	\epsilon^j  \wedge \lneuf i
			+ \lp \bar\kappa^i (\dl s)
				- (\dl\bs) \kappa^i \rp
			 \wedge \epsilon^B  \wedge \lhuit{iB} \rp
		- m\bs s \epsilon^B  \wedge \lneuf B\\
	&\mkern-18mu
	 = -\epsilon^B\wedge \lp
			\frac12 \lp  \bs \gamma^a (\dl s)
						- (\dl\bs) \gamma^a s \rp \wedge \lhuit{aB}
			+\frac{i}2 \lp \bar\kappa^i (\dl s)
							- (\dl\bs) \kappa^i \rp \wedge \lhuit{iB} 
			+ m\bs s \lneuf B
		\rp\\
	& + \epsilon^j \wedge \frac 12 \lp
				\bs \{\sigma_j,\gamma^a\} s \lneuf a 
				+ i \lp \bar\kappa^i (\sigma_j s) - (\bs\bar\sigma_j) \kappa^i \rp \lneuf i
		\rp	
\end{split}\end{equation}

\subsection{The total Euler-Lagrange terms}

Gathering the expressions~(\ref{eqno:grcons},\ref{eqno:Spcons},\ref{eqno:grdyn},\ref{eqno:Spdyn}), the total Euler-Lagrange terms corresponding to the Poincaré-Cartan form~(\ref{PoinCarSp}) are then, using again a vertical vector field $X=\epsilon^A_B \ddl A B$,
\begin{subequations} \label{eqno:ELtotal}
\begin{align}
	(\EL)^A_{\nonbc{BC}} &= \frac12\dll^A \wedge \lhuit{\nonbc{BC}} \label{eqno:ELconsEWC}\\
	\lp \EL \rp_i^\alpha &= \frac{i}2\dl s^\alpha  \wedge \lneuf i \label{eqno:ELconsDira}\\
	\label{eqno:ELepsilon}
	\begin{split}
	\EL_X &=
		\epsilon^A \wedge \left[
					\dll^D \wedge \frac12 p^{BC}_D \lsept{BCA}
				+
					\dl \lp \frac12 p^{BC}_A\lhuit{BC} \rp \right.\\
			&\qquad 	\left. -
					\frac12 \lp  \bs \gamma^b (\dl s)
								- (\dl\bs) \gamma^b s \rp \wedge \lhuit{bA}
				-   \frac{i}2 \lp \bar\kappa^j (\dl s)
									- (\dl\bs) \kappa^j \rp \wedge \lhuit{jA} 
				-  m\bs s \lneuf A
				\right]\\		
		&\qquad + \epsilon^j \wedge \frac 12 \lp
					\bs \{\sigma_j,\gamma^a\} s \lneuf a 
					+ i \lp \bar\kappa^i (\sigma_j s) - (\bs\bar\sigma_j) \kappa^i \rp \lneuf i
			\rp
	\end{split}\\
	\EL^\alpha
		&= (\gamma^a\dl s)^\alpha
			 \wedge \lambda^{(9)}_a
			+  \frac{1}{2} (\gamma^a s)^\alpha
			\dll^c  \wedge \lhuit{ac}
			- m s^\alpha \ldix
			+  \frac{i}{2} \dl \lp \kappa^i
				\lambda^{(9)}_i \rp^\alpha		\label{eqno:ELalpha}
\end{align}
\end{subequations}

%

%
The term $\EL_X$ can be decomposed according to the different components of $\epsilon$ : $\epsilon^a_b$, $\epsilon^i_b$, $\epsilon^a_j$, $\epsilon^i_j$. Each one corresponds to a variation of a different part of the structure of generalised frame bundle : $\epsilon^a_b$ corresponds to variations of the tetrad, $\epsilon^i_b$ corresponds to variations of the connection, $\epsilon^a_j$ corresponds to variations of the orbits and $\epsilon^i_j$ corresponds to variation of the action of $\lor$.

Unfortunately, the very presence of the Lagrange multipliers $p^{\nonbc{BC}}_A$ and $(\kappa^{\alpha i},\bar\kappa_{\alpha}^i)$ 
make the corresponding differential equations $\phi^*\EL = 0$ hard to study beyond the geometric structure of a generalised Cartan connection and a section of the associated spinor bundle. Since the Lagrange multipliers need not be equivariant, they have to be studied on the total bundle space. Note however the dependency on the multipliers : they appear in $\dl$-exact terms
\begin{gather*}
	\dl \lp \frac12 p^{BC}_A\lhuit{BC} \rp\\
	\dl \lp \kappa^i \lambda^{(9)}_i \rp^\alpha
\end{gather*}
and in the following three terms :
\begin{gather*}
	\epsilon^A \wedge
		\dll^D \wedge \frac12 p^{\nonbc{BC}}_D \lsept{\nonbc{BC}A} \\
	\epsilon^A \wedge
		\lp \bar\kappa^j (\dl s) - (\dl\bs) \kappa^j \rp \wedge \lhuit{jA}\\
	\epsilon^j \wedge
		\lp \bar\kappa^i (\sigma_j s) - (\bs\bar\sigma_j) \kappa^i \rp \lneuf i
\end{gather*}
If we assume that~(\ref{eqno:ELconsEWC},\ref{eqno:ELconsDira}) vanish, these three terms are shown to be dependent only on the vertical component $\epsilon^A_k\lambda^k$. This will be useful in te treatment of the Euler-Lagrange equations in Section~\ref{secno:EuclFE}.

\section{Lagrange multipliers as a differential primitive of the Euler-Lagrange form}\label{LagMult}
In this section we take a closer look at the \enquote{non-holonomic} Lagrange multipliers such as the ones used in Section~\ref{secno:FBdyn} Expression \eqref{multvirg} :
\begin{equation*}
\lp \d\lambda + \fwb\lambda\lambda\rp^A  \wedge \frac12 p^{ij}_A \lambda^{(8)}_{ij} 
+ \lp \d\lambda + \fwb\lambda\lambda\rp^A  \wedge p^{aj}_A \lambda^{(8)}_{aj}
\end{equation*}
This section is self-contained so the notations are only meant to be suggestive. We consider a configuration bundle $Q\xrightarrow{\pi}\tot$ over a $n$-dimensional source space $\tot$ and its $1$-jet bundle $\Jun(Q)\xrightarrow{\pi^1} Q \xrightarrow{\pi} \tot$. We also consider local fibration coordinates $(y^A,z^I)$ and use $1$st order jet coordinates $v^A_I$ on the jet bundle $\Jun(Q)$. Suppose we are provided with a Lagrangian $\Lag^0$ and $\tot$ is oriented by a volume form written $\vol$.

We explain in this section how non-holonomic Lagrange multipliers can be used to impose \emph{non-local} Euler-Lagrange equations, namely the \emph{exactness} of Euler-Lagrange terms. The idea is to proceed in two steps, first identifying the \enquote{constraint} equations imposed by the Lagrange multipliers and assuming the dynamical fields to satisfy them. Second, one manipulates the remaining Euler-Lagrange equations in order to gather all Lagrange multipliers in one exact term.

\subsection{Lagrange multipliers}
Standard holonomic Lagrange multiplier terms of the form
\begin{equation*}
	P_af^a(z,y)\vol
\end{equation*}
are added to Lagrangians in order to impose the equations
\begin{equation*}
	f^a(z,y)=0
\end{equation*}
as Euler-Lagrange equations corresponding to the variations of $P_a$.

In the same way, for $n$-forms $(F^a)$ of a more general form -- we will consider elements of $\Omega^n(Q)$ -- one can consider the Euler-Lagrange term associated to
\begin{equation*}
	\Lag^{cons} = P_a F^a(z,y)
\end{equation*}
with $(P_a)_{\lel 1 a m}$ being additional free scalar-valued fields (or one field with value in a suitable vector bundle). We write $p_a$ for the corresponding added coordinate in the configuration space. One performs the Legendre transformation to obtain the corresponding Poincaré-Cartan form :
\begin{equation}
	\thcons = p_aF^a
\end{equation}
as $F^a$ is invariant under the Legendre transformation (see the last comment of Appendix~\ref{annLegTrans}). The corresponding premultisymplectic form is
\begin{equation}
	\d\thcons = \d p_a \wedge  F^a + p_a\d F^a
\end{equation}
The contribution to the Euler-Lagrange forms is
\begin{align}
	\EL^{cons,a} &= \dd_{p_a}\lr \d\thcons = F^a \label{eqno:ELconsa}\\
	\ELcons_{A} &= \dd_{y^A}\lr \d\thcons = -\d p_a  \wedge i_{\dd_A} F^a +  p_a i_{\dd_A}\d F^a
\end{align}
hence for a field $\phi : \tot \to Q\times \setR^m$ the Euler-Lagrange equations associated to the whole Lagrangian $\Lag_0 + \Lag_{cons}$ are
\begin{empheq}[left=\empheqlbrace]{align}
	\phi^*F^a &= 0\\
	\phi^*\EL^0_A &=  \phi^*\lp \d p_a  \wedge i_{\dd_A} F^a -  p_a i_{\dd_A}\d F^a \rp \label{ELp}
\end{empheq}
The first equation gives the constraints one intends to impose on the fields. In the second equation $p_a$ can compensate for some nonzero components of $\EL^0_A$. Indeed in the case $F^a=f^a\vol$, the right-hand term is $-P_a \phi^*\lp \dd_{y^A}f^a\vol \rp$. In the case the $f^a$ have independent differentials (in the vertical $y^A$ directions), $P_a$ parametrise non-trivial components of $\phi^*\EL^0_A$ along the $\d_y f^a$ that are allowed by the variational principle \emph{restricted to sections satisfying the constraint condition} $\phi^*f^a = 0$.

We dealt with $n$-forms so far, we now explain how to use Lagrange multipliers with lower degree forms. For any form $f\in\Omega^k(Q)$ with $k \leqslant n$, assuming we are given a local frame of $\tot$, we can consider the forms
\begin{equation}
	F_I = f \wedge \vol^{(k)}_I
\end{equation}
for $I$ a multi-index of size $k$ parametrising the basis of $(n-k)$-forms $\vol^{(k)}_I$ (as described in Section~\ref{anndual}). We consider free multiplier fields $P^I$ which we gather in a term
\[
	P^I F_I = f\wedge P^I \vol^{(k)}_I = f \wedge P
\]
with $P\in \Omega^{n-k}(\tot)$. The term encodes the constraints 
\begin{equation*}
	(\phi^*f) \wedge \vol^{(k)}_I = 0
\end{equation*}
hence
\begin{equation*}
	(\phi^*f) = 0
\end{equation*}
which is very similar to the mechanism we used in our application with the term \eqref{multvirg}. The difference being that we only required \emph{specific components} of $\d\lambda+\fwb\lambda\lambda$ to vanish, by selecting specific values for the multi-index $I$.

Such constraints can be of non-holonomic nature : consider $F_i = \d f \wedge  \vol_i$ for $f\in\Omega^0(Q)$. Couple them to Lagrange multipliers $p^i$ to obtain a term
\[
	p^i \d f \wedge \vol_i =  \d f \wedge p
\] 
with $p\in\ExT^{n-1}(T^*\tot)$. Considering the Lagrangian $\Lag_0 + \d f \wedge p$ we obtain the following Euler-Lagrange equations on a field $(\phi,P)$ 
with variations $X\in j\phi^*(T\Jun(Q))$ 
\begin{empheq}[left=\empheqlbrace]{align}
	\phi^*\d f &= 0\\
	j\phi^*(\EL_X^0) - \phi^*(\Lie_X(f))\wedge \d P &= 0
\end{empheq}
which is to compare with the equations associated with a single holonomic Lagrange multiplier $\pi f(z,y) \vol$ : 
\begin{empheq}[left=\empheqlbrace]{align}
	\phi^*f &= 0\\
	j\phi^*(\EL_X^0) + \Pi \phi^*(\Lie_X(f)) \vol &=0
\end{empheq}
The constraint enforced by the term $\d f\wedge p$ is
\begin{equation}
	\d\phi^*f = 0
\end{equation}
or in another words \emph{$\phi^*f$ has to be constant}. The difference with the holonomic constraint is that \emph{we do not specify the constant}. Indeed we only require $\phi$ to be tangent to the leaves of the foliation defined by $\d f$, in other words the level hypersurfaces of $f$. Hence the leaf is allowed to change between the fields, only this is a \emph{non-local} variation.

\subsection{Exact Euler-Lagrange terms}

We now explain how using specific non-local variations of the dynamic fields it is possible to gather all Lagrange multiplier contributions in an exact term. Then by performing an integration it is possible to obtain a non-local Euler-Lagrange equation which has no multiplier contribution. Let us start with a simple example.

We consider a real line bundle $Q = \setR\times \tot \to \tot$ with a fibre coordinate $y$ and provided with a Lagrangian $\Lag_0$. We enforce the constraint $\d y = 0$ by adding a term $p\wedge \d y$ to the Lagrangian with $p$ a free variable in $\ExT^{n-1} T^*\tot$ :
\[
	\Lag = \Lag_0 + p\wedge\d y
\]
The equation~(\ref{ELp}) on a field $\phi : \tot\to\setR \times \ExT^{n-1} T^*\tot$ associated to the variation field $\dd_y$ becomes 
\begin{equation}
	\phi^*\EL^0_y = \d P 
\end{equation}
asserting that $\phi^*\EL^0_y$ is an exact form of primitive $P$. In a similar fashion enforcing a constraint $\d y \wedge \vol_i = 0$ with a term $p \d y\wedge \vol_i$, the Euler-Lagrange term would be \enquote{exact in the $i$ direction}, in the sense that $\phi^*\EL^0 = \dd_i P \vol$.

In the case $\tot$ is a compact manifold, the variation of the action $\int_\tot \phi^*\Lag$ under a global translation of $\phi$ by a \emph{constant} value is $\int_\tot \phi^*\EL^0_y$ so that asserting that $\phi^*\EL^0_y$ is exact is equivalent to asserting that the action has a trivial variation under this non-local field variation.

In order to formalise this observation let $F^a$ a family of homogeneous forms of degree $\Omega^{k_a}(Q)$ and consider Lagrange multipliers $p_a\in \ExT^{n-k_a} T^*\tot$. We therefore have a Lagrangian
\[
	\Lag = \Lag_0 + p_a\wedge F^a
\]
Now let $X$ be a vector field on $Q$ (hence which does not act on the multipliers) such that $\Lie_X$ preserves the ideal generated by the $F^a$. Beware that this ideal of constraints is not necessarily generated in degree one, hence is not necessarily generated by the annihilator of a plane distribution. Consider the Euler-Lagrange term corresponding to $X$ :
\begin{equation}\begin{aligned}
	\EL_X &= i_{jX} \d(\Theta^0 + p_a \wedge F^a)
		   = \EL^0_X -\d(p_a \wedge i_X F^a) + \Lie_X (p_a \wedge F^a)\\
		  &= \EL^0_X -\d(p_a \wedge i_X F^a) + P_a  \wedge \Lie_X (F^a)
\end{aligned}\end{equation}
so that, under the constraint Euler-Lagrange equations~(\ref{eqno:ELconsa}) $\phi^*\Lie_X F^a \equiv 0$ and the Euler-Lagrange equation corresponding to $X$ is equivalent to
\begin{equation}
	\phi^*\EL^0_X \equiv \d(P_a  \wedge  i_X F^a) \mod (F^a)
\end{equation}
in a process very reminiscent of Noether's conservation theorem. The degrees of freedom $p_a$ then take the role of coefficients parametrising a differential primitive of $\EL^0_X$. Suppose that it is possible to find a set of such vector \emph{fields} $(X^I)$ preserving the constraints (which is a local property of the first order prolongations of $X^I$) which spans the vertical directions of $Q$. Then we can express the whole Euler-Lagrange equation system as
\begin{empheq}[left=\empheqlbrace]{align}
	\phi^*F^a &= 0\\
	\phi^*\EL^0_{X^I} &\equiv \d(P_a  \wedge F_I^a)
\end{empheq}
in which we wrote $F_I^a := i_{X^I} F^a$. Note a possible gauge freedom as $P_a  \wedge F_A^a$ is only involved through its exterior differential, so that variations of $P_a$ inducing closed variations of all the $P_a  \wedge F^a_I$ are symmetries of the equations.

Now if the source space $\tot$ is compact, it is possible to integrate over $\tot$ to obtain non-local equations
\begin{equation}
	\int_\tot \phi^*\EL^0_{X^I} = 0 
\end{equation}

When $\tot$ is not compact, one may try to find lower dimension compact submanifolds and to build from the Euler-Lagrange lower degree forms which are exact. This is our approach in the next section. We will integrate along $6$-dimensional orbits, and for that purpose we will need to factor out a \emph{closed} $4$-form from both $\phi^*\EL^0_X$ and the exact multiplier term.

%
%

\section{Derivation of the Einstein-Cartan-Dirac equations on spacetime in Riemannian signature}\label{secno:EuclFE}
So far we have derived the following Euler-Lagrange terms :
\begin{align}
	(\EL)^A_{\nonbc{BC}} &= \frac12\dll^A \wedge \lhuit{\nonbc{BC}} \tag{\ref{eqno:ELconsEWC}}\\
	\lp \EL \rp_i^\alpha &= \frac{i}2\dl s^\alpha  \wedge \lneuf i \tag{\ref{eqno:ELconsDira}}
\end{align}
which constrain the coframe $\varpi$ to define a generalised frame bundle structure and the field $\psi$ to define a basic spinor field, as well as the following Euler-Lagrange terms : 
\begin{align}
	\tag{\ref{eqno:ELepsilon}}
	\begin{split}
	\EL_X &=
		\epsilon^A \wedge \left[
					\dll^D \wedge \frac12 p^{BC}_D \lsept{BCA}
				+
					\dl \lp \frac12 p^{BC}_A\lhuit{BC} \rp \right.\\
			&\qquad 	\left. -
					\frac12 \lp  \bs \gamma^b (\dl s)
								- (\dl\bs) \gamma^b s \rp \wedge \lhuit{bA}
				-   \frac{i}2 \lp \bar\kappa^j (\dl s)
									- (\dl\bs) \kappa^j \rp \wedge \lhuit{jA} 
				-  m\bs s \lneuf A
				\right]\\		
		&\qquad + \epsilon^j \wedge \frac 12 \lp
					\bs \{\sigma_j,\gamma^a\} s \lneuf a 
					+ i \lp \bar\kappa^i (\sigma_j s) - (\bs\bar\sigma_j) \kappa^i \rp \lneuf i
			\rp
	\end{split}\\
	\EL^\alpha
		&= (\gamma^a\dl s)^\alpha
			 \wedge \lambda^{(9)}_a
			+  \frac{1}{2} (\gamma^a s)^\alpha
			\dll^c  \wedge \lhuit{ac}
			- m s^\alpha \ldix
			+  \frac{i}{2} \dl \lp \kappa^i
				\lambda^{(9)}_i \rp^\alpha		\tag{\ref{eqno:ELalpha}}
\end{align}
relating geometric quantities depending on $\varpi$, matter quantities depending on $\psi$ and the Lagrange multipliers $p_i^{bc}$ and $\kappa^{\alpha i}$.

We assume in this section that the generalised frame bundle structure is an actual frame bundle structure, that is 
\begin{enumerate}
\item The Lie algebra action on $\tot$ integrates into a Lie group action
\item The orbit space $\ET$ has a Hausdorff quotient manifold structure
\item The fibration $\tot\to\ET$ forms a principal bundle over the orthogonal group or over the spin group.
\end{enumerate}
We will further need to modify our problem by replacing the Lorentzian signature by a \emph{Riemannian signature}. In this case, we show that the Euler-Lagrange equations on $\tot$ imply field equations on $\ET$ which match with the usual Einstein-Cartan-Dirac field equations.

The generalised frame bundle structure cannot always be identified with an actual frame bundle. In order to make it possible, two global properties are required (this is detailed in Appendix~\ref{annCartInt} and in greater length in \cite{CartInt}).

First, the Lie algebra action has to integrate to a group action. If the Lie algebra acts by non-complete vector fields, the manifold needs to be completed. The problem cannot be circumvented by restricting to \enquote{maximal solutions}. The necessary condition is \emph{univalence} which means that the action of an element of the group does not depend on the path leading from identity to the element used to construct it (so that orbital mappings $x\mapsto g\cdot x$ can be unambiguously defined, although partially). The Lie algebra action is then \enquote{globalisable}, which means the manifold can be embedded into a larger (possibly non-Hausdorff) manifold on which the Lie group acts, with an embedding equivariant under the infinitesimal action. On a compact manifold the vector fields are complete and the action readily integrates into a group action of the simply-connected Lie group integration of the Lie algebra.

The second property is \emph{properness} of the group action which ensures sufficient separation on the orbit space. In particular, it holds for all compact Lie group actions.

When these two properties are satisfied, there is an dense open subset of $\tot$ which can be identified with an orthonormal frame bundle, or a spin frame bundle, over its orbit space. In particular, all orbits on this open subset are either isomorphic to the $\Spin$ group or to the $\SOp$ group.

The derivation of the spacetime equations will proceed as follows. First we assume that the \enquote{constraint} Equations (\ref{eqno:ELconsEWC},\ref{eqno:ELconsDira})
are satisfied and that they define a frame bundle structure with a spinor field on the underlying spacetime. Next, we identify variations of the field which preserve the constraints, so that by the mechanism introduced in Section~\ref{LagMult} we obtain Euler-Lagrange equations with all Lagrange multiplier contributions gathered in an exact term. This is to compare with the approach in~\cite{LFB} in which they perform an explicit change in coordinate depending on the choice of a local section on the frame bundle and after some algebraic manipulations manage to identify exact divergence terms in the Euler-Lagrange equations. In particular, the use of explicit coordinate systems make the computations quite heavy.

We want to get rid of the exact term by integrating over the orbits under the $\Spin$ group, which are compact \emph{in Riemannian signature}. For this purpose we need an extra step to obtain equations on $6$-forms along the orbits with all the Lagrange multipliers gathered in an exact $6$-form. Only then we can proceed to integration. Identifying an integrand which is invariant under the action of $\Spin_4$ and has an integral of zero, we conclude that it must identically vanish. We thus obtain an equivariant equation on $\tot$ which we interpret as an equation on $\ET$, from which the Lagrange multipliers are absent. We conclude the paper with an brief analysis of the field equations, which is already standard material in the literature.

As we replace the Lorentzian signature with the Riemannian signature, we need to replace the Poincaré algebra with the Lie algebra of the Euclidean group $\eucl = \soq\ltimes \setR^4$ as well as the metric $\met$ by the Euclidean metric on $\setR^4$. The corresponding spin group is $\Spin_{4,0}$, usually denoted $\Spin_4$. The algebraic manipulations involving the spinors we made in Section~\ref{VEqSp} still hold, see Appendix~\ref{annspin} for more detail. For the sake of brevity we will assume we have a spin frame bundle and will write about the action of $\Spin_4$ but when the spinor field $\psi$ is identically zero the principal bundle can be a mere metric $\SO_4$-structure.

We keep the indices conventions corresponding to the decomposition $\eucl = \soq\ltimes\setR^4$. We will make implicit use of the isomorphism $\so_4\simeq \spin_4$.
\subsection{Exact terms in the Euler-Lagrange equations}

Given a Lagrangian involving general Lagrange multipliers, we explained in Section~\ref{LagMult} how it is possible, using suitable vector fields, to obtain Euler-Lagrange equations which have all the dependency in Lagrange multipliers gathered in an exact term. In our case, we have Lagrange multipliers $p_A^{\nonbc{BC}}$ involved in a Lagrangian term
\[
	p_A^{\nonbc{BC}} \dll^A \vhuit{\nonbc{BC}}
\]
as well as Lagrange multipliers $\kappa^{\alpha i}$ involved in a term
\[
	\frac i 2 \lp \bar\chi \dl s - \dl\bs \chi \rp \lneuf i
\]
We want to study solutions to Equations~(\ref{eqno:ELtotal}). For this we want to get rid of the non-physical Lagrange multipliers fields. The solution is to gather them in an exact term and to make it vanish by integration. The integration cannot be made on the total space $\tot$ which is not assumed to be compact. We need to find suitable compact submanifolds on which the Euler-Lagrange Equations~(\ref{eqno:ELepsilon},\ref{eqno:ELalpha}) can be \enquote{restricted} in a meaningful way while preserving the exactness of the Lagrange multipliers term. But before this, let us find those suitable vector fields. We want vertical variations of the field which preserve the constraints~(\ref{eqno:ELconsEWC},\ref{eqno:ELconsDira}) on-shell.

\subsubsection{Infinitesimal variations preserving the constraints}

Once a frame bundle structure is given by a field $\phi=(\varpi,P,\psi,K) : \tot\to Q$ satisfying the constraint equation~(\ref{eqno:ELconsEWC}), it is natural to consider variations of the structure corresponding to variations of the tetrad or of the connection in the usual spacetime formalism. These are given by \emph{equivariant} variations of $\omega^i=\phi^*\lambda^i$ and $\alpha^a = \phi^*\lambda^a$ as we will show. In a similar way, we will consider equivariant variations of the spinor field $\psi^\alpha = \phi^* s^\alpha$. Note however the difference with the situation described in Section~\ref{LagMult} : equivariance is formulated using the principal bundle structure hence only makes sense on $\tot$ and is a notion dependent on the field $\phi$. Instead of assuming our variations equivariant from the start we will derive this condition.

\subsubsection*{Variations of the coframe}

Let us start with variations of the coframe $\varpi^A$. The field $\phi$ provides us with a nondegenerate equivariant $1$-form : the coframe itself $\varpi^A = \phi^*\lambda^A$. The variation of $\varpi$ will be given by a vertical vector field $X$ on the image of $\varpi$ :
\[
	X\in \Gamma \lp \tot,\varpi^*\left[ V(\Isototp)  \right] \rp 
\]
Using here again the isomorphism
\[
	V(T^*\tot\otimes \eucl) \simeq T^*\tot\otimes \eucl \times_\tot T^*\tot\otimes \eucl
\]
a vertical variation of $\varpi$ is equivalent to a $1$-form $\epsilon^A\in\Omega^1(\tot,\eucl)$. We will use for convenience Lie derivatives $\Lie_X$ but they will not depend on the chosen $1$st order extension of $X$ as we will work on the image of $\phi$. The main property we will use is
\begin{equation}
	\phi^*{\Lie_X \lambda}^A = \epsilon^A
\end{equation}
Let us decompose $\epsilon^A$ into a $\soq$-valued component $\tau^i\in\Omega(\tot,\so_4)$ and a $\setR^4$-valued component $\beta^a$ : 
\[
	\epsilon = \tau\oplus\beta 
\]

To study the action of $X$ on the constraints, we will have it act on $\lp \d\lambda +\fwb\lambda\lambda \rp\lhuit{BC}$.
We can then compute
\begin{align*}
	\phi^*\Lie_X \lp \d\lambda+\fwb\lambda\lambda \rp^i &= \d\phi^*\Lie_X \lambda^i + 2\phi^*\fwb{\Lie_X\lambda}\lambda^i = \d\tau^i + \wb\omega\tau^i\\
	\phi^*\Lie_X \lp \d\lambda+\fwb\lambda\lambda \rp^a  &= \d\phi^*\Lie_X \lambda^a + 2\phi^* \fwb{\Lie_X\lambda}\lambda^a = \d\beta^a + \wb\omega\beta^a + \wb\tau\alpha^a 
\end{align*}
and
\begin{equation}
	\phi^*\Lie_X \lhuit{BC} = \phi^*\lp (\Lie_X\lambda^D)\wedge \lsept{BCD}\rp  = \epsilon^D\wedge \vsept{BCD}
\end{equation}
which gather to give
\begin{subequations}\label{eqno:phiLieX}
\begin{align}
	\phi^*\Lie_X \lp \lp \d\lambda+\fwb\lambda\lambda \rp^i \wedge \lhuit{\nonbc{BC}} \rp
		&= \lp \d\tau^i + \wb\omega\tau^i \rp \wedge \lhuit{\nonbc{BC}} + \Omega^i\wedge \epsilon^D\wedge \vsept{\nonbc{BC}D}
	\label{eqno:varcour}\\
	\phi^*\Lie_X \lp \lp \d\lambda+\fwb\lambda\lambda \rp^a \wedge \lhuit{\nonbc{BC}} \rp
		&= \lp \d\beta^a + \wb\omega\beta^a + \wb\tau\alpha^a \rp \wedge \lhuit{\nonbc{BC}} + \Omega^a\wedge \epsilon^D\wedge \vsept{\nonbc{BC}D}
	\label{eqno:vartor}		
\end{align}
\end{subequations}
In order for these terms to vanish under the constraint equation~\eqref{eqno:ELconsEWC}, we will require the following three conditions on $\epsilon$ :
\begin{subequations}
\begin{enumerate}
\item $\epsilon$ to be purely horizontal : $\epsilon^A = \epsilon^A_b\alpha^a$. This will prevent $\vsept{\nonbc{BC}D}$ from having purely horizontal components (of type $\lhuit{bc})$ so that the term 
\[ \Omega^A\wedge \epsilon^D\wedge \vsept{\nonbc{BC}D} \]
necessarily vanishes.

\item There exists coefficients $r^i_{bc}$ such that
\begin{equation} \d\tau^i + \wb\omega\tau^i = \frac12 r^i_{bc}\alpha^b \wedge \alpha^c  \label{eqno:rab} \end{equation}
So that the term $\lp \d\tau^i + \wb\omega\tau^i \rp \wedge \lhuit{\nonbc{BC}}$ vanishes. But now that we assumed that $\tau$ is horizontal, this equation exactly means that \emph{$\tau$ is equivariant}

\item There exists coefficients $t^a_{bc}$ such that
\begin{equation} \d\beta^a + \wb\omega\beta^a = \frac12 t^a_{bc}\alpha^b \wedge \alpha^c \label{eqno:tab} \end{equation}
As $\wb\tau\alpha$ is now assumed to be purely horizontal, the term $\lp \d\beta^a + \wb\omega\beta^a + \wb\tau\alpha^a \rp \wedge \lhuit{\nonbc{BC}}$ vanishes. But this is exactly requiring that \emph{$\beta^a$ is equivariant}.
\end{enumerate}\end{subequations}
Under these three conditions, the terms of \eqref{eqno:phiLieX} vanish.

We also need to check whether such variations preserve the constraint \eqref{eqno:ELconsDira} : 
\begin{equation}\begin{aligned}
	\Lie_X \lp \lp \d s + \omega^i\sigma_i s \rp \wedge \lneuf i \rp
		&= \lp \d s + \omega^i\sigma_i s \rp \wedge \Lie_X \lambda^B \wedge \lhuit{iB}
		= \lp \d s + \omega^i\sigma_i s \rp \epsilon^B_c\lambda^c \wedge \lhuit{iB}\\
		&= \lp \d s + \omega^i\sigma_i s \rp \epsilon^c_c\lneuf i
\end{aligned}\end{equation}
Thus the constraint is preserved without condition.

To conclude, any equivariant horizontal $1$-form $\epsilon$ represents a variation preserving the constraints (\ref{eqno:ELconsEWC},\ref{eqno:ELconsDira}). They can be identified with families of equivariant coefficients $\epsilon^A_b$ on $\tot$ with value in ${\setR^4}^*\otimes\eucl$.

Unfortunately, we cannot find coefficients $\epsilon^A_i$ such that \eqref{eqno:phiLieX} vanish in a similar manner for two reasons.
First, to a nonzero $\epsilon^i$ will correspond a term
\[ \Omega^A\wedge \epsilon_i^D\varpi^i\wedge \vsept{\nonbc{BC}D}
	= \Omega^A \lp 
		\epsilon^i_i \vhuit{BC}
	 	- \delta_i^C \epsilon_i^D \vhuit{BD}
		+ \delta_i^B \epsilon_i^D \vhuit{CD}
	 	\rp
	\]
which can contain nonzero components of $\Omega^A$ when there are non-vanishing components $\epsilon^d_i$.

Second, for a non-horizontal $1$-form $\epsilon$, equivariance is no longer equivalent to 
\[
	\d\epsilon^A + \wb\omega\epsilon^A = \frac12 E^A_{bc}\alpha^b\wedge\alpha^c 
\]
as the following equation holds, with $\epsilon_i=\epsilon(\h_i)\in\p$ : 
\[
	\lp \Lie_{\bh_i} + \h_i \cdot \rp \epsilon
		= i_{\h_i} \lp \d\epsilon + \wb\omega\epsilon \rp + \d\epsilon_i + \wb\omega{\epsilon_i} \]
Thus arbitrary equivariant $1$-forms are not necessarily solutions of $i_{\h_i} \lp \d\epsilon + \wb\omega\epsilon \rp = 0$. Indeed, $\omega$ itself is equivariant but obeys 
\[
	\d\omega^i + \wb\omega\omega^i
		= \Omega^i +\fwb\omega\omega^i
		= \frac12\Omega^i_{bc}\alpha^a\wedge\alpha^b + \frac12 c^i_{jk}\omega^j\wedge\omega^k
\]

We therefore proceed only with the horizontal variations of $\varpi$ and general variations of $\psi$. Insofar as the fibration above the spacetime does not vary, these variations can be viewed as moving the direct orthonormal frame bundle inside the general linear frame bundle of the spacetime. Note however that beside metric and connection variations, there is an extra gauge freedom as the spin group $\Spin_4$ acts on $T^*\tot\otimes \eucl$ by bundle automorphisms \cite{DynPrincBundle}.

\subsubsection*{Variations of the spinor field}

If the structure group is $\Spin_4$, the same computation can be done with variations $X$ of $\psi$, identified with $\Sp$-valued fields $\xi$ over $\tot$ (while keeping the generalised frame structure unvaried). In this case, the compatibility with the constraint \eqref{eqno:ELconsDira} requires
\begin{equation}
	\Lie_X (\d s + \omega^i \sigma_i\cdot s) = \d\xi + \omega^i\sigma_i \xi = \varsigma_a\alpha^a
\end{equation}
which here again expresses the local equivariance of $\xi$ over $\tot$.

It is obvious that variations of $\psi$ preserve that constraint term $\lp \d\lambda +\fwb\lambda\lambda \rp\lhuit{BC}$.

\subsubsection{A basis for the constraint-preserving variations}\com{Nécessaire ?}

We want to construct a local basis of equivariant $\soq$-valued horizontal forms, as well as equivariant $\Sp$-valued fields. Here we will make use of the structure of principal bundle of $\tot$, or more exactly the existence of slices. A slice $S$ is the image of a local section $\tot/\Spin_4 \to \tot$. Thus equivariant fields on $\Spin_4\cdot S$ are uniquely identified by their value on $S$. 

We write $\beta^{\ba,\mu}$ (resp. $\tau^{\bi,\mu}$) for a basis of horizontal $\setR^4$-valued forms (resp. horizontal $\so_4$-valued forms) on $S$ : the superscript $\mu$ corresponds to a basis of horizontal scalar $1$-forms on $S$, while $\ba$ and $\bi$ correspond to a basis of $\soq$ (resp. $\setR^4$)-valued maps on $S$. 
Similarly we index equivariant spinor variations by their value on $S$ and write $\bar\xi_\balpha$ for a basis of such vectors (variations of the adjoint spinor giving the spinor field equation), which can be identified with $\bSp$-valued maps over $S$.

\subsubsection{Euler-Lagrange terms corresponding to the variations}

Recall the decomposition of the Poincaré-Cartan form $\bar\theta = \theta + \thcons$ :
\begin{align}
	 \theta &= \frac12 2\rho_{i,c}^a\met^{cb}\dll^i \wedge  \lhuit{ab}
	 + \frac{1}2 \lp
	 \bs \gamma^a (\dl s)
	 - (\dl\bs) \gamma^a s \rp
	  \wedge \lambda^{(9)}_a
	 -m\bs s \lambda^{(10)}
	 \label{eqno:PC0}\\
	\thcons &= \frac12 p^{\nonbc{BC}}_D\dll^D \wedge \lhuit{\nonbc{BC}}
			+ \frac i 2\lp \bar\kappa^i (\dl s)
			- (\dl\bs) \kappa^{i} \rp	 \wedge \lambda^{(9)}_i
	\label{eqno:PCcons}
\end{align}

Let $X$ a vertical variation of $\phi$ which consists in an equivariant horizontal variation $\epsilon$ of $\varpi$ and an equivariant variation $\xi$ of $\psi$. According to the principle presented in Section~\ref{LagMult}, the Euler-Lagrange term corresponding to a variation $X$ and Poincaré-Cartan form $\thcons$ is exact up to a term which vanish under the constraint : 
\begin{equation}
\begin{aligned}
	i_X \d\thcons = \Lie_X (\thcons) + \d i_X\thcons \equiv \d i_X\thcons \equiv 0 
\end{aligned}
\end{equation}

We define the unconstrained Euler-Lagrange forms :
%
\begin{align}
	(\EL^0)_X &= i_X \d\theta
\end{align}
The form $\EL^0$ has no dependency in $P_A^{\nonbc{BC}}$ nor in $K^{\alpha i}$.

Then a field $\phi$ satisfying the Euler-Lagrange equations (\ref{eqno:ELconsEWC},\ref{eqno:ELconsDira}) satisfies the following : 
\begin{equation}\label{eqno:phiEL0exact}
	\phi^*(\EL^0)_X = - \phi^*\d i_X\thcons
\end{equation}
The dependency in $P_A^{\nonbc{BC}}$ and in $K^{\alpha i}$ is gathered in the exact term. In order to make use of the exactness, we want to perform an integral.

\subsection{Integration into variational equations on the spacetime}

\subsubsection*{Unconstrained Euler-Lagrange terms}

The unconstrained Euler-Lagrange form corresponds to the terms in (\ref{eqno:ELconsEWC}-\ref{eqno:ELalpha}) not involving the Lagrange multipliers $P^{\nonbc{BC}}_A$, $K^{\alpha i}$ :

\begin{equation}\label{eqno:ELexact}\begin{aligned}
	\phi^*\EL^0_X =
		&\epsilon^A \wedge \left[
							\dll^i \wedge \frac12 p^{bc}_i \lsept{bcA}
							- \frac12 \lp  \bs \gamma^b (\dl s)
										- (\dl\bs) \gamma^b s \rp \wedge \lhuit{bA}
						-  m\bs s \lneuf A
						\right]\\		
				& + \epsilon^j \wedge \frac 12 \lp
							p^{bc}_j\dll^D \wedge \lsept{bcD}
							+ \bs \{\sigma_j,\gamma^a\} s \lneuf a 
					\rp\\
		& + \bar\xi_\alpha \lp
			\frac{1}{2} \lp
						2 (\gamma^a\dl s)^\alpha
						 \wedge \lambda^{(9)}_a
						+ (\gamma^a s)^\alpha
						\dll^c  \wedge \lhuit{ac} \rp
					- m s^\alpha \ldix \rp
\end{aligned}\end{equation}

The idea now is to perform a \enquote{partial integration}, or fibre integration, of these Euler-Lagrange equations over the orbits under $\Spin_4$. As they are compact, the exact terms will vanish and we will be left with equations involving only $\EL_0$. Furthermore, $\Spin_4$-equivariance of the variation $X$ implies $\Spin_4$-\emph{invariance} of the form $\phi^*\EL_X$, so that an vanishing integral over orbits implies that $\phi^*\EL_X$ vanishes at each point. But for this, we need to transform \eqref{eqno:ELexact} into an equation on $6$-forms so that it can be integrated along the $6$-dimensional orbits of $\Spin_4$.

To this end we want to \enquote{factor out} a factor $\aquatre$ while keeping the exactness of the right-hand term in \eqref{eqno:phiEL0exact}. The computation will be easier if we use explicit $10$-forms on $\tot$. Doing so we will not need to keep track of contact terms and constraint terms.

Let us thus reexpress the different terms in \eqref{eqno:ELexact} : 
\begin{subequations}\label{eqno:EL0Xvdix}
\begin{align}
	&\phi^*\lp \dll^i\wedge \epsilon^A\wedge\lsept{bcA} \rp
		= \Omega^i\wedge\epsilon^A\wedge\vsept{bcA}
		= \frac12 \Omega^i_{de}\alpha^d\wedge\alpha^e\wedge\epsilon^A_f \alpha^f \vsept{bcA}
		= \frac12\Omega^i_{de}\epsilon^a_f \delta^{[def]}_{bca}\vdix\\
%
%
%
%
%
%
%
			&\phi^*\lp \epsilon^A \wedge \lp  \bs \gamma^b (\dl s)
				- (\dl\bs) \gamma^b s \rp \wedge \lhuit{bA} \rp
			\begin{aligned}[t]
			  			&= - \lp \bp \gamma^b \dom \psi - \dom \bp \gamma^b \psi \rp \epsilon^A \wedge \vhuit{bA}\\
			 			&= - \lp \bp \gamma^b \dom \psi - \dom \bp \gamma^b \psi \rp \epsilon^A_c
							\wedge \lp \delta^c_A \vneuf{b} - \delta^c_b \vneuf A \rp\\
						&= - \epsilon^a_c
					  		\lp \bp \gamma^b \dd_d \psi - \dd_d \bp \gamma^b \psi \rp 
					  		\lp \delta^c_a \delta^d_b - \delta^c_b \delta^d_a \rp \vdix
			\end{aligned}\\
	&\phi^*\lp \epsilon^A \wedge\bs s \lneuf A \rp
		= \epsilon^A_A \bp  \psi \vdix = \epsilon^a_a \bp \psi \vdix \\
	&\phi^*\lp \epsilon^j \wedge \dll^D \wedge \lsept{bcD} \rp
		 = \frac12 \epsilon^j_e \Omega^d_{fg} \delta^{[efg]}_{bcd} \vdix\\
	%
	&\phi^*\lp \epsilon^j \wedge \bs \{\sigma_j,\gamma^a\} s \lneuf a \rp = \epsilon^j_a \bp \{\sigma_j,\gamma^a\} \psi \vdix\\
	&\bar\xi_\alpha \phi^*\lp
			(\gamma^a\dl s)^\alpha \wedge \lambda^{(9)}_a
				+ \frac{1}{2} (\gamma^a s)^\alpha \dll^c  \wedge \lhuit{ac} \rp	
		= \bar\xi_\alpha \lp \gamma^a\dd_a \psi^\alpha + \frac12 \gamma^a\psi^\alpha \Omega^c_{ac} \rp \vdix
\end{align}
\end{subequations}
The non-normalised \emph{antisymmetric Kronecker symbol} $\delta^{[efg]}_{bcd}$ is defined as follows :
\[
	  \delta^e_b\lp \delta^f_c\delta^g_d - \delta^f_d \delta^g_c \rp
	+ \delta^e_c\lp \delta^f_d\delta^g_b - \delta^f_b \delta^g_d \rp
	+ \delta^e_d\lp \delta^f_b\delta^g_c - \delta^f_c \delta^g_b \rp
\]
\com{acceptable d'utiliser un crochet non-normalisé ici mais normalisé plus tard ?}

From these it is straightforward to factorise by $\aquatre$ all the terms. We will write 
\[
	\phi^*\EL^0_X = \phi^*E_X \wedge \aquatre
\]
with $\EL^0_X$ defined as a purely horizontal $6$-form on the image of $\phi$ (but it can also be expressed as a form on the jet bundle : $\EL^0_X\in \ExT^6 T^*\tot\otimes\Jun(Q)$).

\subsubsection*{Factorisation and exactness of the constraint terms}

We need to factorise the exact term $\phi^* \lp \d i_X \thcons \rp$. Extracting the relevant terms from Equations~\eqref{eqno:ELtotal} we have
\[\begin{aligned}
	i_X\thcons 
		&= i_X \lp \frac12 p^{\nonbc{BC}}_D\dll^D \wedge \lhuit{\nonbc{BC}}
				+ \frac i 2\lp \bar\kappa^i (\dl s)
				- (\dl\bs) \kappa^{i} \rp	 \wedge \lambda^{(9)}_i\rp\\
		&= \frac12 p^{\nonbc{BC}}_D\epsilon^D \wedge \lhuit{\nonbc{BC}}
						- \frac i 2 \bar\xi_\alpha \kappa^{\alpha i}  \wedge \lneuf i\\
		&=  p^{jc}_D  \epsilon^D_c \lneuf{j} 
				- \frac i 2 \bar\xi_\alpha \kappa^{\alpha i}  \wedge \lneuf i\\
		&= 	\lp p^{jc}_D  \epsilon^D_c \lambda^{\soq(5)}_j 
						- \frac i 2 \bar\xi_\alpha \kappa^{\alpha i} \lambda^{\soq(5)}_i
			\rp \wedge \lmquatre
\end{aligned}\]
Note how essential it is here again that \emph{$\epsilon$ is a purely horizontal form}. This is what allows us to factor 
$\epsilon^D\wedge \lhuit{\nonbc{BC}}$ by $\lmquatre$.

Considering the exterior differential of the pullback, we obtain
\begin{equation*}
\begin{aligned}
	\phi^*\d i_X \thcons
		&= \d\phi^*i_X\thcons\\
		&= \d 	\lp \lp P^{jc}_D  \epsilon^D_c \omega^{(5)}_j 
								- \frac i 2 \bar\xi_\alpha K^{\alpha i} \omega^{(5)}_i
					\rp
				\wedge \aquatre \rp\\
		&=\d 	\lp P^{jc}_D  \epsilon^D_c \omega^{(5)}_j 
										- \frac i 2 \bar\xi_\alpha K^{\alpha i} \omega^{(5)}_i
				\rp
						\wedge \aquatre
		+	\lp P^{jc}_D  \epsilon^D_c \omega^{(5)}_j 
										- \frac i 2 \bar\xi_\alpha K^{\alpha i} \omega^{(5)}_i
			\rp
						\wedge \d \aquatre\\
		&= \d 	\lp P^{jc}_D  \epsilon^D_c \omega^{(5)}_j 
												- \frac i 2 \bar\xi_\alpha K^{\alpha i} \omega^{(5)}_i
				\rp
								\wedge \aquatre
			+ 0
\end{aligned}
\end{equation*}
We used $\d\aquatre = 0$ which is a consequence of Equation~\eqref{eqno:ELconsEWC}
\[ \d\alpha^a + \wb\omega\alpha^a = \frac12 \Omega^a_{bc} \alpha^b\wedge\alpha^c \]
We call $\Econs_X$ the term which is differentiated : 
\begin{equation}
	\phi^*\Econs_X := P^{jc}_D \epsilon^D_c \omega_j^{(5)} - \frac i 2 \bar\xi_\alpha K^{\alpha i} \omega_i^{(5)}
\end{equation}

We finally arrive to the following equation, which holds for each $X$ vertical variation of $\phi$ which is equivariant : 
\begin{equation}\label{eqno:ELaquatre}
\boxed{
	\phi^*\EL^0_X \wedge \aquatre = \d\phi^*\Econs_X \wedge\aquatre
}
\end{equation}

\subsubsection*{Integration along the orbits}

The tangent space to the orbits under $\Spin_4$ is exactly the kernel of $\aquatre$. This means that on any orbit $\Spin_4\cdot x\subset \Spin_4\cdot S$, Equation~\eqref{eqno:ELaquatre} implies
\begin{equation}
	\phi^*\EL^0_X|_{\Spin_4\cdot x} = \d\phi^*\Econs_X |_{\spin_4\cdot x}
\end{equation}

The orbit being compact, we can integrate along the orbit : 
\begin{equation}
	\int_{\Spin_4\cdot x} \phi^*\EL^0_X|_{\Spin_4\cdot x}
		= \int_{\Spin_4\cdot x} \d\phi^*\Econs_X |_{\spin_4\cdot x}
		= 0
\end{equation}
by virtue of the Stokes theorem.

Now we want to conclude $\phi^*\EL^0_X|_{\Spin_4\cdot x} = 0$. It will follow if we can show that $\phi^*\EL^0_X$ is $\Spin_4$-invariant. 
Indeed as $\Spin_4$ preserves the orientation of the orbit and acts transitively, any invariant form with vanishing integral is necessarily identically zero.

But Equations \eqref{eqno:EL0Xvdix} express $\phi^*\EL^0_X$ as $\vdix$, which is $\Spin_4$-invariant by construction, with a factor which is $\Spin_4$-invariant as a complete contraction of equivariant quantities (including $\epsilon$ and $\bar\xi$). Thus factoring out an equivariant $\aquatre$ lefts us with an equivariant $\phi^*E^0_X$. 

Thus we proved that 
\begin{equation}\label{eqno:phiEL0X0}
\boxed{
	\phi^*\EL^0_X = 0
}
\end{equation}
for $X$ which are variations of $\psi$ which are equivariant and variations of $\varpi$ which are \emph{horizontal and} equivariant, over $\Spin_4\cdot x$ for any orbit. Hence it holds at each point.

Equation~\eqref{eqno:phiEL0X0} a priori only holds for $1$-forms $\epsilon$ which are \emph{constant} linear combinations of the fields $\beta^{\ba,\mu}$, $\tau^{\bi,\mu}$ and $\bar\xi_{\balpha}$. However it is a tensorial equation which holds at each point. Thus it still holds if $X$ was multiplied by any real function on $\Spin_4\cdot S$. One concludes that Equation~\eqref{eqno:phiEL0X0} holds for $X$ which is \emph{any variation of $\psi$} and \emph{any horizontal variation of $\varpi$}. Let us recall the coefficient of $\omsix$ in $\phi^*E^0_X$, as expressed in~\eqref{eqno:EL0Xvdix} :

\begin{equation}\begin{aligned}
	&\epsilon^a_b \left(
			 \frac12 p_i^{cd} \frac12 \Omega^i_{ef} \delta^{[bef]}_{cda}
			+\frac12\lp \delta^b_a \delta^d_c - \delta^b_c \delta^d_a \rp
							\lp \bp \gamma^c \dd_d \psi - \dd_d \bp \gamma^c \psi \rp 
			- m \delta^b_a \bp  \psi
		\right)\\
	+{}&\epsilon^i_b \left(
			\frac12 p_i^{ef} \frac12 \Omega^g_{cd} \delta^{[bcd]}_{efg}
			+ \bp \{\sigma_i,\gamma^b\} \psi
		\right)
	+ \bar\xi_\alpha \lp \gamma^a\dd_a \psi^\alpha + \frac12\gamma^a\psi^\alpha \Omega^c_{ac} - m \psi^\alpha \rp
\end{aligned}\end{equation}
Since it has to vanish for \emph{any} $\epsilon^A_b$ and $\bar\xi_\alpha$, it means that each coefficient vanishes : 
\begin{subequations}\label{eqno:FEtot}
\begin{gather}
		\frac14 p_i^{cd} \Omega^i_{ef} \delta^{[bef]}_{cda}
		+ \frac12 \lp \delta^b_a \delta^d_c - \delta^b_c \delta^d_a \rp
				  \lp \bp \gamma^c \dd_d \psi - \dd_d \bp \gamma^c \psi \rp 
		- m \delta^b_a \bp  \psi
		= 0 \label{eqno:FEtotab}\\
		\frac14 p_i^{ef} \Omega^g_{cd} \delta^{[bcd]}_{efg}
		+ \bp \{\sigma_i,\gamma^b\} \psi
		= 0 \label{eqno:FEtotib}\\
		\gamma^a\dd_a \psi^\alpha + \frac12 \gamma^a\psi^\alpha \Omega^c_{ac} - m \psi^\alpha
		= 0 \label{eqno:FEtotalpha}
\end{gather}\end{subequations}

We obtained \emph{tensorial equations} on $\tot$. By assumption, $\tot$ is a (spin) frame bundle above a spacetime $\ET$. We now express these tensorial equations on $\ET$.

\subsubsection*{Expression in spacetime coordinates}

As all fields involved in Equations~\eqref{eqno:FEtot} are equivariant, they can be pushed forward to sections of the associated principal bundles on spacetime. Thus the equations can be formulated on spacetime. We need a local system of coordinates, with a local trivialisation of the spinor frame bundle. We use greek indices $\mu,\nu\dots$ for coordinates. We keep the notation $\psi$ for the spinor field as well as the superscript $\alpha$ for spinor fields.\com{Changer la notation ?} There is on $\ET$ a metric $g$ corresponding to $\met_{ab}\alpha^a\otimes \alpha^b$ ; it is compatible with the $\Spin_4$-structure defined by $\tot\to \ET$.

Taking care of the factors $p^{ab}_i$ requires some care. Let us recall their definition :
\[
	p_i^{bc} = 2\rho^b_{i,d}\met^{dc}
\]
We also stated that they correspond to an isomorphism 
\[\eucl\xrightarrow{\sim} \ExT^2 {\setR^4}^*\]
First, 
\[\frac12 p_i^{de}\Omega^i_{ab} \text{ corresponds to } \Riem^{o}{}_{\mu\nu\pi}g^{\pi\xi} \]
(with indices corresponding in alphabetic order) as explained in Appendix~\ref{annRic}.

The $\Omega$ term in Equation~\eqref{eqno:FEtotab} becomes :
\begin{equation*}
	\frac12 \Riem^\xi{}_{\pi\rho \upsilon} g^{\upsilon o} \delta^{[\nu\sigma\tau]}_{\xi o \mu}
	 	= 	  \Riem^\nu{}_{o\mu \upsilon} g^{uo} 
	 		- \Riem^\xi{}_{\xi \mu \upsilon} g^{u\nu} 
	 		+ \Riem^\xi{}_{\xi o \upsilon} g^{uo} \delta^\nu_\mu
		= \delta^\nu_\mu \Scal - 2 g^{\nu\pi}\Ric_{\mu\pi}
\end{equation*}
For Equation~\eqref{eqno:FEtotib} we recall as well $\sigma_i$ can be written as a function of $\gamma^a$ in the following way : 
\[
	\sigma_i = \frac14 p_i^{ab}\gamma_a\gamma_b = \frac18 p_i^{ab}[\gamma_a,\gamma_b]
\]
as explained in Appendix~\ref{annPinSpin} ($p$ is twice the map $\rho$ described there). Equation~\eqref{eqno:FEtotib} is thus equivalent to
\[
	0 = \frac12 \Omega^g_{cd} \delta^{[bcd]}_{efg}
			+ \frac14 \bp \left\{ [\gamma_e,\gamma_f],\gamma^b \right\} \psi
	  = \Omega^g_{fg} \delta^b_e + \Omega^g_{ef} \delta^b_g - \Omega^g_{eg} \delta^b_f
	  	+ \frac14 \bp \left\{ [\gamma_e,\gamma_f],\gamma^b \right\} \psi
\]
The spacetime tensor corresponding to $\Omega^a_{bc}$ is the torsion tensor $T^\mu_{\nu\xi}$. We define its trace 
\[
	\tr(T)_\rho = T^\sigma_{\sigma\rho}
\]
The equation on spacetime corresponding to \eqref{eqno:FEtotib} is :
\begin{equation}
	\lp -\tr(T)_\rho \delta^\mu_\nu+ \tr(T)_\nu \delta^\mu_\xi + T^\mu_{\nu\xi} \rp
		 + \frac14 \bp \{[\gamma_\nu,\gamma_\xi],\gamma^\mu\} \psi
\end{equation}

Last, the horizontal derivatives $\dd_a$ turn to covariant derivatives on spacetime, as they correspond to deriving the fields on the frame bundle along horizontal directions. In particular, $\gamma^a\dd_a$ corresponds to the (covariant) Dirac operator. We summarize the correspondence in a table :
\begin{table}[h]
\centering
\begin{tabular}{|c|c|}
	\hline
	Frame bundle & Spacetime\\
	\hline
	$\dd_a$		&	$\nabla_a$\\
	\hline
	$\gamma^a\dd_a$	&	$\Dirac$\\
	\hline
	$\met_{ab}$	&	$g_{\mu\nu}$\\
	\hline
	$p_i^{cd}\Omega^i_{ef}$ &	$\Riem^{\xi}{}_{\mu\nu}{}^o$	\\
	\hline
	$\Omega^a_{bc}$	&	$T^\mu_{\nu\xi}$\\
	\hline
	$\psi^\alpha$ & $\psi^\alpha$\\
	\hline
\end{tabular}
\end{table}

For convenience, we convert~\eqref{eqno:FEtotab} to a totally covariant equation. Separating geometry terms and matter terms in Equations~\eqref{eqno:FEtot}, we obtain : 
\begin{subequations}\begin{empheq}[left=\empheqlbrace]{align}
	2\Ric_{\mu\nu} - g_{\mu\nu} \Scal
		&= 	g_{\mu\nu} \lp \frac12 \bp \DirGD \psi - m \bp\psi \rp
		 - \frac12 \bp \gamma_\mu \nabGD_\nu \psi
	\label{eqno:EqRicSp} \\
	T^\mu_{\nu\xi} - \tr(T)_\xi \delta^\mu_\nu + \tr(T)_\nu \delta^\mu_\xi
		&= - \frac14 \bp \left\{ [\gamma_\nu,\gamma_\xi],\gamma^\mu \right\} \psi
	\label{eqno:EqTorsSp}\\
	\Dirac\psi - \frac12 \tr(T)_\mu \gamma^\mu\psi -m\psi &= 0
	\label{eqno:EqRGDirac}
\end{empheq}\end{subequations}
\begin{sloppypar}
These are the Einstein-Cartan-Dirac field equations (Equation~\eqref{eqno:EqRicSp} is usually presented with an extra factor $1/2$). Equation~(\ref{eqno:EqTorsSp}) defines algebraically the tensor $T - \Id\wedge \tr T$ as a function of the spinor field, hence the torsion as well (as ${\tr(T - \Id\wedge\tr T) = (1 - 4 + 1) \tr(T)}$).\end{sloppypar}

\subsubsection*{Untreated variational equations}

We obtained equations which correspond to the Euler-Lagrange equations for equivariant variations of $\psi$ and horizontal equivariant variations of $\varpi$. We proved that in this case the term $\phi^*\EL^0_X$ has to vanish. This is only implied by the Euler-Lagrange equations and by no means equivalent. There are two parts of the Euler-Lagrange which we do not use.

First, for \emph{non-equivariant variations} of $\psi$ and \emph{non-equivariant horizontal variations} of $\varpi$, we proved that $\phi^*\EL^0_X$. But the Euler-Lagrange equation~\eqref{eqno:phiEL0exact} is 
\[
	\phi^*(\EL^0_X + i_X\d\thcons) = 0 
\]
Thus the Euler-Lagrange equation is equivalent to 
\begin{equation}
	 \phi^*(i_X\d\thcons) = 0
\end{equation}
which is an equation involving the Lagrange multipliers $p_i^{ab}$ and $\kappa^{\alpha i}$ as well as the fields $\varpi^A$ and $\psi^\alpha$.

Second, we did not consider at all \emph{vertical} variations of $\varpi$. This is because we could not find such variations which preserve the constraints. Furthermore, verticality would also prevent factorising an $\aquatre$ factor out of the exact term in \eqref{eqno:ELaquatre}. Vertical variations of $\alpha^a$ correspond to variations of the vertical distribution which integrates into the orbits. Vertical variations of $\omega^i$ correspond to variations of the vectors fields representing the Lie algebra $\spin_4$.

We only note that these equations are subject to some degeneracy, according to Noether's second theorem. Indeed the Poincaré-Cartan form (\ref{eqno:PC0},\ref{eqno:PCcons}) is invariant by diffeomorphisms of $\tot$ \cite{DynPrincBundle}. 

As the method that allowed us to obtain variational equations on spacetime does not apply to these equation, their study falls outside of the scope of the present paper.

\subsection{Decomposing the field equations}\label{secno:DecompFE}
\com{Dans une section à part ?}

For completion, we present a brief analysis of the Einstein-Cartan-Dirac equations. It is classical and can be found in the literature, as well as its physical implications~\cite{ECD,ECT,STTors}. In order to carry out the analysis, we want to decompose the tensor equations inti components with different index symmetries (sub-representations under $\SO_4$).
\subsubsection*{Pure axiality of the torsion}

Starting with the torsion, it is convenient to have set all index at the same type (covariant) : we write $T_{\tau\mu\nu} = g_{\tau\pi}T^\pi{}_{\mu\nu}$.
The torsion can be decomposed as \cite{STTors,TPGravity,TG}
\begin{equation}
	T_{\tau\mu\nu} = \frac13 \lp \tr(T)_\nu g_{\tau\mu} - \tr(T)_\mu g_{\tau\nu} \rp + \Ax_{\tau\mu\nu} + \mathcal T_{\tau\mu\nu}
\end{equation}
with $\tr(\mathcal T) = 0$, the component $\Ax$ purely antisymmetric and $\mathcal T_{\tau\mu\nu} + \mathcal T_{\mu\nu\tau} + \mathcal T_{\nu\tau\mu} = 0$ ($\mathcal T$ is called the \emph{pure torsion} part). We then express the torsion term of~(\ref{eqno:EqTorsSp}) in terms of these components :
\begin{equation}
	T_{\tau\mu\nu} - \lp g_{\tau\mu} \tr(T)_\nu - g_{\tau\nu} \tr(T)_\mu \rp
	= -\frac23 \lp \tr(T)_\nu g_{\tau\mu} - \tr(T)_\mu g_{\tau\nu} \rp + \Ax_{\tau\mu\nu} + \mathcal T_{\tau\mu\nu}
\end{equation} 

Now the matter term $-\frac 1 2 \bp \{\sigma_{\mu\nu},\gamma_\tau \} \psi$ is totally antisymmetric (see Appendix~\ref{annCommSigGam}). Equation~(\ref{eqno:EqTorsSp}) hence decomposes into the following three equations
\begin{subequations}\begin{align}
	\tr(T) &= 0\\
	\Ax_{\tau\mu\nu} &= - \frac 1 4 \bp \{\sigma_{\mu\nu},\gamma_\tau \} \psi  \label{eqno:TorsAx}\\
	\mathcal T &= 0
\end{align}\end{subequations}
The equations require the torsion to be reduced to its so-called \emph{axial} part. Notably, the trace term appearing in~(\ref{eqno:EqRGDirac}) has to be vanish~\cite{TG} so that
\begin{equation}
	T^\mu_{\nu\xi}
			= - \frac14 \bp \left\{ [\gamma_\nu,\gamma_\xi],\gamma^\mu \right\} \psi
\end{equation}
The $3$-form $\Ax$ which is part of the spacetime geometry is (algebraically) coupled to the spinor field ; one can say that this degree of freedom is what separates Einstein-Cartan theory from Einstein's theory of General Relativity. Equation~(\ref{eqno:EqTorsSp}) corresponds to variations of the connection and as such the matter term (the $\psi$ part) is identified with the angular momentum current~\cite{EMTens}. Torsion hence couples the angular momentum current with the various fields of the theory ($\psi$ in the present case).

As a purely antisymmetric $(4-1)$-form, we can express $\Ax$ by its dual (pseudo-)vector $A$, also called \emph{axial vector} :
\[
	\Ax_{\tau\mu\nu} = (A\lr \vol)_{\tau\mu\nu} = A^\xi\epsilon_{\xi\tau\mu\nu}
\]
with $\epsilon_{\xi\tau\mu\nu}$ the Levi-Civita symbol, which is the components of a volume form in a basis of determinant $1$.
Equation~(\ref{eqno:TorsAx}) can be reexpressed using the chirality operator \[\gamma_5 = \gamma_0 \gamma_1 \gamma_2 \gamma_3\] defined in Appendix~\ref{annCommSigGam} : 
	\[A^\xi\epsilon_{\xi\tau\mu\nu} = -\frac12 \epsilon_{\xi\mu\nu\tau}\bp \gamma^\xi\gamma_5 \psi\]
from which we obtain
\begin{equation}\begin{aligned}
	A^\xi &= - \frac12\bp\gamma^\xi\gamma_5 \psi
\end{aligned}\end{equation}
The equation of motion of the spinor~(\ref{eqno:EqRGDirac}) and its conjugate then take the form
\begin{subequations}\label{seqno:SpEOM}\begin{align}
	\lp \Dirac - m \rp \psi &= 0\\
	\lp \bar \Dirac - m \rp \bp &= 0
\end{align}\end{subequations}
and one concludes that the term
	$ - g_{\mu\nu} \bp (\frac 1 2 \DirGD  - m) \psi$
in~(\ref{eqno:EqRicSp}) has to vanish. This is a general property of homogeneous Lagrangians (as is the Dirac Lagrangian) : they take the value zero on-shell.

\subsubsection*{Symmetric and antisymmetric parts of the curvature-energy relation}

The first equation~(\ref{eqno:EqRicSp}) corresponds to Equation \eqref{eqno:FEtotab} hence to the coefficient $\epsilon^a_b$ in variations, that is to say it is the Euler-Lagrange term corresponding to horizontal variations of $\alpha$. It can be decomposed into a symmetric and an antisymmetric parts. We use the parenthesis $(\mu\nu)$ notation for symmetrisation and bracket notation $[\mu\nu]$ for antisymmetrisation (normalized by a $1/2$ factor) :

\begin{subequations}\begin{align}
	2\Ric_{(\mu\nu)}
	- g_{\mu\nu} \Scal
	&= g_{\mu\nu} \lp \frac12 \bp \DirGD \psi - m \bp\psi \rp 
	- \frac12\bp \gamma_{(\mu}\nabGD_{\nu)} \psi 
	  \label{eqno:RicSym}\\
	2\Ric_{[\mu\nu]}
	&= - \frac12 \bp \gamma_{[\mu}\nabGD_{\nu]} \psi \label{eqno:RicAsym}
\end{align}\end{subequations}

We first look at Equation~(\ref{eqno:RicAsym}). It is the variational term corresponding to \enquote{antisymmetric} variations in the solder form, that it to say variations which preserve the metric. In other words, they correspond to infinitesimal automorphisms of the frame bundle.

The Bianchi identity on the Ricci curvature~(\ref{eqno:RicBianchiBase}) (in Appendix \ref{annRicBianchi}) relates the asymmetric part of the Ricci curvature to the exterior divergence of the torsion
\begin{equation}\label{eqno:RicBianchimunu}
	2\Ric_{[\mu\nu]}
	= \nabla_\pi T^\pi_{\mu\nu} - \d\tr (T)_{\mu\nu}
	= \nabla_\pi T^\pi_{\mu\nu}
\end{equation}
It allows to rewrite \eqref{eqno:RicAsym} as an equation on torsion : 
\begin{equation}
	\nabla_\pi T^\pi_{\mu\nu} = -\frac12 \bp \gamma_{[\mu}\nabGD_{\nu]} \psi
\end{equation}
The right-hand term, which corresponds to (twice) the antisymmetric part of the \enquote{canonical energy-momentum tensor} (the terms in $\psi$ in~(\ref{eqno:EqRicSp})), is the so-called \emph{Belinfante (improvement) tensor}~\cite{WeinbergI,MetricAffine2,STandFields}.

The Cartan geometry, through the Bianchi identity, then imposes the following equation
\begin{equation}\label{eqno:ConsBianchitmp}
	\boxed{
	\frac12 \bp \gamma_{[\mu}\nabGD_{\nu]} \psi
	= \frac14\nabla_\pi \bp \{\sigma_{\mu\nu},\gamma^\pi \} \psi }
\end{equation}
which relates the Belinfante tensor to the \emph{covariant} divergence of the angular momentum current (also called \emph{spin density}~\cite{ECT} or \emph{spin current}~\cite{MetricAffine1}). In theories without torsion, the connection has to follow the solder form variations, so that Equation~\eqref{eqno:ConsBianchitmp} would directly take the place of~\eqref{eqno:RicAsym} as in~\cite{WeinbergI} (as variational equation for variations of the solder form which preserve the metric).

The symmetric component~\eqref{eqno:RicSym} corresponds to complementary variations of the solder form hence to variations of the metric (symmetric variations of a frame have been considered as the natural complementary to isometric variations~\cite{MetVar}). The corresponding matter term is then identified with the symmetric energy-momentum tensor~\cite{EMTens,MetricAffine1}. It is Einstein's field equation binding spacetime's (Ricci symmetric) curvature to the distribution of energy-momentum.

Taking into account the fact that the term corresponding to the Dirac Lagrangian vanishes due to~\eqref{seqno:SpEOM}, Equations~\eqref{eqno:RicSym} simplifies to
\begin{equation}\boxed{
	2\Ric_{(\mu\nu)} - g_{\mu\nu} \Scal
	= \frac 12 \bp \gamma_{(\mu}\nabGD_{\nu)} \psi 
}\end{equation}

\subsection{Expression in terms of the Levi-Civita connection}

To compare the Einstein-Cartan theory with Einstein's General Relativity, we relate the connection to the Levi-Civita connection by means of its \emph{contorsion}, which is defined as its difference to the Levi-Civita connection :
\begin{equation}\label{KContors}
	\nabla = \nabla^{LC} + K
\end{equation}
with $K$ a $1$-form on $\tot$ with values in $\so(T\tot)$. According to the previous section, the field equations~(\ref{eqno:EqTorsSp}) require the torsion to be purely axial. As contorsion is uniquely defined by the torsion (assuming metricity) \cite{TPGravity,STTors}, in our case $K$ has to be $\frac12 T$ :
\[  K_{\mu\nu}^\pi = \frac12 T^\pi_{\mu\nu}  \]

\subsubsection*{Ricci and scalar curvatures}

According to Equation~\eqref{VarRic} from Section~\ref{RicContors}, the Ricci curvature of the connection can be related to the Ricci curvature of the Levi-Civita connection by the following equation :
\begin{equation}\begin{aligned}
	\Ric_{\mu\nu}
	&= \Ric^{LC}_{\mu\nu} + \delta^\tau_\pi \lp
		\nabla^{LC}_\tau K^\pi_{\mu\nu} - \nabla^{LC}_\mu K^\pi_{\tau\nu}
		+ K^\pi_{\tau \kappa}K^\kappa_{\mu\nu} - K^\pi_{\mu \kappa}K^\kappa_{\tau\nu} \rp\\
	&= \Ric^{LC}_{\mu\nu} + \frac12\operatorname{div}^{LC}T_{\mu\nu} - 0 + 0 - \frac14 T^\pi_{\mu \kappa}T^\kappa_{\pi\nu}
\end{aligned}\end{equation}
We see that the Ricci curvatures difference have an antisymmetric term $\operatorname{div}^{LC}T_{\mu\nu}$ and a symmetric term $-\frac14 T^\tau_{\mu \kappa}T^\kappa_{\tau\nu}$, as the torsion is purely axial. Since the Ricci curvature of the Levi-Civita connection is symmetric (due to the Bianchi identity \eqref{eqno:RicBianchimunu} for a torsion-free connection), the antisymmetric difference between the Ricci curvatures is exactly the antisymmetric part of the Ricci curvature of the connection, as expressed in the Bianchi identity. 

We express the Ricci curvature difference in term of the axial vector. Start with the quadratic term :
\begin{equation*}\begin{aligned}
	T^\pi_{\mu\kappa}T^\kappa_{\pi\nu} 
		&= g^{\pi\tau}g^{\kappa\rho} T_{\tau\mu\kappa} T_{\rho\pi\nu} 
		= g^{\pi\tau}g^{\kappa\rho} A^\xi A^\upsilon \ve_{\xi \tau\mu\kappa} \ve_{\upsilon \rho\pi\nu}\\
		&= A^\xi A^\upsilon (g_{\xi\upsilon}g_{\mu\nu} - g_{\xi \nu}g_{\mu\upsilon})\\
		&= A^\xi A_\xi g_{\mu\nu} - A_\nu A_\mu
\end{aligned}\end{equation*}
The change in scalar curvature is directly derived :
\begin{equation} 
	g^{\mu\nu} \lp - \frac14 T^\pi_{\mu\kappa}T^\kappa_{\pi\nu} \rp = -\frac{3}4 A^\xi A_\xi
\end{equation}
Identifying the totally antisymmetric torsion with a $3$-form, its divergence for the torsion-free Levi-Civita connection corresponds to (minus) the codifferential for the Hodge duality structure \cite{Petersen}. It can hence be identified with the Hodge dual of the exterior differential of the axial $1$-form $A_\pi = g_{\pi\xi}A^\xi$ :
\begin{equation} \operatorname{div}^{LC} T_{\mu\nu} = -(\star \d A)_{\mu\nu}
\end{equation}

The left-hand terms of~(\ref{eqno:RicSym}-\ref{eqno:RicAsym}) are expressed as
\begin{equation}
\begin{aligned}	
	2\Ric_{(\mu\nu)} - g_{\mu\nu}\Scal
	&=	2\Ric^{LC}_{\mu\nu} - g_{\mu\nu} \Scal^{LC}
	 -	\frac24 (A^\xi A_\xi g_{\mu\nu} - A_\mu A_\nu) +g_{\mu\nu} \frac34 A^\xi A_\xi\\
	&= 2\Ric^{LC}_{\mu\nu} - g_{\mu\nu} \Scal^{LC}
	 +	\frac14 \lp A^\xi A_\xi g_{\mu\nu} + 2 A_\mu A_\nu \rp 
\end{aligned}\end{equation}
and
\begin{equation}
	2\Ric_{[\mu\nu]} = -(\star \d A)_{\mu\nu}
\end{equation}

\subsubsection*{The spinor connection}

We also need to express the covariant derivative of spinor fields in terms of $K$. It is straightforward from~(\ref{KContors}) : 
\begin{equation}
	\nabla_\mu\psi
	 = \nabla_\mu^{LC} \psi + \frac12 K_{\mu\nu}^\pi \frac12\sigma_\pi{}^{\nu}\psi
	 = \nabla_\mu^{LC} \psi + \frac18 T_{\tau\mu\nu}\sigma^{\tau\nu}\psi
\end{equation}
We can re-express the kinetic term $\bp \gamma_\mu\nabGD_\nu \psi$ : 
\begin{equation}\label{eqno:ConnToLC}\begin{aligned}
	\bp \gamma_\mu\nabGD_\nu \psi 
		&= \bp \gamma_\mu \lp \nabla^{LC}_\nu + \frac18 T_{\nu\pi\tau}\sigma^{\tau\pi} \rp\psi
		- \lp \nabla^{LC}_\nu + \frac18 T_{\nu\pi\tau}\bar\sigma^{\tau\pi} \rp \bp \gamma_\mu \psi\\
		&= \bp \lp \gamma_\mu\nabGD_\nu^{LC} +\frac18 T_{\nu\pi\tau}\{\sigma^{\pi\tau},\gamma_\mu\} \rp \psi
\end{aligned}\end{equation}
and in terms of the axial (pseudo)-vector, using the formulae~(\ref{eqno:CommChir}-\ref{eqno:LCcontr}) from Appendix~\ref{annCommSigGam} :  
\begin{gather}
	\frac12 \{\sigma_{\pi\tau},\gamma_\mu\} = \ve_{\upsilon\pi\tau\mu}\gamma^\upsilon\gamma^5
	\label{eqno:CommChir}\\
	\frac12 \ve^{\xi\nu\pi\tau}\ve_{\upsilon\pi\tau\mu}
	= \lp \delta^\xi_\upsilon \delta^\nu_\mu - \delta^\xi_\mu\delta^\nu_\upsilon \rp
	\label{eqno:LCcontr}
\end{gather}
to obtain :
\begin{equation}\label{eqno:ConnToLCAx}\begin{aligned}
	\bp \gamma_\mu\nabGD_\nu \psi 
		&= \bp \lp \gamma_\mu\nabGD_\nu^{LC} +\frac14
		T_{\nu\tau\kappa}
		\ve^{\upsilon\tau\kappa}{}_\mu{}
		\gamma_{\upsilon} \gamma_5\rp \psi\\
		&= \bp \lp \gamma_\mu\nabGD_\nu^{LC} +\frac14
		A^\xi\ve_{\xi\nu\tau\kappa}
		\ve^{\upsilon\tau\kappa}{}_\mu{}
		\gamma_{\upsilon} \gamma_5\rp \psi\\
		&= \bp \lp \gamma_\mu\nabGD_\nu^{LC}
			+ \frac12 \lp A^\xi\gamma_\xi g_{\mu\nu} - A_{\mu}\gamma_\nu \rp
			\gamma_5 \rp \psi
\end{aligned}\end{equation}
The Dirac operator is readily rewritten as well, using~(\ref{eqno:CommChir}) and the total antisymmetry mentioned right above:
\begin{equation}\begin{aligned}
	\Dirac &= \Dirac^{LC} + \frac18\gamma^\mu T_{\tau\mu\nu}\sigma^{\tau\nu}
		   = \Dirac^{LC} + \frac18 A^\xi\ve_{\xi\tau\mu\nu}\frac12\{\gamma^\mu,\sigma^{\tau\nu}\}
		   = \Dirac^{LC} - \frac18 A^\xi 3!\gamma_\xi\gamma_5\\
		   &= \Dirac^{LC} - \frac34 A^\xi\gamma_\xi\gamma_5
\end{aligned}\end{equation}

The field equations~(\ref{eqno:EqRicSp}-\ref{eqno:EqRGDirac}) can then be reformulated as a Levi-Civita connection + axial (pseudo)-vector theory :
\begin{subequations}\begin{empheq}[left=\empheqlbrace]{align}
	2\Ric^{LC}_{\mu\nu} -g_{\mu\nu} \Scal^{LC}
	 +\frac14 \lp A^\xi A_\xi g_{\mu\nu} + 2 A_\mu A_\nu \rp 
	&= 	\frac12 \bp \lp \gamma_{(\mu}\nabGD^{LC}_{\nu)}
			+ \frac12 \lp A^\xi\gamma_\xi g_{\mu\nu} - A_{(\mu}\gamma_{\nu)} \rp \gamma_5 
		\rp \psi		\label{eqno:LCRicAx}\\
	(\star \d A)_{\mu\nu}
	&= \frac12 	\bp \lp
			\gamma_{[\mu}\nabGD^{LC}_{\nu]}
			- A_{[\mu}\gamma_{\nu]}  \gamma_5
		\rp \psi		\label{eqno:LCTorsAx}\\
	A^\xi &= -\frac12\bar\psi \gamma^\xi\gamma_5 \psi   \label{eqno:AxSpin}\\
	\Dirac^{LC} \psi  - \frac34 A^\xi\gamma_\xi\gamma_5\psi - m \psi &= 0 \label{eqno:AxDir}
\end{empheq}\end{subequations}
The axial pseudo-vector is algebraically defined by~(\ref{eqno:AxSpin}) according to which it corresponds the chiral currents. It can be integrated away from the field equations, giving for example in~(\ref{eqno:AxDir}) a cubic interaction term
\[+ \frac38 \lp \bp \gamma^\xi\gamma_5\psi \rp \gamma_\xi\gamma_5\psi \]
This being said, bringing back the dimensional constants in the equations, one notice that the right-hand side of~(\ref{eqno:LCRicAx}) has the gravitational constant as a factor~\cite{GaugeGrav}. Hence the cubic term tends to be very weak in standard matter.
In~(\ref{eqno:LCRicAx}), the (axial) torsion has a contribution to both the Einstein curvature term and the (matter) energy-momentum term.

\appendix\clearpage

\section{Conventions for dual forms}\label{anndual}
Let $E$ a $n$-dimensional $\setK$-vector space with $n\in\mathbb N$. Provide it with a volume element $\vol\in\ExT^n E$. The volume element defines an isomorphism $\ExT^n E \simeq \setK$. Note that in the article we would use a space (bundle) of \emph{linear forms} as $E$. This Appendix will make use of the Einstein summation convention for repeated indices (as stated in~\ref{secno:notconv}).

We define here the notation for interior products between $p$-covectors and $q$-vectors. We will write $\wedge_{i:1 \to q} \alpha^i$ for the wedge product $\alpha^1\wedge\alpha^2\dots\wedge\alpha^q$. Let $\alpha\in\ExT^q E$ and $(X_j)\in (E^*)^p$. Our convention will be the following :
\begin{equation}
\boxed{
	(X_1\wedge X_2\wedge \dots \wedge X_p)\lr \alpha
	= i_{X_1\wedge X_2\wedge \dots \wedge X_p} \alpha
	= i_{X_p}i_{X_{p-1}}\cdots i_{X_1}\alpha}
\end{equation}
The volume element provides isomorphisms $\ExT^p E^* \to \ExT^{n-p} E$ under the following contraction
\begin{equation}
	X\in\ExT^p E^* \mapsto X\lr\vol
\end{equation}
Note that if $E$ is provided with an inner product then precomposing the isomorphisms by the induced inner product on the exterior powers $\ExT^p E \to \ExT^p E^*$ gives the \emph{Hodge duality operator} $\star$.

Let $(e^i)_{\lel 1 i n}$ be a \emph{direct} basis of $E$ and $(u_i)$ its dual basis. Let $I$ a sequence of $p$ indices in $\llbracket 1,n\rrbracket$. Define 
\[ e^I = \operatornamewithlimits{\bigwedge}_{i : 1 \to p} e^{I_i} \]
and dually
\[ u_I = \operatornamewithlimits{\bigwedge}_{i : 1 \to p} u_i \]
We will explicitly use the map $\ExT^p E^* \to \ExT^{n-p} E$ and adopt the notation
\begin{equation}
	e^{(n-p)}_I := u_I\lr\vol
\end{equation}

In components, $\vol$ is represented by the \emph{Levi-Civita symbol} commonly written $\ve_ {i_1\dots i_n}$, interpreted as a \emph{completely antisymmetric} rank $n$ tensor \emph{in a basis of determinant $1$}. For example, the duality between $E^*$ and $\ExT^{n-1} E$ is expressed, for $A=A^iu_i$ as
\begin{align}
	(A^iu_i)\lr \vol &= A^i(u_i\lr\vol) = A^ie^{(n-1)}_i\\ 
	(A\lr \vol)_{i_1\dots i_{n-1}} &= A^{i_0}\ve_{i_0i_1\dots i_{n-1}}
\end{align}
which generalizes to the case of $p$-forms on $E$ in a straightforward manner.

From now on we assume that $I$ does not contain repeated indices of the basis so that $u_I$ and $e^I$ are nonzero. The conventions have been chosen such that for two increasing multi-indices $I$ and $J$,
\begin{equation}
u_I\lr e^J = \delta^J_I
\end{equation}
Define $I^c$ as the set of indices absent from $I$, identified with the corresponding increasing sequence. We will write $\epsilon(I)$ for the signature of the permutation which has $I$ as its 1st $p$ values and then $I^c$ (the permutation is a shuffle in the case $I$ follows the increasing order) so that
\begin{equation}\label{eqno:eIIcvol}
	e^I \wedge e^{I^c} = \epsilon(I)\vol
\end{equation}
We can then express the contraction in the following way
\begin{equation}\label{eqno:eInp}
	e^{(n-p)}_I = u_I\lr \vol = u_I\lr \epsilon(I)e^I\wedge e^{I^c}
	 = (u_I\lr e^I)\epsilon(I)\wedge e^{I^c} = \epsilon(I)e^{I^c}
\end{equation}
Note that this formula can act as a definition in the case $(e^i)$ form a general family of vectors (not assumed to be a basis).
Combining (\ref{eqno:eIIcvol},\ref{eqno:eInp}) obtain the following formula stating that $e^{(n-p)}_I$ represent a (twisted) $1$-form on $\ExT^p E$ : for $f_Je^J\in\ExT^p E$ 
\begin{equation}\boxed{
	\lp f_Je^J \rp \wedge e^{(n-p)}_I = f_I\vol}
\end{equation}
which justifies placing $I$ as a subscript in $e^{(n-p)}_I$.

Note that 


\section{Spinors and Clifford algebra}\label{annspin}
\subsection{Clifford modules}\label{annCMod}

The spinors we consider are the so-called \emph{Dirac spinors}. They are elements of a complex Clifford module. Set a signature $(p,q)$, which will be either Euclidean $(4,0)$ or $(0,4)$ or Lorentzian $(3,1)$ or $(1,3)$ for our purposes (see~\ref{annSignConv} for details about the signature alternatives).

The Clifford algebra $\Cl_{p,q}$ is defined as the quotient of the tensor algebra of $\Rpq$ by the ideal generated by the (even) elements $v\otimes v + \sprod{v}{v}_{p,q}$ for $v\in\Rpq$. It is the universal algebra satisfying ${v_1 \cdot v_2 + v_2\cdot v_1 = -2\sprod{v_1}{v_2}_{p,q}}$. It has a natural structure of super-algebra ($\setZ/2\setZ$-grading). The Euclidean (4-dimensional) Clifford algebras are $\Cl_{4,0}\simeq \Cl_{0,4}\simeq\Mat_2(\setH)$ while the Lorentzian Clifford algebras are $\Cl_{1,3}\simeq\Mat_4(\setR)$ and $\Cl_{3,1}\simeq \Mat_2(\setH)$ \cite{MajSpin}.

Our spinors are subject to the Dirac equation $\Dirac \psi - m\psi = 0$ which requires (with these conventions) a structure of \emph{real} Clifford module. Nonetheless we will consider complex spinors, so as to get a Clifford algebra in which Wick rotations between all signatures are straightforward to implement (as well as flipping the sign in the Clifford algebra definition). The complexified Clifford algebras of dimension $2k$ are all isomorphic to $\Mat_{k}(\setC)$, hence the irreducible module (class) is of complex dimension $k$. We will write $\Sppq$ for the irreducible (complex) module. It admits (for each signature) an hermitian form which is compatible with the \emph{real} Clifford module structure in the following sense :
\begin{equation}\label{eqno:CliffEquiv}
	\forall \gamma_a\in\Rpq\subset\Cl_{p,q},(s_1,s_2)\in\Sppq^2,\quad
	\sprod{\gamma_a\cdot s_1}{s_2} + \sprod{s_1}{\gamma_a\cdot s_2} = 0
\end{equation}
It is uniquely defined up to a nonzero real factor. It is defined so that the Dirac operator $\gamma^a\nabla_a$ is hermitian. We will fix such a \emph{spinor metric} and use the same normalization everywhere. Normed vectors of positive (squared) norm act by unitary transformations while those of negative norm act by involutive anti-isometries of the metric. Note that the signature of the hermitian form is either $(4,0)$, $(0,4)$ or $(2,2)$, depending on the spacetime signature \cite{SpinHerm}.

We will make use of the implicit notation $\bar{s_1}s_2$ for the product $\sprod{s_1}{s_2}$ of elements of $\Sppq$. In this notation $s_1$ is to be understood as an element of $\overline\Sppq$ identified with $\Sppq^*$ through the hermitian form. Elements of $\Cl_{p,q}$ acting on $\overline\Sppq$ will be often represented \emph{as acting on the right} for notational convenience : $\bar{s_1} \gamma_a = - \bar{\gamma_a} \cdot \bar{s_1}$.

\subsection{\texorpdfstring{$\Pin$ and $\Spin$ groups}
			{Pin and Spin groups}}\label{annPinSpin}

The invertible elements of the Clifford algebra act by conjugation (one also considers twisted conjugation, but it will not be relevant for our purposes). The subgroup of elements preserving the vector space $\Rpq\subset\Cl_{p,q}$ will be written $G$. It can be identified as the group composed of elements of $\Clpq$ that can be expressed as a product of non-isotropic vectors of $\Rpq$ (in a non-unique way).

For practical purposes, it is more convenient to reduce this group to the products of \emph{normed} vectors of $\Rpq$. The group $G$ has a morphism $G\xrightarrow{N} \setR^*$ which corresponds to the product of the (non-negated) squared norms of the vectors which compose the element \cite{NotesSpin}. The hermitian form is then $G$-equivariant seen as a morphism $\overline\Sppq\otimes\Sppq\to \setC$. The inverse image of $\{\pm1\}$ by $N$ is called the (Cliffordian) \emph{Pin group}, written $\Pin_{p,q}$. Its elements act on $\Rpq$ by isometries hence it is provided with a natural morphism to $\operatorname{O}_{p,q}$, which is a two-fold covering. The subgroup composed of even elements is called the \emph{Spin group}, written $\Spin_{p,q}$ and has a natural two-fold covering map to $\SO_{p,q}$. 

In Euclidean signature, $\Spin_{p,0}$ is identified with the kernel of $G\to\setR^*$. The group $\Pin_{p,0}$ is a maximal compact subgroup of $G$ and $\Spin_{p,0}$ its principal component.
In non-Euclidean signature the kernel of $G\to\setR^*$ is a different subgroup. It corresponds to products of normed vector with an even number of spacelike (negative norm) vectors. They act on $\Rpq$ by isometries preserving the orientation in time. The intersection of the kernel with the even subalgebra defines the \emph{orthochronous $\Spin$ group}, written $\Spinp_{p,q}$. It acts on $\Rpq$ by isometries preserving both orientations in space and in time. Note that some authors (as in~\cite{NotesSpin}) take this group as the $\Spin$ group.

Note that due to Equation~(\ref{eqno:CliffEquiv}) $\Spinp_{p,q}$ acts by isometries of $\Sppq$, and $\sopq$ by infinitesimal isometries. The hermitian form being invariant under $\Spinp_{p,q}$ can be interpreted as the form establishing an equivariant isomorphism $\overline\Sppq \to \Sppq^*$ from the complex conjugated $\Spin$ module to the dual complex $\Spin$ module.

The Lie algebra of $\SO_{p,q}$ and $\Spin_{p,q}$ are isomorphic and they turn out to be realized in $\Clpq$ by the space (linearly) spanned by commutators of vectors, for the algebraic (ungraded) commutator bracket \cite{NotesSpin} :
\[ \forall (a,b)\in\Rpq,\quad  \frac12(ab-ba)\in\spin_{p,q}\subset \Clpq \]

The isomorphism with $\sopq$ uses the standard representation as anti-symmetric operators $\sopq\subset \End(\Rpq)\simeq \Rpq\otimes\Rpq^*$ and composes with the inverse metric $\Rpq^*\to\Rpq$ to obtain a \emph{linear} isomorphism $\sopq\overset\rho\simeq \ExT^2 \Rpq$. It is then sent onto $\spin_{p,q}$ by the mapping
\[ a\wedge b \to \frac14(a\cdot b - b\cdot a) \]
For this reason, when using indices $i$ for $\sopq$ and $a,b$ for $\Rpq$, we will be using both notations $1/2\sigma_i$ and $1/2\sigma_{ab}$ for a basis of $\spin_{p,q}$ embedded in $\Clpq$, with a notation reminiscent of the Pauli matrices (which represent a 3-dimensional Clifford algebra), using the $1/2$ factor but not the $i$ factor common in the physics literature (used to turn them into hermitian operators). Explicitly, writing in components $\rho_{i}^{ab}$ the morphism $\sopq\to\ExT^2\Rpq$ the two notations are related by
\begin{equation} \frac12 \rho_{i}^{ab}\sigma_{ab} = \sigma_i	\end{equation}
We also use the common notation $\gamma : \Rpq \to \End(\Sppq)$.

The \emph{chirality} operator is defined as the image of the volume element of $\Rpq$ : for an oriented orthonormed basis $(e_1,e_2,e_3,e_4)$
\begin{equation} \gamma_5 := \gamma_1\gamma_2\gamma_3\gamma_4 \end{equation}
which as the notation suggests defines a morphism from a higher dimensional Clifford module.
Aiming for a unified treatment, we do not add the usual $i$ factor in the case of a Lorentzian signature. It satisfies $(\gamma_5)^2 = (-1)^q$ and lies in the supercenter of the algebra in even dimension. Its eigenspaces on $\Sppq$ are irreducible representations of $\Spin_{p,q}$ and spinors with values in these representations are called \emph{Weyl spinors}. For our purposes, it will implement the duality between codegree $1$ forms and vectors (as discussed in Section~\ref{anndual}).

\subsection{Chiral current}\label{annCommSigGam}

Seen as a morphism $\Rpq \to \End(\Sppq)$, $\gamma$ is $\Spin_{p,q}$-equivariant ($\Spin_{p,q}$ being represented by $\SO_{p,q}$). The hermitian metric allows us to define a \enquote{dual} $\Gamma : \overline\Sppq\otimes \Sppq \to \Rpq^*$ which is equivariant as well :
\begin{equation} (a,s_1,s_2)\in\Rpq\otimes\overline\Sppq\otimes\Sppq\mapsto \sprod{s_1}{\gamma(a)\cdot s_2}
\end{equation}
Notice how the action of the $\Pin_{p,q}$ group is \emph{twisted} under the morphism, as $\gamma$ takes antihermitian values. In a similar way, one can define tensor-valued hermitian forms by using products of $\gamma$.

We are interested in the element $\{\sigma_{\mu\nu},\gamma_\tau\}$ as it appears in the equation~(\ref{eqno:ConnToLC}). By definition $\sigma_{\mu\nu} = \frac12[\gamma_\mu,\gamma_\nu]$. Given that the commutator bracket is a Poisson bracket, one has a Jacobi-like (derivation) identity with the anti-commutator, so that
\begin{equation}
	\{[\gamma_\mu,\gamma_\nu],\gamma_\tau\}
	= [\gamma_\mu,\{\gamma_\nu,\gamma_\tau\}] - \{\gamma_\nu,[\gamma_\mu,\gamma_\tau]\}
	= \{\gamma_\nu,[\gamma_\tau,\gamma_\mu]\}
\end{equation}
hence the $\End(\Sppq)$-valued $3$-form $\{\sigma_{\mu\nu},\gamma_\tau\}$ is antisymmetric in two pairs of indices, namely totally antisymmetric as two transpositions span the whole symmetric group. It can be expressed using the chirality element and the Levi-Civita symbol (see Appendix~\ref{anndual}), using the method described in \cite{GammaSigma} (with a chirality element different by an $i$ factor) :
\begin{equation}\tag{\ref{eqno:CommChir}}
	\frac12 \{\sigma_{\mu\nu},\gamma_\tau\} = \ve_{\upsilon\mu\nu\tau}\gamma^\upsilon\gamma^5
\end{equation}
As $\gamma_\mu$ have antihermitian values, $\frac12 \{\sigma_{\mu\nu},\gamma_\tau\}$ takes value in \emph{hermitian} operators.

We also record the following formula, also proved (using the Lorentzian signature) in \cite{GammaSigma} :
\begin{equation}\tag{\ref{eqno:LCcontr}}
	\frac12 
	\epsilon_{\xi\nu\tau\chi}
	\epsilon^{\upsilon\tau\chi\mu}
	= (-1)^q \lp \delta_\xi^\upsilon \delta_\nu^\mu - \delta_\xi^\mu\delta_\nu^\upsilon \rp
\end{equation}
with $(-1)^q$ corresponding to the norm of the positive volume element $\vol$.

\subsection{Signature and conventions}\label{annSignConv}

There exists two different conventions for the Clifford algebra of a given signature, namely to choose either $v_1 \cdot v_2 + v_2\cdot v_1 = -2\sprod{v_1}{v_2}_{p,q}$ or $v_1 \cdot v_2 + v_2\cdot v_1 = 2\sprod{v_1}{v_2}_{p,q}$. Going from one convention to the other is equivalent to consider the opposite metric ; as noted in~\cite{NotesSpin} Clifford algebras of opposite signature are ($\setZ/2\setZ$-graded) opposite algebras of each other. As $\Spin$ groups are composed of even elements of the Clifford algebra, $\Spin$ groups of opposite signature are \emph{isomorphic}.

As illustrated in Section~\ref{annCMod} real Clifford algebras of different signature can be non-isomorphic. Complex Clifford algebras, on the other hand, are all isomorphic as all (real) bilinear metrics are congruent under the action of the complex linear group. Note however that the spinorial metric, as we defined it in~\ref{annCMod}, is dependent on the \emph{real} Clifford algebra.

Physical theories require an action of the $\Pin$ group on \enquote{spinors} -- they are then sometimes called \emph{pinors}~\cite{MajSpin,PinGroups}. The catch is that $\Pin$ groups of opposite signature are \emph{non-isomorphic}~\cite{PinGroups} (and may actually need to be specified~\cite{PinGR}). In this respect, the exact choice of a signature sign (along with a sign convention) does matter. The structure needed on a pseudo-riemannian manifold to carry (real) pinors is a $\Pin$ structure. That being said, a $\Spinp$ structure naturally induces $\Spin$ and $\Pin$ structures of the corresponding signatures. As $\Spinp_{1,3}\simeq\Spinp_{3,1}$ (lifting an isomorphism $\SOp_{1,3}\simeq\SOp_{3,1}$), a given $\Spinp_{1,3}$ structure can induce both $\Pin_{1,3}$ and $\Pin_{3,1}$ structures (for opposite metrics).

\section{Universal formulas for differentials of canonical forms}\label{anncalc}
Define $Q:=T^*\tot\otimes\p$. It is provided with a tautological $\p$-valued 1-form $\lambda$. Let in this section $\Sp$ be a $\p$-module (as a trivial vector bundle over $\tot$), possibly a $\l$-module, with $s^a$ as fibre coordinates and $\bs_\alpha$ coordinates for the dual module. Recall that $\dl s:= \d s + \lambda \cdot s$ with an implicit wedge product. 
We will establish the following formulae in a \enquote{universal} framework, without the structure of a principal connection:
\begin{subequations}\begin{align}
	\d\ldix &= \dll^A\wedge \lneuf A	 			\\
	\d\lneuf A &= 		\dll^C\wedge \d\lhuit {AC} = \dl \lneuf A	\label{dlneuf}\\
	\d\lhuit {AB} &= 	\dll^C\wedge \d\lsept{ABC}-c^C_{AB}\lneuf C	\\
	\d \lp \bs_\alpha s^\alpha \rp &= (\dl \bs)_\alpha s^\alpha + \bs_\alpha \dl s^\alpha \\
	\dl \dll &= 0	\\
	\dl(\dl s)^\alpha &=	\dll \cdot s^\alpha	
\end{align}\end{subequations}
with $\lambda$ the canonical form on $\T^*\tot\otimes\p$.
%
%

%
%
%
The first computation goes
\begin{equation*}
\begin{aligned}
	\d\ldix
		&= (\d\lambda^A) \wedge \lneuf A\\
		&= \lp \dll-\fwb\lambda\lambda \rp^A \wedge \lneuf A\\
		&= \dll^A \wedge \lneuf A
\end{aligned}\end{equation*}
In a similar fashion
\begin{equation*}
\begin{aligned}
	\d\lneuf A
		&= (\d\lambda^B) \wedge \lhuit{AB}\\
		&= \lp \dll-\fwb\lambda\lambda \rp^B \wedge \lhuit{AB}\\
		&= \dll^B \wedge \lhuit{AB} - \frac12 c^B_{CD}\lambda^C \wedge \lambda^D \wedge \lhuit{AB}\\
		&= \dll^B \wedge \lhuit{AB} - c^B_{AB}\ldix\\
		&= \dll^B \wedge \lhuit{AB}
\end{aligned}\end{equation*}
with $c^B_{AB}=0$ by \emph{unimodularity} of the Poincaré algebra. To relate $\d\lneuf A$ to $\dl \lneuf A$ one just has to notice that 
\[ \lambda\cdot \lneuf A = -c^{C}_{BA}\lambda^B \wedge \lneuf C = -c^B_{BA}\ldix = 0\]
according to the same argument, so that
\[ \dl\lneuf A = \d\lneuf A \]
The next computation is similar :
\begin{equation*}
\begin{aligned}
	\d\lhuit {AB}
		&= (\d\lambda^C) \wedge \lsept{ABC}\\
		&= \lp \dll-\fwb\lambda\lambda \rp^C \wedge \lsept{ABC}\\
		&= \dll^C \wedge \lsept{ABC} -\frac12 c^C_{DE}\lambda^D \wedge \lambda^E \wedge \lsept{ABC}\\
		&= \dll^C \wedge \lsept{ABC} - \lp c^C_{BC}\lneuf A - c^C_{AC}\lneuf B + c^C_{AB}\lneuf C \rp \\
		&= \dll^C \wedge \lsept{ABC} - c^C_{AB}\lneuf C 
\end{aligned}\end{equation*}
Next we check the compatibility with equivariant contractions :
\begin{equation*}
\begin{aligned}
	\d(\bs_\alpha s^\alpha)
		&= (\d \bs_\alpha) s^\alpha + \bs_\alpha \d s^\alpha\\
		&= (\dl \bs-\lambda \bs)_\alpha s^\alpha + \bs_\alpha (\dl s-\lambda \cdot s)^\alpha\\
		&= (\dl \bs_\alpha) s^\alpha + \bs_\alpha (\dl s)^\alpha
			- (\lambda \cdot \bs)_\alpha s^\alpha) + \bs_\alpha (\dl s-\lambda \cdot s)^\alpha \\
		&= (\dl \bs_\alpha) s^\alpha + \bs_\alpha (\dl s)^\alpha
\end{aligned}\end{equation*}
In particular, for $\psi$ a $\Sp$-valued $k$-form over $Q$ and $\bp$ a $\bSp$-valued $l$-form, the following holds
\begin{equation}
	\d(\bp_\alpha \wedge \psi^\alpha) = 
	(\dl \bp_\alpha) \wedge \psi^\alpha + (-1)^l \bp_\alpha \wedge (\dl\psi)^\alpha
\end{equation}

Next is the \emph{Bianchi identity}, which will make use of the \emph{Jacobi identity} $\wb{\wb{\lambda}{\lambda}}{\lambda} = 0$ :
\begin{equation*}
\begin{aligned}
	\dl\dll^A
		&= \d \lp \d\lambda + \frac12\wb\lambda\lambda \rp^A + \wb{\lambda}{\d\lambda + \frac12\wb\lambda\lambda}^A\\
		&= \frac12\lp \wb{\d\lambda}{\lambda} - \wb\lambda{\d\lambda} \rp^A + \wb{\lambda}{\d\lambda}^A\\
		&= 0
\end{aligned}\end{equation*}
And finally,
\begin{equation*}
\begin{aligned}
	\dl(\dl s)^\alpha
		&= (\d+\lambda\cdot) (\d s + \lambda \cdot s)^\alpha\\
		&= \lp \d(\lambda \cdot s) + \lambda\cdot \d s + \lambda\wedge\lambda \cdot s\rp^\alpha\\
		&= \lp (\d\lambda)\cdot s + \fwb\lambda\lambda \cdot s \rp^\alpha\\
		&= \dll \cdot s^\alpha
\end{aligned}\end{equation*}
Note that the two last equations can as well be written with $\d$ : 
\begin{align*}
	\d \dll &= \wb{\dll}{\lambda}	\\
	\d(\dl s)^\alpha &=	(\dll \cdot s - \lambda\cdot \dl s)^\alpha
\end{align*}

\section{Frame bundle, Ricci curvature and torsion}\label{annRic}
In this section we recall the framework of frame bundles and establish some common identities concerning the Ricci curvature in presence of torsion, which are used in Section~\ref{secno:DecompFE}.
%

\subsection{Structure of the frame bundle}\label{annFrameB}
\subsubsection*{Frame bundle, Spin structures and connections}

We start with a brief reminder on the structure of the bundle of orthonormal frames. A general reference is~\cite{KobaNomi1}. Everything done in this section applies in a straightforward manner to any metric signature and to unoriented frame, time-oriented frame and space-oriented bundles as well as to spin (and pin) frame bundles.

Start with a $n$-manifold $M$, provided with a metric of signature $(p,q)$ as well as an orientation. The bundle of orthonormal frames $\SOp(M)\overset{\pi}{\to} M$ is a principal $\SOp_{p,q}$-bundle equipped with a solder form $\alpha\in\Omega^1(\SOp(M),\setR^{p,q})$ which is $\SOp_{p,q}$-equivariant as well as horizontal (vanishes on vertical vectors). It establishes a $\SOp_{p,q}$-equivariant mapping 
\begin{center}
\begin{tikzcd}
	T\SOp(M)/V\SOp(M)\simeq \pi^*TM
		\ar[r] &
	\Rpq
	\\
	T\SOp(M)
		\ar[u]
		\ar[ur,"\alpha"] &
\end{tikzcd}
\end{center}

A \emph{$\Spinp_{p,q}$}-structure is given by a \enquote{lifting} of the so-called \emph{structure group} $\SOp_{p,q}$ to $\Spinp_{p,q}$. In other words, given a metric and a space-and-time orientation, it is defined by a principal bundle $P$ with a $\Spinp_{p,q}$-equivariant bundle map $P\to \SOp(M)$. Note that a similar lifting of the \emph{linear} frame bundle to a principal bundle with the connected double cover of $\SL_n$ as structure group induces a $\Spinp_{p,q}$-structure for every metric and space-and-time orientation. Alternatively, the mapping to $\SOp(M)$ is equivalent to the data of the solder form pulled back to $P$ : the $\Spinp_{p,q}$-structure is equivalently a principal $\Spinp_{p,q}$-bundle equipped with an nondegenerate horizontal $\Rpq$-valued $1$-form which is equivariant for the action $\Spinp_{p,q}\to \SOp_{p,q}$.

A connection 1-form on $\SOp(M)$ (hereafter \emph{connection 1-form}) is given by an $\sopq$-equivariant $\sopq$-valued 1-form $\omega$ on $\SOp(p,q)$, which is normalized for the action of $\sopq$ : for $\h\in\sopq$, writing its action by $\bh\in\Gamma(T\SOp(M))$, the normalization condition is
\begin{equation} \omega(\bh) = \h 	\label{eqno:omnormcond}\end{equation}
A connection 1-form defines an Ehresmann connection given by its kernel. The equivariance of the form ensures that the horizontal distribution is equivariant. The combined data of 
\begin{equation}
	\omega\oplus\alpha\in\Omega^1(\SOp(M),\so_{p,q}\ltimes\Rpq)
\end{equation}
is called a \emph{Cartan connection} $1$-form, or \emph{affine connection}.

\subsubsection*{Tensorial forms}

As the frame bundle trivialises the (pullback of the) tangent bundle of $M$, any tensor-valued differential form on $M$ pulls back to $\SOp(M)$ to a differential form with values in a trivialised bundle, which is \emph{horizontal} (contracts to $0$ with any vertical vector) and \emph{equivariant} (the identification $\pi^*\lp TM^{\otimes k}\otimes T^*M^{\otimes l} \rp \to \Rpq^{\otimes k}\otimes \Rpq^{*\otimes l}$ being equivariant). Such forms on $\SOp(M)$ are called \emph{basic} or \emph{tensorial} and are in bijection with forms of the corresponding type on $M$. We write 
\[
	\Omega^\bullet_h \lp \SOp(M),\, \Rpq^{\otimes k}\otimes \Rpq^{*\otimes l} \rp^{\SOpL}
\]
for the space of $\Rpq^{\otimes k}\otimes \Rpq^{*\otimes l}$-valued tensorial forms.
Equivariance under the connected group $\SOpL$ is equivalent to equivalence under $\soL$, which can be written as follows for a $\Rpq^{\otimes k}\otimes \Rpq^{*\otimes l}$-valued form $\sigma$ on $\SOp(M)$ is 
\[ (\d i_\bh + i_\bh \d) \sigma + \h\cdot \sigma = 0
\] 
for all $\h\in \so_{p,q}$. As $\sigma$ is horizontal, it can be written
\[ (i_\bh \d + \h\cdot)\sigma = 0 \]
Since $\omega$ vanishes on the horizontal directions, is normalized~(\ref{eqno:omnormcond}) and since $\alpha^a$ span the horizontal forms, the infinitesimal equivariance  can be equivalently written as
\begin{equation}\label{eqno:equivFrameB}
	\d\sigma + \omega\cdot \sigma = S_{\mathcal I}\alpha^{\mathcal I}
\end{equation}
where ${\mathcal I}$ are multi-indices, $\alpha^{\mathcal I}$ form a basis of horizontal forms and $S_{\mathcal I}$ are coefficients determined by $\sigma$.

\subsubsection*{Curvature and Torsion forms}

To a connection on $\SOp(M)$ are associated its torsion 2-form $\Theta$ (with values in $\setR^{p,q}$) and its curvature 2-form $\Omega$ (with values in $\so_{p,q})$) defined by
\begin{align}
	\Theta &:= \d\alpha + \omega \cdot \alpha\\
	\Omega &:= \d\omega + \fwb\omega\omega
\end{align}
with $\cdot$ denoting the (tensor) combined product of the wedge product on forms and the action of $\sopq$ on $\Rpq$ and $\fwb\cdot\cdot$ similarly with the adjoint action. They both are horizontal and equivariant, and as such are associated to tensor fields on $M$. Note that it has nothing to do with the Poincaré-Cartan form defined in Section~\ref{annLegTrans}.

From the curvature $2$-form is constructed the Ricci curvature form $\Ric\in \Omega^1(\SOp(M),\Rpq^*)$. It is obtained from $\Omega$ by representing $\sopq$ in $\End(\Rpq)$ then taking the trace \emph{with respect to the first 2-form index}, using the connection to identify $\Rpq$ with the horizontal space. Write \[\rho : \sopq\to\End(\Rpq)\] the natural representation of $\sopq$ and \[u : \Rpq \to \Gamma(T\SOp(M))\] the horizontal vector fields such that at each point of $\SOp(M)$ (frames of $T_x M$ for $x\in M$), $u_a$ is the horizontal lift to $T\SOp(M)$ of the vector in $T_x M$ corresponding to $e_a\in\Rpq$. In other words, $\sprod{\alpha}{u} : \Rpq \to\Gamma(\SOp(M),\Rpq)$ is the natural embedding into constant sections.

Then the Ricci curvature form can be expressed by the following formula, using $I,J$ for coordinates of $\SOp(M)$, $i$ for indices in $\sopq$ and $a,b$ for indices in $\Rpq$ :
\begin{equation}
	\Ric_{J,b} = \Omega^i_{IJ}\rho^a_{i,b} u_a^I
\end{equation}
As a tensorial $\Rpq^*$-valued $1$-form on $\SOp(M)$, it is associated to a bilinear form on $M$.

\subsection{The Bianchi identity}\label{annRicBianchi}

The curvature obeys the so-called (\emph{algebraic}, or \emph{first}) \emph{Bianchi identity} :
\begin{equation}\label{eqno:Bianchi}
	\d\Theta + \omega\cdot \Theta = \Omega\cdot\alpha
\end{equation}
We are interested in its consequences for the Ricci curvature, so we compute the contraction
\begin{equation}\begin{aligned}
	(\Omega\cdot\alpha)_{IJK}^a u_a^I
	&= (\Omega^i\rho_{i b}^a\wedge\alpha^b)_{IJK}u_a^I\\
	&= u_a^I\lp \Omega^i_{IJ}\rho_{i b}^a\wedge\alpha^b_K
	+ \Omega^i_{JK}\rho_{i b}^a\wedge\alpha^b_I
	+ \Omega^i_{KI}\rho_{i b}^a\wedge\alpha^b_J \rp\\
	&= \Ric_{J,b}\wedge\alpha^b_K
	+ \Omega^i_{JK}\rho^a_{i a}
	- \Ric_{K,b}\wedge\alpha^b_J\\
	(\Omega\cdot\alpha)_{IJK}^a u_a^I
	&=\Ric_{J,b}\wedge\alpha^b_K
	- \Ric_{K,b}\wedge\alpha^b_J
\end{aligned}\end{equation}
in which $\Omega^i\rho^a_{i a}$ vanished because $\sopq$ acts by traceless endomorphisms ; this term corresponds to the action of $\Omega$ on the determinant line bundle of $M$. Applying the same contraction to the left side of~(\ref{eqno:Bianchi}) we obtain
\begin{equation}\label{eqno:divtors}
	\lp \d\Theta + \omega\cdot \Theta\rp_{IJK}^a u_a^I
	 = \Ric_{J,b}\wedge\alpha^b_K
	 - \Ric_{K,b}\wedge\alpha^b_J
\end{equation}
which can be read as : the antisymmetric part of the Ricci curvature is equal to the covariant exterior divergence of the torsion. 

%

We want to express the left hand term with a covariant divergence, working in the vector bundle trivializing \emph{vector-valued $2$-forms}. Let us state the final formula beforehand. We write it over the base manifold, using $\tr(T)_\mu = T^\nu{}_{\nu\mu}$ and $\div^\nabla T_{\mu\nu} = \nabla_\rho T^\rho{}_{\mu\nu}$ :
\begin{equation}\label{eqno:RicBianchiBase}
	\Ric_{\mu\nu} - \Ric_{\nu\mu} = \div^\nabla(T)_{\mu\nu} - \d\tr(T)
\end{equation}

\subsubsection*{Preliminary definitions}

We need to introduce a somewhat cumbersome notation in order to differentiate between \emph{horizontal forms} and $\ExT^\bullet \Rpq^*$-valued forms on the frame bundle. This is because although they are identified by the solder form the covariant exterior differential acts differently on each of them. This distinction is harder to keep track of when working on the base manifold, but working on the frame bundle makes it clearer.

We now define the morphism relating forms of both kind. Let $E$ be an arbitrary (finite-dimensional) representation of $\sopq$. Let $\sigma$ to an equivariant horizontal $k$-form with values in $E$. We define $\hat \sigma$ the equivariant section of $\ExT^k \Rpq^*\otimes E$ defined by the isomorphism that the connection establishes between the horizontal distribution and $\Rpq$ : 
\begin{equation}
	\hat\sigma_{ab\dots k}^A := \sigma^A_{I_1 I_2\dots I_k} u^{I_1}_a u^{I_2}_b \dots u^{I_k}_k
\end{equation}

We introduce two avatars of the covariant differential : the first one is the standard covariant exterior differential
\[\dom A =\d A + \omega\cdot A\]
for $A$ an equivariant horizontal differential form with values in a representation $E$ of $\sopq$. For example the definition of the torsion can be written as :
\begin{equation}
	\Theta = \dom \alpha
\end{equation}

The second one is an antisymmetrised covariant derivative. We write it $\dom\wedge$ and \emph{only applies to equivariant sections of representations $\ExT^k \Rpq^*\otimes E$ of $\sopq$} ($0$-forms) with $E$ a representation of $\sopq$. Let $\hat \sigma$ be an equivariant section of $\ExT^k \Rpq^*\otimes E$ (corresponding to an equivariant section $\sigma\in \Omega^k(\SOp(M),E)$). 
We consider 
\[\dom \hat \sigma \in\Omega^1(\SOp(M),\ExT^k\Rpq^* \otimes E)
	\xrightarrow[\cdot\mapsto \hat{\cdot}]{\sim}
	\Gamma(\SOp(M),\Rpq^*\otimes\ExT^k\Rpq^* \otimes E)\]
and compose with the antisymmetrisation $\Rpq^*\otimes \ExT^k\Rpq^*\to \ExT^{k+1}\Rpq^*$ to obtain 
\begin{equation}
	\dom\wedge : \Gamma(\SOp(M),\ExT^k\Rpq^* \otimes E) \to \Gamma(\SOp(M),\ExT^{k+1}\Rpq^* \otimes E)
\end{equation}
We denote $\hwedge$ for the product in the exterior algebra $\ExT^* \Rpq^*$, to distinguish it from the wedge product $\wedge$ of $\Omega^*(\SOp(M))$. We also introduce the following \emph{trace} of vector-valued horizontal $k$-forms :
\begin{equation}
	\tr(\sigma) := u^I_a \sigma^a_{IJK\dots} \in\Omega^{k-1}(\SOp(M))
\end{equation}
and will also use the natural trace on $\ExT^k\Rpq^*\otimes \Rpq$. 
The contracted Bianchi identity~(\ref{eqno:divtors}) takes the form
\begin{equation}\label{eqno:trRic}
	\widehat{\tr \lp {\dom\Theta}\rp_{ab}} = \widehat\Ric_{ab} - \widehat\Ric_{ba}
\end{equation}
The last definition we need is that of the contraction of a $E$-valued $k$-form with $\Theta$. Let $A\in\Omega^k(\SOp(M),E)$. We define the following equivariant $E$-valued $(k+1)$-form
\[
	(\Theta\lr A)_{I_0\cdots I_k} := \sum_{\lel{0}{i<j}{k}} (-1)^{i+j} \Theta^b_{I_i,I_j} u^J_b A_{J,I_0,\cdots \hat{I_i}\cdots \hat{I_i} \cdots I_k}
\]
Alternatively, the contraction can be defined using the trace :
\[ \hTh\lr \hat A = \tr\lp \hTh \hwedge \hat A\rp - \tr(\hTh) \hwedge \hat A \]  

\subsubsection*{The contracted Bianchi identity}

We want to re-express $\tr\circ\dom$ in
\begin{equation}
	\tr\lp \dom \Theta \rp_{JK}
	= u^I_a \lp \d\Theta_{IJK} + \omega\cdot \Theta \rp
\end{equation}
We will need the following formula
\begin{lemma}\label{lmno:CovExtDiffTors}
	Let $A$ be a $E$-valued equivariant horizontal $k$-form. Then
	\[\widehat{\dom A} = \dom\wedge\hat A + \widehat{\Theta \lr \hat A} \]
\end{lemma}
The proof simply uses the fact that $u_a\lr \omega = 0$ and one computes the components of the exterior differential :
\begin{equation*}\begin{aligned}
	\dom A (u_{a_0},u_{a_1}\dots u_{a_k}) 
	&= \d A (u_{a_0},u_{a_1}\dots u_{a_k})\\
	&= \begin{multlined}[t]
		\sum_{i} (-1)^i u_{a_i} ( A(u_{a_0}\dots \hat u_{a_i} \dots u_{a_k} ) )\\
			+ \sum_{i<j} (-1)^{i+j}
				A([u_{a_i},u_{a_j}], u_{a_0}\dots \hat{u_{a_i}} \dots \hat{u_{a_j}} \dots u_{a_k} )
		\end{multlined}\\
	&= \begin{multlined}[t]
		\sum_{i} (-1)^i u_{a_i} ( \hat{A}_{a_0\dots \hat{a_i} \dots a_k}  )\\
		+ \sum_{i<j} (-1)^{i+j}
			A(u_b\Theta^b(u_{a_i},u_{a_j}), u_{a_0}\dots \hat{u_{a_i}} \dots \hat{u_{a_j}} \dots u_{a_k} )
		\end{multlined}\\
	&= \sum_{i} (-1)^i \dom \hat{A}_{a_0\dots \hat{a_i} \dots a_k}(u_{a_i})
	+ \sum_{i<j} (-1)^{i+j}
	\Theta^b(u_{a_i},u_{a_j}) \hat{A}_{b,a_0\dots \hat{a_i} \dots \hat{a_j} \dots {a_k}}
\end{aligned}\end{equation*}
which we write as
\begin{equation}
	\widehat{\dom A} = \dom \wedge \hat A + \hTh\lr \hat{A}
\end{equation}
We can now rewrite~(\ref{eqno:trRic}) :
\begin{equation}
	\tr(\dom \wedge \hTh + \widehat{\Theta\lr \Theta} )_{ab} = \widehat\Ric_{ab} - \widehat\Ric_{ba}
\end{equation}
We extract a divergence term with the help of the following lemma:
\begin{lemma}\label{lmno:CovExtDiv}
	Let $\hat A$ be an equivariant section of $\Rpq\otimes\ExT^k\Rpq^*$.
	Then $\tr\lp \dom\wedge \hat A \rp$ decomposes as follows
	\[\tr\lp \dom\wedge \hat A \rp = \tr \dom \hat A - \dom\wedge\tr \hat A  \]
\end{lemma}
It follows from decomposing the indices over which the trace is taken :
\begin{equation}\begin{aligned}
	\tr \lp \dom\wedge \hat A \rp_{a_1\dots a_k}
	&= \delta^{a_0}_b \lp  \sum_{i} (-1)^i \dom \hat{A}^b_{a_0\dots \hat{a_i} \dots a_k}(u_{a_i}) \rp\\
	\tr \lp \dom\wedge \hat A \rp_{a_1\dots a_k}
	&= \dom \hat A^b_{a_1\dots \hat{a_i} a_k}(u_b) - \sum_{i} (-1)^{i-1} \dom \hat{A}^b_{b a_1\dots \hat{a_i} \dots a_k}(u_{a_i})
\end{aligned}\end{equation}
Using Lemma~\ref{lmno:CovExtDiv} then once again Lemma~\ref{lmno:CovExtDiffTors}, we do the following computation
\begin{equation*}
\begin{aligned}
	\tr(\dom\wedge \hTh) 
		&= \tr \dom\hTh - \dom\wedge \tr\hTh\\
		&= \tr \dom\hTh - \lp \widehat{\dom\tr\Theta} - \widehat{\Theta\lr\tr\Theta} \rp
\end{aligned}
\end{equation*}
but $\tr\Theta$ is simply an (equivariant) $1$-form so that
\[
	\dom\tr\Theta = \d\tr\Theta
\]
and Equation~\eqref{eqno:trRic} now takes the form:
\begin{equation}
	 \lp \tr(\dom \hTh) -\widehat{\d\tr\Theta} + \widehat{\tr(\Theta\lr \Theta)} + \widehat{\Theta\lr\tr\Theta} \rp_{ab}= \widehat\Ric_{ab} - \widehat\Ric_{ba}
\end{equation}
with $\tr \dom \hTh$ corresponding to the usual covariant divergence.
%
%
Finally we prove that the quadratic term in $\Theta$ vanishes :

\begin{lemma}
	\[
		\tr(\hTh\lr\hTh) + \widehat{\Theta\lr\tr\Theta} = 0
	\]
\end{lemma}
The computation is straightforward :
\begin{equation*}
\begin{aligned}
	\tr(\hTh\lr\hTh)_{ab}
		&= \delta_e^c \lp \hTh^e_{dc} \hTh^d_{ab} + \hTh^e_{da} \hTh^d_{bc} + \hTh^e_{db} \hTh^d_{ca} \rp\\
		&= \hTh^c_{dc} \hTh^d_{ab} + \hTh^c_{da} \hTh^d_{bc} + \hTh^c_{db} \hTh^d_{ca}\\
		&= -\tr(\hTh)_d \hTh^d_{ab} + \hTh^c_{da} \hTh^d_{bc} - \hTh^d_{ca} \hTh^c_{bd}\\
		&= - \widehat{\Theta\lr\tr\hTh} + 0
\end{aligned}
\end{equation*}

Our final rewriting of~(\ref{eqno:trRic}) is
\begin{equation}\label{eqno:RicBianchiFrame}\boxed{
	\tr(\dom \hTh)_{ab} - \d(\tr\hTh)_{ab}
	= \widehat\Ric_{ab} - \widehat\Ric_{ba}	
}\end{equation}
which corresponds on $M$ to Equation~\eqref{eqno:RicBianchiBase}.

\subsection{Variation of the Ricci curvature}\label{RicContors}

We want to compare the Ricci curvature of two connections. Let $\omega$ and $\omega+\tau$ be two connection $1$-forms : $\tau\in\Omega_h^1(\SOp(M),\sopq)^{\SOp_{p,q}}$. Note their respective curvature $2$-forms $\Omega$ and $\Omega^\tau$. They are related by the following equation
\begin{equation}
	\Omega^\tau
	= \d\omega + \d\tau + \fwb\omega\omega + \wb\omega\tau + \fwb\tau\tau
	= \Omega + \dom\tau + \fwb\tau\tau
\end{equation}
Using the representation embedding : $\sopq\hookrightarrow \Rpq\otimes\Rpq^*$ we can take the trace we obtain
\begin{equation*}
	\Ric(\omega + \tau)_{J,b}
	= \rho^a_{i,b} u_a^I (\Omega + \dom\tau + \fwb\tau\tau)^i_{I,J}
	= \Ric(\omega)_{J,b} + \tr \lp \dom \tau + \fwb\tau\tau \rp_{J,b}
\end{equation*}
or in terms of $\Rpq^*\otimes\Rpq$-valued field on the frame bundle
\begin{equation}
	\widehat\Ric(\omega + \tau)
	= \widehat\Ric(\omega) + \widehat{\tr \lp \dom \tau + \fwb\tau\tau \rp} \label{VarRic}	
\end{equation}
Using the lemmas, it is also possible to reformulate it in a similar way to Equation~\eqref{eqno:RicBianchiFrame} : 
\begin{equation}
	\widehat\Ric(\omega + \tau)
	= \widehat\Ric(\omega)
		+ \tr\dom\hat\tau - \dom \widehat{\tr\tau}
		+ \tr \lp \widehat{\Theta^\omega\lr\tau} + \frac12\widehat{\wb\tau\tau} \rp
\end{equation}

\section{Maurer-Cartan forms, G-torsors and Frame bundles}\label{annMC}
In this section, we consider Lie algebra actions defined from Lie-algebra-valued differential forms, and how they can be integrated into global group actions. The manifold $X$ will be assumed connected throughout. Useful references are \cite{DiffGeoCartan} for the construction of a Lie algebra action from a $1$-form, \cite{LieActionInt,GlobalLie} for the problem of globalising a Lie algebra action into a group action and \cite{slice,CompactTransGroups} for the problem of obtaining a principal fibration over the orbit space. This is also the topic of a forthcoming paper by the author \cite{CartInt}

We start the discussion with Maurer-Cartan forms to illustrate how a $1$-form can encode the action of a Lie algebra on manifold.

\subsection{Maurer-Cartan forms}
Let $G$ a \emph{simply-connected} Lie group and $\Glie$ its Lie algebra, with underlying space isomorphic to $T_eG$. The invariant vector fields can be identified with the tangent space at the identity in two ways. The \emph{left-invariant} vector fields, which correspond to the \emph{right} action of $\Glie$ and the \emph{right-invariant} vector fields, which correspond to the \emph{left} action of $\Glie$. Writing $L_g$ and $R_g$ for the respective left and right actions of $G$, the following $T_eG$-valued $1$-forms are defined:
\begin{subequations}\begin{align}
	\omega_L(X)_{|g} &=\d R_{g^{-1}} X_{|g}\\
	\omega_R(X)_{|g} &=\d L_{g^{-1}} X_{|g}
\end{align}\end{subequations}
These forms establish isomorphisms $TG\to G\times \Glie$. We call $\omega_L$ the \emph{left Maurer-Cartan form} and $\omega_R$ the \emph{right Maurer-Cartan form} of $G$. They are subject to the following structure equations:
\begin{subequations}\label{MCeq}
\begin{align}
	\d\omega_L -\fwb{\omega_L}{\omega_L}&=0\\
	\d\omega_R +\fwb{\omega_R}{\omega_R}&=0
\end{align}\end{subequations}
with the wedge bracket defined as $\wb{\alpha}{\beta}(X,Y)=2 \lp \wb{\alpha(X)}{\beta(Y)} - \wb{\beta(X)}{\alpha(Y)} \rp$.
Writing the inversion $i:G\to G$, the two $1$-forms are related by
\begin{equation}
	 i^*\omega_R = -\omega_L
\end{equation}
By the isomorphisms $\Glie\to \Gamma(TG)$ they establish, these forms can be read as presenting the actions of the Lie algebra of $G$. Indeed Equations~(\ref{MCeq}) imply that the right action of $\Glie$ is a Lie algebra morphisms (this is artificial to the extent that the Lie algebra is often directly defined as the algebra of (left-)invariant vector fields) while the left action of $\Glie$ is a Lie algebra morphism for the opposite bracket $-[\cdot,\cdot]_\Glie$. The form $\omega_L$ is \emph{right-invariant} while $\omega_R$ is \emph{left-invariant}.

Now let $X$ be a differentiable manifold on which the Lie group $G$ acts on the right. By differentiation there is a Lie algebra action $\rho : \Glie\to \Gamma(TX)$, which we also write $\h\in\Glie\mapsto \bh\in \Gamma(TX)$. If $X$ is a \emph{principal homogeneous space}, that is to say the action is both transitive and free, then the corresponding morphism $X\times\Glie\to TX$ is an isomorphism. We write $\omega: TX\to \Glie$ its inverse: $\omega$ is a $\Glie$-valued $1$-form on $X$. Take two elements $\h_1,\h_2$ in $\Glie$ and write the compatibility to the bracket:
\begin{equation}\begin{aligned}
	\omega([\h_1,\h_2])
		&= \bh_1(\omega(\bh_2)) - \bh_2(\omega(\bh_1)) - \d\omega(\h_1,\h_2)\\
		&= \bh_1(\h_2) - \bh_2(\h_1) - \d\omega(\h_1,\h_2)\\
	\omega([\h_1,\h_2])
		&= - \d\omega(\h_1,\h_2)\end{aligned}
\end{equation}
usually written under the form
\begin{equation}
	\d\omega(\h_1,\h_2) + [\omega(\h_1),\omega(\h_2)] =0  \label{MCact}
\end{equation}
which then holds for all vectors of $X$ as the $(\bh)_{\h\in\Glie}$ generate all the vector fields over $\mathcal{C}^\infty(X)$.

Hence the \emph{Lie algebra action} data is contained in $\omega$; such a form satisfying the Maurer-Cartan equation (\ref{MCact}) is called a \emph{(right) Maurer-Cartan form (with value in $\Glie$)}. In a similar fashion one can consider left Maurer-Cartan forms, with a different sign in the Maurer-Cartan equation.

An immediate question is the following: \emph{what about a non-transitive action ?} Can we find an equivalent to the Maurer-Cartan form? The case we have in mind is that of a principal bundle. The Lie algebra action then only defines a distribution, which integrates to the orbits of the (tentative) group action. The morphism $X\times\Glie\to TX$ is not surjective. We can then ask for a section $\omega : TX\to \Glie$ that is normalised on the $\bh$ : 
\[\omega(\bh) = \h\]
Such a $\Glie$-valued $1$-form is then of maximal rank and its kernel defines a distribution which is an \emph{Ehresmann connection} in the case $X$ is the total space of a principal bundle. But $\omega$ is not sufficient to construct back the Lie algebra action  $X\times\Glie\to TX$. A solution would be to have the data of a whole coframe which includes the vector fields representing the action of $\Glie$. This is the situation we consider in the following sections.

\subsection{Cartan $1$-forms}\label{annCartanConnForm}

We are looking for a frame field on a manifold $X$ such that part of the vectors form an action of $\Glie$ on $X$. We want the bracket compatibility property to be deduced from equation on the dual coframe field which is formulated in terms of the wedge product and the exterior differential, similarly to the Maurer-Cartan Equations \eqref{MCeq}. Let us choose a specific linear representation of $\Glie$ which we will write $\setR^n$. Let $\omega\oplus\alpha$ a $\Glie\ltimes\setR^n$-valued coframe on $X$; we will also use the notation $\varpi = \omega\oplus\alpha$. For $\h\in\Glie$ and $\xi\in\Glie\ltimes \setR^n$ we write \[\bh:=\omega^{-1}(\h)\] and \[\bar\xi:=\lp \varpi \rp ^{-1}(\xi)\]
Then the following equation :
\begin{equation}\label{eqno:equivframe}
	\forall(\h,\xi)\in\Glie\times(\Glie\ltimes\setR^n),\quad 
		[\bh,\bar\xi] = \overline{[\h,\xi]}
\end{equation}
which implies that $\h\in\Glie\to\bh\in\Gamma(TX)$ is a Lie algebra action, is equivalent to the existence of (variable) coefficients $\Omega^a_{b,c}, \Omega^i_{b,c}$ such that
\begin{equation}\label{eqno:omalcomp}\begin{cases}
	\d\omega^i + \fwb{\omega}{\omega}^i &= \frac12 \Omega^i_{b,c} \alpha^b\wedge \alpha^c\\
	\d\alpha^a + \wb{\omega}{\alpha}^a &= \frac12 \Omega^a_{b,c} \alpha^b\wedge \alpha^c	
\end{cases}\end{equation}
Indeed Equations~\eqref{eqno:omalcomp} is equivalent to
\begin{equation}\label{eqno:equivcontract}
	\forall \h\in \Glie, \quad i_{\bh} \lp \d\varpi + \fwb\varpi\varpi \rp = 0
\end{equation}
hence to 
\[
	\forall (\h,\xi)\in \Glie\times(\Glie\ltimes\setR^n), \lp \d\varpi + \fwb\varpi\varpi \rp (\bh,\bar\xi) = 0
\]
To obtain~\eqref{eqno:equivframe} one simply has to compute explicitly : 
\begin{equation*}\begin{aligned}
	\d\varpi(\bh,\bar\xi) + \fwb\varpi\varpi(\bh,\bar\xi) 
	 &= \Lie_\bh(\varpi(\bar\xi)) - \Lie_{\bar\xi}(\varpi(\bh)) - \varpi([\bh,\bar\xi]) + [\varpi(\bh),\varpi(\bar\xi)]\\
	 &= \Lie_\bh(\xi) - \Lie_{\bar\xi}(\h) - \varpi([\bh,\bar\xi]) + [\h,\xi]\\
	 &= 0 - 0 - \varpi([\bh,\bar\xi]) + [\h,\xi]
\end{aligned}\end{equation*}
Let us also mention that \eqref{eqno:omalcomp} is equivalent to the equivariance of $\varpi$ under the action of $\Glie$ : the term vanishing in \eqref{eqno:equivcontract} can be written as
\[
	 i_{\bh} \lp \d\varpi + \fwb\varpi\varpi \rp
	 	= (\Lie_\bh -\d i_\bh)\varpi + 2\fwb{\varpi(\bh)}{\varpi}
	 	= \Lie_\bh \varpi -\d \h + \wb{\h}{\varpi}
	 	= \Lie_\bh \varpi - 0 + \h\cdot\varpi
\]

The prime example for $1$-forms satisfying \eqref{eqno:omalcomp} is that of \emph{Cartan connection $1$-forms} on frame bundles, defined in Appendix~\ref{annFrameB}. These are coframe fields defined on frame bundles, combining a vertical connection $1$-form and a horizontal solder form. 

A $\Glie\ltimes\setR^n$-valued coframe satisfying \eqref{eqno:omalcomp} thus defines an action of the Lie algebra $\Glie$ on $X$ but as the equation is entirely local it cannot be sufficient to encode the whole structure of a principal bundle with connection. The first step, which is tackled in the next section, is to study the conditions under which the Lie algebra action integrates to a Lie group action.

\subsection{Integrating a Lie algebra action into a group action}\label{annIntAct}

Let $X$ a connected manifold on which a Lie algebra $\Glie$ of dimension $r$ acts \emph{on the right}. $X$ is called a $\Glie$-manifold~\cite{LieActionInt}.
The vector fields representing the Lie algebra are commonly called \emph{fundamental vector fields}.
We want to integrate the action into a Lie group action. In other words, we want to define an action of a connected Lie group $G$ integrating
$\Glie$ to which the associated infinitesimal action corresponds with the existent $\Glie$ action.

A first obstruction is that the vector fields might not be complete. An immediate example is a open strict subset of $G$.
The problem will hence be to \emph{globalise} the Lie algebra action, by which we mean embedding $X$ into a larger manifold on which $G$ acts, such that the Lie algebra actions correspond.

The property identified as necessary and sufficient is the \emph{univalence} of the infinitesimal action~\cite{GlobalLie}. It roughly means that the action of an element $g$ of $G$ on a point of $X$ should not depend on the smooth path from $e$ to $g$ in $G$ used to construct it. We now define it in more detail.

Consider the product $X \times G$. The Lie algebra $\Glie$ acts \emph{on the left} on $G$ and on the right on $X$. Considering the \emph{opposite left action} on $X$, the Lie algebra left action on $X\times G$ define a Lie subalgebra of $\Glie\hookrightarrow \Gamma(T (X\times G))$ of constant rank $r$. Indeed the $T G$ component is already of rank $r$. Furthermore as a Lie subalgebra it is closed under vector bracket, which means it defines an involutive distribution. Invoking the Frobenius theorem, it integrates into a foliation the leaves of which project to $G$ by local diffeomorphisms (the differential of the projection being the parallelism $\Glie\hookrightarrow T (X\times G) \to T G$). However the projections of the leaves on $G$ do not have to be onto, nor one-to-one	. The action is called \emph{univalent} if the projection $G\times X\to G$ is injective on each leaf.

If the fundamental vector fields on $X$ are \emph{complete} then from each point $(x,g)\in X\times G$ belonging to a leaf $L$, the flows of the fundamental fields of $X\times G$ define a mapping $\Glie \to L\subset X\times G$ the image of which projects in $G$ to the neighbourhood $\exp(\Glie)g$ of $g$. Hence the image of the projection of $L$ to $G$ is invariant under multiplication by $\exp{\Glie}$ which generates the whole (connected) group $G$, and the projection has to be onto. It turns out to be a covering map~\cite{GlobalLie}. If we choose $G$ to be the simply-connected integration of $\Glie$ then each leaf has to be diffeomorphic to $G$ under the projection $X\times G\to G$. The action is then readily globalisable on $X$ to a group action : the embedding $X\simeq X\times \{e\} \hookrightarrow X \times G$ identifies $X$ as the \emph{leaf space} of $X \times G$. As the \emph{right} action of $G$ on $X\times G$ preserves the fundamental vector fields, it preserves the foliation and factors to the leaf space, identified with $X$.

Now even when the fundamental vector fields are not complete, the leaf space construction still produces a $G$-space. Only, there is no guarantee the leaf space is Hausdorff. Univalence is what ensures us that no two different elements of $X \times \{e\}$ can belong to the same leaf, so that $X$ embeds into the leaf space. The leaf space can fail to have a natural manifold structure, but is provided with a smooth structure~\cite{LieActionInt}.

Even in the case of a non-univalent action, the leaf space construction produces a $G$-space, whether the $\Glie$-manifold $X$ is complete or not. However, when forming the leaf space, the manifold $X$ will reduced to a quotient so that the action becomes univalent (effectively identifying points that are connected by a loop based at identity in $G$). This construction can be applied with any Lie integration of $\Glie$, but the larger the fundamental group is, the smaller (and possibly more singular) the quotient of $X$ will be. The leaf space of $X\times G$ is called the \emph{$G$-completion of $X$} and is proved to satisfy a universal property for $\Glie$-equivariant maps from $X$\cite{LieActionInt}.

\subsection{Integration of a Cartan connection into a frame bundle with connection}\label{annCartInt}

Assume that $X$ is provided with a nondegenerate $\spin_{p,q}\ltimes\Rpq$-valued $1$-form $\varpi = \omega\oplus\alpha$ which satisfies
\begin{equation*}
	\d\varpi + \frac12\wb{\varpi}{\varpi} = \frac12\Omega_{ab}\alpha^a\wedge\alpha^b
\end{equation*}
As justified in Section~\ref{annCartanConnForm}, it defines a right action of $\spin_{p,q}$ on $X$, for which $\varpi$ is equivariant.

\subsubsection*{Globalising the Cartan $1$-form}

We apply the construction described in Section~\ref{annIntAct} : we consider the integration $\Spinp_{1,3}$ of $\spin_{p,q}$ (simply connected for $p\geqslant3$ and $q\leqslant 1$ or the opposite condition) and we equip $X\times \Spinp_{p,q}$ with the foliation integrating the distribution spanned by $-\bh \oplus \h\in\Gamma(T(X\times \Spinp_{p,q}))$ for $\h\in\spin_{p,q}$ represented by the right-invariant vector fields. We extend the notation $\varpi$ for $\omega\oplus\alpha$ to $X\times \Spinp_{p,q}$ by $(x,g)\mapsto \Ad_g^{-1} \varpi_{|x}$. It is preserved under holonomy : for $\h\in\spin_{p,q}$, 
\begin{equation}\begin{aligned}
	\Lie_{-\bh\oplus\h} \lp \Ad_g^{-1}\varpi \rp 
		= \Lie_\h \lp \Ad_g^{-1} \rp \varpi + \Ad_g^{-1}\Lie_{-\bh}\varpi
		= \Ad_g^{-1}(-\operatorname{ad}_\h)\varpi - \Ad_g^{-1} \operatorname{ad}_{-\h} \varpi
		= 0 
\end{aligned}\end{equation}

It is also manifestly equivariant for the right action of $\Spinp_{p,q}$ on $X\times \Spinp_{p,q}$ (acting trivially on the $X$ factor); note that this extension of $\varpi$ is only possible using the group $\Spinp_{p,q}$ or $\SOp_{p,q}$ (or non-connected extensions thereof) as it requires an action on $\so_{p,q}\ltimes \Rpq$.

\emph{From now on, we require that the leaf space's smooth structure is that of a Hausdorff differentiable manifold} (more about this in \cite{CartInt}).

This hypothesis allows us to factorise $\varpi$ to the $\Spinp_{p,q}$-completion $\SpX$ of $X$ to a $\Spinp_{p,q}$-equivariant $\spin_{p,q}\ltimes\Rpq$-valued $1$-form (as the pullback to $X\times \Spinp_{p,q}$ of the tangent bundle of the leaf space identifies with the normal bundle of the foliation). As $\varpi$ is required to be nondegenerate on $TX$ which is a sub-bundle of $T(X\times \Spinp_{p,q})$ supplementary  to the tangent bundle of the foliation, $\varpi$ is nondegenerate on the normal bundle of foliation hence its factorisation to $\SpX$ is nondegenerate. We will write this factorisation $\uvpi$.

The space $\SpX$ is hence provided with an action of $\Spinp_{p,q}$ and a $\spin_{p,q}$-equivariant map from $X\simeq X\times\{e\}$ which maps $\varpi$ to $\uvpi$. As $\varpi$ and $\uvpi$ are nondegenerate, the mapping $X\to\SpX$ is a local diffeomorphism. It satisfies a universal property as a $\Spinp_{p,q}$-completion and if the action of $\spin_{p,q}$ integrates to an action of $\Spinp_{p,q}$ on $X$ then the mapping $X\to\SpX$ is a global diffeomorphism: the leaves of $X\times \Spinp_{p,q}$ \emph{are} the orbits of the action of $\Spinp_{p,q}$ and the orbit space naturally identifies with $X$.

\subsubsection*{Principal bundle structure}

We will need to make a further assumption in order to obtain a principal bundle structure : the action of $\Spinp_{p,q}$ needs to be \emph{proper}. It means that $\SpX$ has a basis of neighbourhoods $\{\mathcal U \}$ that are \emph{small}, that is~\cite{slice}
\begin{equation}
	\forall x\in\mathcal U, \exists V_x\in\mathcal V(x), \{ g\in \Spinp_{p,q} | g\mathcal V_x \cap \mathcal U \neq \emptyset
	\}\text{ is relatively compact in }\Spinp_{p,q}
\end{equation}
writing $\mathcal V(x)$ for the neighbourhoods of x. Note that in Euclidean signature the group $\Spinp_p$ is compact and this property is trivially satisfied.

Properness ensures the existence of \emph{slices} for the action of $\Spinp_{p,q}$: a \emph{slice at $x$} is a $\Stab_{\Spinp_{p,q}}(x)$-stable subset $S$ of $\SpX$ which contains $x$ such that $\Spinp_{p,q}\cdot S$ is open, $S$ is closed in $\Spinp_{p,q}\cdot S$ and we have $g\cdot S\cap S = \emptyset$ for $g\in \Spinp_{p,q}\setminus \Stab(x)$\cite{GSpaces}. Another equivalent definition is the existence of an equivariant map $\Spinp_{p,q}\cdot S \to \Spinp_{p,q} / \Stab(x)$ such that $S$ is the inverse image of the class $[\Stab(x)]$\cite{slice}. It is proved in~\cite{slice} that a proper Lie group action admits slices at each point.

The \emph{type} of an orbit is defined as its isomorphism class as a transitive $\Spinp_{p,q}$-space. It is equivalent to consider the conjugacy class of its isotropy subgroups. The existence of slices implies that the orbits of points in a neighbourhood of $x$ are of larger type than the orbit of $x$ : all $x$ in $\SpX$ admit a neighbourhood $\mathcal U_x$ such that for $y$ in $\mathcal U_x$, $\Stab_y$ is conjugated to a subgroup of $\Stab_x$. One then concludes the existence of a \emph{principal orbit type}: a maximal orbit type, of which the union of the orbits is a dense open submanifold of $\SpX$ \cite{CompactTransGroups} (in which it is stated for a compact group but only uses the existence of a slice and the compactness of isotropy subgroups).

Furthermore, $\uvpi$ identifies the tangent bundle of $\SpX$ with the trivial bundle of fibre $\spin_{p,q}\ltimes \Rpq$ in a $\Spinp_{p,q}$-equivariant way. In particular, it provides a trivialisation of the normal bundle of the orbits (the quotient of the tangent bundle by its sub-bundle generated by the fundamental vector fields of $\spin_{p,q}$). The principal orbit type is then characterised by that the fact that its isotropy subgroups act trivially on the slices generated by the flow of the vectors corresponding to $\Rpq$ starting from $x$ (this slice is called \emph{linear} as the action of $\Stab_x$ on it is diffeomorphic to its linear action on a neighbourhood of the origin). In particular, the isotropy subgroup at $x$ has a trivial action on $\Rpq\subset T_x \lp \SpX \rp$. There are only two possibilities : either $\Stab_x$ is trivial and the principal orbit type corresponds to the free action of $\Spinp_{p,q}$, or $\Stab_x$ is the $\{\pm 1\}$ subgroup and the principal orbit type corresponds to $\SOp_{p,q}$.

Note that the union of the orbits of principal type does not have to be the whole manifold. Consider for example a \enquote{higher Moebius strip} of the type $\SOp_{p,q}\times \Rpq / (\setZ/2\setZ)$ with $\setZ/2\setZ$ acting by parity simultaneously on both factors. The trivial product $\SOp_{p,q}\times \Rpq$ has an action of $\so_{p,q}$ and is provided with a tautological equivariant $\Rpq$-valued $1$-form $\alpha$ orthogonal to the $\SOp_{p,q}$ factor, which is given by the embedding
\[\SOp_{p,q}\hookrightarrow \operatorname{Hom}(\Rpq,\Rpq) \simeq \Hom(T_x\Rpq,\Rpq)\]
for all $x\in\Rpq$.
The form $\alpha$ and the action of $\so_{p,q}$ are preserved by the action of $\setZ/2\setZ$ hence we have a $\so_{p,q}\ltimes\Rpq$-valued $1$-form on the quotient, satisfying Equation~(\ref{eqno:omalcomp}) (the right-hand term is zero : there is no curvature). But as the only point of $\Rpq$ invariant under parity is its origin $0$, the orbit type of $[g,0]$ in the quotient is $\operatorname{PSO}^\uparrow_{p,q}=\SOp_{p,q}/\{\pm1\}$ while each other orbit is of type $\SOp_{p,q}$.

We conclude that over an open dense submanifold of $\SpX$, either $\SOp_{p,q}$ or $\Spinp_{p,q}$ acts freely and properly. This is enough to ensure that the orbit space has a differentiable manifold structure and the fibration is a locally trivial principal bundle (with local parametrisations given by the linear slices)~\cite{slice}. Now, the principal bundle is provided with a Cartan connection $1$-form $\uvpi$ which identifies it with a frame bundle or a spin bundle over the orbit space (as described in Section \ref{annFrameB}).

Suppose $X$ is provided with a field $\psi$ with value in a $\so_{p,q}$-module $\Sp$ and which satisfies the equivariance condition~(\ref{eqno:equivFrameB}) \[\d \psi + \varpi \cdot \psi = \varsigma_a\varpi^a\]
Then in a similar fashion to $\varpi^a$ the field $\psi$ can be extended to a field $g^{-1}\cdot \psi$ over $X\times \Spinp_{p,q}$ which is equivariant and factors to $\SpX$ to a field $\underline\psi$ which satisfies \[\d \underline\psi + \varpi \cdot \underline\psi = \underline\varsigma_a\uvpi^a\]
The section $\underline\psi$ is then $\Spinp_{p,q}$-equivariant and on the orbit subspace on which $\SpX$ is a principal bundle, $\underline \psi$ has an associated section of the associated bundle. In particular, if $\Sp$ is a \emph{spinorial module} of $\so_{p,q}$ (meaning that the action of $\Spinp_{p,q}$ does not factor to $\SOp_{p,q}$) the orbit types of the points in the support of $\underline\psi$ have to be $\Spinp_{p,q}$ (as the isotropy subgroup has to act trivially on the module). One conclude that if $\psi$ is not identically trivial (and assumed smooth), the principal orbit type is $\Spinp_{p,q}$ and the union of the principal orbits provides a $\Spinp_{p,q}$-structure to its orbit (sub)space.

\subsubsection*{Weaker orientation structures}

As we have $\SOp_{p,q}$-structures, the spacetimes we obtain are necessarily time and space-oriented. If we want to consider spacetimes with a weaker (or no) orientation, we have two options. Call $G$ the group extension of $\SOp_{p,q}$ preserving the structure we want to consider, and $\bar G$ the extension of $\Spinp_{p,q}$ (which is yet extra data~\cite{PinGroups,8Spin}). To forget the orientation data is straightforward: it is enough to extend the total space $\SpX$ for example by constructing the associated bundle (for the action of $\Spinp_{p,q}$ on $\SpX$) over the $G$ (or $\bar G$ for a spin frame bundle). It is equivalent to construct $\SpX$ by taking the non-connected group $\bar G$ instead of $\Spinp_{p,q}$. This allows us to have non-connected $\SpX$, while we initially restricted our attention to connected $X$.

To obtain genuinely non-orientable spacetimes, we have to interpret the spacetime $\ET\subseteq\SpX/\Spinp_{p,q}$ as a space-and-time orientation cover. We are then looking for quotients of $\ET$ by time and/or space parity acting on the orientation cover. The group $G$ is an semi-direct product of $\SOp_{p,q}$ with a finite group $K$, which lifts to $\bar G$ as a semi-direct product of $\Spinp_{p,q}$ with a finite group $\bar K$. To obtain a corresponding frame bundle structure, one needs to extend the free action of $\Spinp_{p,q}$ (resp. $\SOp_{p,q}$) on $\SpX$ by a free action of $\bar K$ (resp. $K$) which has suitable intertwining relations with the action of $\Spinp_{p,q}$ (resp. $\SOp_{p,q}$). For the resulting action to be considered as defining a frame bundle, $\uvpi$ has to be equivariant under the action of $K$ (resp. $\bar K$) as well. In the case the spinor field $\psi$ is nonzero, for it to live on the unoriented manifold $\ET/K$, $\psi$ has to be equivariant under $\bar K$. In this sense, unoriented spacetimes relativity theories can be interpreted as theories on oriented spacetime with extra $P$/$T$/$PT$ symmetry. Note also that the space and time-orientation cover is the natural framework to formulate non-parity-invariant and non-time reversal-invariant field theories.

\renewcommand{\l}{\lbar}
\emergencystretch=1em
\printbibliography[heading=bibintoc]

\end{document}